\documentclass[12pt]{iopart}

\pdfoutput=1
\usepackage{graphicx}
\usepackage{caption}
\usepackage{subfig}

\begin{document}

\title[Hyperviscosity and statistical equilibria of Euler turbulence]{Hyperviscosity and statistical equilibria of Euler turbulence on the torus and the sphere}

\author{W. Qi and J. B. Marston}

\address{Department of Physics, Brown University, Providence, Rhode Island 02912, USA}
\ead{\mailto{wanming@brown.edu} and \mailto{marston@brown.edu}}

\begin{abstract}
Coherent structures such as jets and vortices appear in two-dimensional (2D) turbulence. To gain insight into both numerical simulation and equilibrium statistical mechanical descriptions of 2D Euler flows, the Euler equation with added hyperviscosity is integrated forward in time on the square torus and on the sphere.  Coherent structures that form are compared against a hierarchy of truncated Miller-Robert-Sommeria equilibria. The energy-circulation-enstrophy MRS-2 description produces a complete condensation of energy to the largest scales, and in the absence of rotation correctly predicts the number and polarity of coherent vortices. Perturbative imposition of the quartic Casimir constraint improves agreement with numerical simulation by sharpening the cores and transferring some energy to smaller-scale modes. MRS-2 cannot explain qualitative changes due to rotation, but descriptions that conserve higher Casimirs beyond enstrophy have the potential to do so. The result is in agreement with the somewhat paradoxical observation that hyperviscosity helps to remedy the non-conservation of the third and higher Casimirs in numerical simulation. For a rotating sphere, numerical simulation also demonstrates that coherent structures found at late times depend on initial conditions, limiting the usefulness of statistical mechanics.

\end{abstract}


\maketitle

\section{Introduction\label{intro}}

Two-dimensional (2D) turbulence at high Reynolds numbers can organize spontaneously into large-scale and long-lived coherent structures of jets and vortices. Coherent structures are ubiquitous in nature with well-known examples such as the Great Red Spot on Jupiter, the Great Dark Spot on Neptune, jets in the atmosphere and the Gulf Stream and Kuroshio Current in the oceans. Numerical simulations and experiments show that despite the co-existing presence of complicated turbulent behavior, many features of coherent states are independent of details of the initial conditions, suggesting the possibility of interpreting coherent states as statistical equilibria. Several statistical mechanical formulations of 2D inviscid flows have been constructed that treat the fluid system either as a collection of point vortices \cite{onsager49} or more satisfactorily as a continuous field \cite{miller90,  RS91, robert91,MWC92}. See references \cite{KM80, Chavanis02, ES06, Majda, Chavanis09, BouchetReview} for reviews. 

Direct numerical simulation of high-Reynolds-number turbulence suffers from the problem that a formidably large range of length scales must be resolved -- often beyond the capacity of the most powerful supercomputers currently available. Numerical simulations that can only resolve structures that are larger than the dissipative scale typically use subgrid models of the unresolved fluid motion such as hyperviscosity \cite{Borue1995,Frisch2008}.  Equilibrium statistical mechanics offers an alternative to numerical simulation.  Among several such descriptions, the continuous-field approach proposed by Miller \cite{miller90, MWC92} and Robert and Sommeria \cite{RS91, robert91}, henceforth designated MRS, is the most satisfactory.  In this paper, we attempt to gain deeper insight into both numerical simulation and MRS descriptions of evolving 2D Euler flows by comparing coherent states found in numerical simulation with MRS equilibria. 

The MRS equilibrium statistical mechanical description is based on the observation that inviscid 2D Euler flows develop finer and finer structures as time evolves.  MRS coarse-grains the vorticity field below a small length scale $a$, which represents spatial averaging that removes fine structures that drop below resolution scale $a$ in simulations. Continuous space is discretized into small cells of side $a$, with each cell labeled by a position vector $\bi{r}$.  Only the information about the local probability distribution $\rho(\bi{r}, \sigma)$ of the vorticity $\sigma$ remains after the coarse-graining operation, and what is observed is the mean field 
\begin{eqnarray}
\bar{\omega}(\bi{r}) \equiv  \int \rho(\bi{r}, \sigma) \sigma \rmd\sigma.
\end{eqnarray}
The assumption that the system is described by one-point statistics $\rho(\bi{r}, \sigma)$ neglects any statistical correlations between the fine-grained fields in different cells. Equilibrium is determined by the least-biased local probability distribution that is consistent with the prior knowledge that the fine-grained flow is the 2D Euler flow conserving all the invariants of the dynamics. That goal is achieved by maximizing the Shannon information entropy
\begin{eqnarray}
S[\rho]=-\int \rho(\bi{r}, \sigma) \ln\rho(\bi{r}, \sigma)~ \rmd^2 \bi{r} \rmd\sigma
\end{eqnarray}
while conserving mean-field energy per unit mass $ E[\bar\omega] \equiv \frac{1}{2} \langle  \bi{u}^2 \rangle$ ($\langle \dots \rangle$ denotes a spatial average) and an infinite number of fine-grained Casimirs
\begin{eqnarray}
\Gamma_n^{f.g.} \equiv \int \overline{\omega^n} \rmd^2 \bi{r} = \int \rho(\bi{r}, \sigma)~ \sigma^n~ \rmd\sigma \rmd^2 \bi{r},\ n=1, ~2, ~3, \cdots. 
\end{eqnarray}
Here $\bi{u}(\bi{r})$ is the mean velocity field determined by the curl of the mean streamfunction $\bar\psi$. The first ($n=1$) and the second ($n=2$) Casimirs are called the circulation $\Gamma \equiv \Gamma_1^{f.g.}$ and the enstrophy respectively.  The conservation of the Casimirs reflects the invariance of 2D Euler equation under the group of area-preserving diffeomorphisms \cite{Arnold, ArnoldBook}.  The exactness of the mean-field approach to the microcanonical measure of the 2D Euler equation can be established by large-deviation theory \cite{MR94}.  The mean-field energy is exact due to the long-range nature of the interaction between vorticity at different positions (see \cite{Bouchet} and references therein).   There have also been attempts to use the MRS theory outside of the context of purely 2D flows, such as quasi-geostrophic flows (see \cite{VVG12, Herbert14} for example) or the shallow-water equations (see \cite{CS02} for example).

As explained below, MRS descriptions that impose conservation upon only a finite subset of the infinite number of Casimirs are of both theoretical and practical interest. The idea was first clearly stated by Naso \etal \cite{NCD09}. We introduce the notation MRS-N to refer to descriptions that maximize the entropy subject to a finite number of constraints, namely that all Casimirs up to $\Gamma_N^{f.g.}$ are conserved. Constraints on higher order Casimirs are not imposed. 

It is frequently hypothesized that linear invariants such as circulation and momentum, and quadratic invariants such as energy and fine-grained enstrophy, suffice to describe coherent structures in 2D high-Reynolds-number turbulence, whereas the third and higher fine-grained Casimirs are irrelevant for constraining large-scale behavior. Some earlier attempts to describe coherent structures used the spectrally-truncated Euler equation that respected only the conservation of the linear and quadratic invariants. Kraichnan constructed a statistical mechanics for such a truncated system conserving only energy and enstrophy, and found an infrared divergence in the energy spectrum for negative temperatures  \cite{Kraichnan67, Kraichnan}. That hints at the main feature: the inverse cascade \cite{batchelor} leads to the formation of coherent structures, and energy condensates into the largest scales. 
Note that for simplicity in this paper, we do not worry about distinguishing the concepts of the infrared divergence in the energy spectrum, the condensation of energy at the largest scales, and the inverse cascade of energy, because they roughly describe the same evolution picture of 2D turbulence. Similar phenomenon that energy condensates at the largest scales is also predicted using an entirely different approach, the variational principle of minimum enstrophy (ME) as developed by Bretherton and Haidvogel \cite{BH76} and Leith \cite{leith84}. The phenomenological approach minimizes enstrophy while holding energy fixed and captures the physics of the inverse energy cascade process as well as selective decay \cite{MM80}. Recent work by Naso, Chavanis and Dubrulle \cite{NCD09} demonstrated the equivalence between the two approaches: the MRS-2 description that conserves energy, circulation, and fine-grained enstrophy is equivalent to the phenomenological ME principle, when there are no other nontrivial conservation laws. The present paper considers 2D flows both on the torus and the sphere, for which in addition to energy and the Casimirs there exist further conservation laws that reflect the symmetry of the domain, namely, linear momentum on the torus and angular momentum on the sphere. Spherical geometry can lead to new physics due to the additional invariant of angular momentum and the properties of spherical harmonics; however in the past, only a few studies have addressed statistical equilibrium on the sphere \cite{FS80, lim2, VL09, Majda, HDCP12short, HDCP12, Herbert13, lim12, lim13}, while most others have focused on the planar geometry. Here by generalizing the proof of Naso \etal \cite{NCD09} to the sphere, we demonstrate that MRS-2 with these additional constraints is still equivalent to the ME principle. That MRS-2 yields the same type of solution as ME under the additional conservation of the $\bi{z}$-component and the norm of the angular momentum on the sphere was also noted recently by Herbert \cite{Herbert13}. MRS-2 thus captures the physics of the inverse energy cascade and describes a complete condensation of energy into the lowest eigenmodes of the Laplacian, in agreement with Kraichnan's energy-enstrophy theory, Lim's spherical model of energy-enstrophy-circulation theory \cite{lim1, lim2, lim3}  and also Bouchet and Corvellec's microcanonical energy-enstrophy theory \cite{Bouchet}. The inverse cascade tendency of energy can also be derived from just the conservation of energy and enstrophy of the 2D inviscid flows, but under an additional assumption that the variance of the energy spectrum will increase as nonlinear interaction continually spreads energy over a greater range of wavenumbers (for example see Pages 580 -- 581 of \cite{davidson}). This is reflected in the MRS-2 description, because both the broadening of the energy spectrum and the maximization of the mixing entropy are only different manifestations of the same irreversibility. 

Whether or not MRS-2 can also describe other features of coherent structures remains to be investigated.  The question may be framed in a different way: What are the large-scale statistical effects of imposing higher fine-grained Casimir constraints? Previous theoretical studies mostly either conserve all the Casimirs such as MRS or discard the higher Casimir constraints beyond enstrophy $\Gamma_2^{f.g.}$, partly for the sake of mathematical simplicity. Although several justifications can be made to motivate the idea of considering MRS-2 or equivalently a linear vorticity-streamfunction relationship instead of the full MRS description with various shapes of the vorticity-streamfunction curve (see \cite{HDCP12short, HDCP12} for a good summary), it is of theoretical interest to understand the information contained in the higher Casimir constraints. The idea of maximizing the entropy while conserving only a finite number of Casimirs was first clearly stated by Naso \etal, and they suggested keeping more and more fine-grained Casimirs among the constraints as a practical way to go beyond MRS-2 \cite{NCD09}. The present paper follows this approach and studies a first-order perturbation theory that weakly imposes the fine-grained cubic and quartic Casimir constraints in MRS-2. (The highest order $N$ of Casimir constraint in MRS-N has to be even for the equilibrium local vorticity probability distribution $\rho(\bi{r}, \sigma)$ to be normalizable.) Perturbative imposition of the $\Gamma_4^{f.g.}$ constraint in MRS-2 will be called perturbative MRS-4. There also exists a different perturbative approach to go beyond MRS-2, namely the strong mixing expansion where the energy constraint is weakly imposed in the MRS theory \cite{CS96}. 
The statistical importance of the cubic and quartic Casimirs in the strong mixing regime is shown in equation (A 11) of the reference \cite{CS96}.
Some numerical experiments have also addressed the effects of the cubic and quartic Casimirs \cite{AM03, DF10}.  Abramov and Majda \cite{AM03} investigated how large-scale equilibrium statistical behavior changes as the higher Casimirs are varied using a numerical algorithm that conserves many Casimirs, confirming the importance of the cubic Casimir.  They further speculated that the fourth and higher Casimirs are less important. A similar study carried out by Dubinkina and Frank \cite{DF10} used a different numerical algorithm that conserves the infinite set of Casimirs points to the relevance of both the cubic and the quartic Casimirs.  (To distinguish such conservative simulations from numerical simulation with hyperviscosity, we refer to the former as ``conservative simulation'' and the latter as ``non-conservative simulation''.)

Recently an alternative statistical mechanical formalism, the generalized-entropy description, has been proposed by Ellis \etal \cite{eht02} and further studied by Chavanis \cite{chavanis05, chavanis08} and other authors. These authors argued that for real situations with small-scale forcing and dissipation, the conservation of all the invariants of the 2D Euler equation is abusive. There are two categories of conserved quantities in MRS. The first category only receive contribution from the coarse-grained mean field, and we will refer to these as ``mean-field conserved quantities''. The kinetic energy and circulation belong to this category.  The second category of conserved quantities depend both on the coarse-grained mean field, and on the fine-grained fluctuations, and thus are affected by the small-scale processes. These will be referred to as the ``fluctuation-dependent conserved quantities'' with examples such as the fine-grained enstrophy $\Gamma_2^{f.g.}$ and the higher fine-grained Casimirs. Ellis \etal \cite{eht02} suggested conserving only the robust mean-field constraints such as energy and circulation, while treating canonically the fragile fluctuation-dependent constraints by fixing the Lagrange multipliers instead of the fine-grained Casimirs $\{ \Gamma_n^{f.g.},~ n \geq 2 \}$ themselves. Assumptions about the infinite set of multipliers may be made to reduce the complexity of the problem. Equivalently a judicious choice for the form of the prior probability distribution $\chi(\sigma)$ may be made to represent the effects of forcing and dissipation. The approach maximizes the relative entropy with respect to the prior distribution
\begin{eqnarray}
S_\chi [\rho] = -\int  \rho \ln (\rho / \chi)~ \rmd^2 \bi{r} \rmd\sigma
\end{eqnarray}
under robust mean-field constraints such as energy and circulation. This is the same as maximizing a generalized entropy in $\bar\omega$-space as proposed by Chavanis \cite{chavanis03}
\begin{eqnarray}
C_s[\bar\omega] \equiv -\int \rmd^2 \bi{r} s(\bar\omega),
\label{generalizedS}
\end{eqnarray}
while conserving the robust mean-field constraints \cite{bouchet08}. Here $s(\bar\omega)$ is a convex function determined by the prior $\chi(\sigma)$ \cite{chavanis05, chavanis08}. Bouchet and Simonnet approximated it as
\begin{eqnarray}
s(\bar\omega) = \frac{\bar\omega^2}{2} - \frac{a_4 \bar\omega^4}{4}
\label{smallS}
\end{eqnarray}
in the limit of small energy when there is no asymmetry between positive and negative vorticity.   They studied bifurcations in the flow topology in the limit of weakly forced and dissipative turbulence using a first-order perturbation theory that assumes that the parameter $a_4$ is small \cite{BS09}.  When $a_4$ vanishes, the generalized-entropy approach is the same as the ME principle. First-order perturbative MRS-4 as investigated here is the microcanonical version of this first-order perturbative generalized-entropy method. The two descriptions, when applied to 2D flows on the square torus, agree on the vorticity-streamfunction relationship but disagree on lifting of degeneracy by the first-order correction. A thorough discussion of the generalized-entropy description in the limit of small energy can be found in the reference \cite{CB12arXiv}. A similar perturbation theory has also recently been studied by Loxley and Nadiga in the context of freely-decaying turbulence \cite{LN13}.

Euler flows also serve as a test problem for the development of methods to address the problem of modeling high-Reynolds-number atmospheric and oceanic turbulence.  Such flows develop finer and finer structures by stretching and straining vortices. Vorticity filaments quickly drop below the resolution scale as time evolves. That the subgrid physics couples to the resolved structures calls for proper subgrid modeling in the numerical scheme. Hyperviscosity is the simplest and most widely used subgrid model\cite{Borue1995,Frisch2008}. Other high-wavenumber filters are also used (see \cite{HRS80} for example). These high-wavenumber dissipative models absorb enstrophy near the grid scale to mimic the loss of resolved enstrophy as enstrophy-containing filaments become too fine to resolve. They are necessary to maintain stability of numerical integration in some cases \cite{HRS80}. The main advantage of using these artificial dissipative models instead of the ordinary Newtonian viscosity is that they are more scale selective, dissipating only at the smallest scales, and thus effectively increase the operational range of resolved length scales. The inertial-range features and the large-scale structures of high-Reynolds-number simulations are insensitive to the form of the small-scale dissipation.  For simplicity the present paper focuses on hyperviscosity rather than other forms of small-scale dissipation.  Note that different powers of the Laplacian operator may be chosen: higher exponents are more scale selective, while an exponent of 1 is equivalent to ordinary Newtonian viscosity.  In the present paper we only study biharmonic dissipation (exponent 2) but we have verified that the large-scale coherent structures that form are insensitive to the precise exponent.  A comparison of the behavior of different exponents can be found in the reference \cite{cho}. 

Although hyperviscosity is widely used, it is still unclear whether it correctly models the conservative properties of the exact flow. Real-space or spectral inviscid simulations without subgrid modeling fail to conserve the third and higher Casimirs, and hyperviscosity may help restore the dynamical effects of these conservation laws.   We adopt the working hypothesis that a subgrid model that properly models the conservative properties of 2D Euler flows is equivalent to MRS coarse-graining. The coarse-grained mean field and the fine-grained fluctuations in MRS correspond to the resolved and unresolved structures in the simulations respectively.  If hyperviscosity models the correct conservative properties, and if the flow is also ergodic, coherent states produced in numerical simulation with hyperviscosity should agree with those found in MRS equilibria.  Now in the absence of hyperviscosity, an inviscid spectral simulation that fails to conserve the third and higher Casimirs may still approach energy-enstrophy equilibrium (see \cite{Shepherd87} and references therein).  If numerical simulation with hyperviscosity agrees with MRS-4 better than MRS-2, that may be taken as evidence in support of the idea that hyperviscosity helps to restore some of the conservative properties.  

The approach of comparing the simulation coherent structures with statistical mechanics encounters two practical difficulties, namely the initial value problem and the breakdown of ergodicity. Calculation of MRS-N equilibrium requires as input information the values of the conserved quantities. The initial value problem refers to the impossibility of measuring the fluctuation-dependent conserved quantities of the underlying Euler flow because the flow is only partly resolved in non-conservative simulations.  Mean-field conserved quantities such as the kinetic energy and circulation of the underlying Euler flows are directly measurable and their conservation is approximately respected in the non-conservative simulations, whereas the fluctuation-dependent conserved quantities such as the second and higher fine-grained Casimirs cannot be directly related to observation in non-conservative simulations. To determine the values of the fluctuation-dependent quantities, the corresponding coarse-grained observables $\{\Gamma_n^{c.g.} \}$ may be used as an approximation. The coarse-grained Casimirs
\begin{eqnarray}
\Gamma_n^{c.g.} \equiv \int {\bar\omega}^n \rmd^2 \bi{r} = \int \rmd^2 \bi{r} (\int \rmd\sigma ~ \rho \sigma)^n, ~ n=1, ~2, ~3, \cdots
\end{eqnarray}
are the mean-field part of the fine-grained Casimirs and can significantly differ from the fine-grained Casimirs.  How to initialize the values of fluctuation-dependent conserved quantities poses a serious problem for higher-order MRS-N descriptions.  Intuitively one expects smaller difference between fine-grained and coarse-grained quantities when there is a larger scale separation between the energy containing scale and the grid scale. Since the energy containing scale increases over time by the inverse cascade, the coarse-grained Casimirs better approximate the fine-grained ones at later times.  Brands \etal \cite{brands97} showed how MRS equilibrium varies with the integration time in one simulation of 2D evolving turbulence on the square torus. They integrated the 2D Euler equation with added viscosity forward in time, calculated the mean fields of MRS equilibria using the resolved coarse-grained vorticity fields at different integration times and compared these fields to the late-time coherent state obtained in the same simulation. They found that the MRS mean field strongly depends upon the time at which the values of the conserved quantities are measured, and the agreement becomes good after a relatively short initial period of violent mixing has passed.  Low-order MRS-N suffers less from the initial value problem than full MRS because differences between the fine-grained and coarse-grained Casimirs are amplified at higher order.  As explained below, MRS-2 is unaffected by the initial value problem because its equilibrium mean field is independent of the fine-grained enstrophy values, but MRS-4 depends on the values of $\Gamma_2^{f.g.}$ and $\Gamma_4^{f.g.}$. 

The other difficulty is the breakdown of ergodicity. To alleviate this problem, only features of coherent structures that are insensitive to details of the initial conditions are studied. Ensemble averages over various initial conditions can be used to ascertain the robust features \cite{YY93, yihy99, tyh071, tyh07}, but here we simply compare different runs without doing a formal ensemble average. The breakdown of ergodicity also indicates that the power of statistical mechanical descriptions to predict late-time quasi-stable states in numerical simulation is limited.  Real flows are inevitably subject to forcing and dissipation and are out-of-equilibrium to varying degrees. In some cases fluids may be sufficiently close to the equilibrium of a conservative system to be explained by statistical mechanics, but the question of how to best incorporate into equilibrium statistical mechanics the effects of forcing and dissipation is not resolved. Non-equilibrium statistical mechanics may be required \cite{marston11,marston12,marston13}.  We discuss this point further at the end of the paper.  

Numerical simulations of the 2D Euler equation with hyperviscosity on the square torus and the sphere (both non-rotating and rotating) are performed starting from random states with approximate symmetry between positive and negative vorticity. Numerical simulation reveals three equilibrium-like features:
\begin{enumerate}
   \item A dipole of opposite vortices is found on the torus.  On the non-rotating sphere, a quadrupole of 2 positive and 2 negative vortices appears \cite{cho}. \label{obs1}
   \item On both surfaces the radial vorticity profile of each coherent vortex, $\omega(r)$, is sharply peaked at the vortex center.  The peak is a manifestation of a nonlinear $\sinh$-like relationship between the vorticity and streamfunction. \label{obs2}
   \item On the rotating sphere, the degree of anisotropy and the most energetic wavenumbers depend on the rate of rotation \cite{YY93, cho, yihy99, tyh071, tyh07}. \label{obs3}
\end{enumerate}
Similar findings have been obtained in previous work on the torus \cite{rhines,matthaeus91,mat912, MMSMO92, MSM93, brands97} and on the sphere \cite{cho}. The above three observations may be compared with MRS-2 and perturbative MRS-4.  MRS-2 equilibria on the torus and the sphere, as solved by Herbert \cite{Herbert13}, Majda and Wang \cite{Majda}, describe the inverse energy cascade and agree with observation (\ref{obs1}) apart from this: MRS-2 permits all combinations of the lowest modes, but observation (\ref{obs1}) shows that only symmetric coherent states with equal numbers and magnitudes of positive and negative vortices appear.  We show below that MRS-4 on the torus improves agreement with numerical simulation in this regard by partly lifting the degeneracy of MRS-2. 
MRS-2 also conflicts with observation (\ref{obs2}) because the complete condensation of energy at the largest possible scales as described by MRS-2 manifests itself as a linear vorticity-streamfunction relationship, and the vortices do not have sharp cores.  The higher fine-grained Casimir constraints yield nonlinear vorticity-streamfunction relationships that are consistent with observations (\ref{obs2}) and (\ref{obs3}). First-order perturbative MRS-4 on the square torus and the non-rotating sphere produces a vorticity-streamfunction relationship that is either $\sinh$-like or $\tanh$-like depending on the values of conserved quantities. Using the values of the coarse-grained quantities at different times in a simulation, the $\sinh$-like vorticity-streamfunction relationship,  observation (\ref{obs2}), is reproduced after a short initial period of filament development and before the perturbation assumption breaks down.  Perturbative imposition of the $\Gamma_3^{f.g.}$ constraint at first order accounts for asymmetry between positive and negative vorticity that is approximately absent in the systems studied by numerical simulation here. 

To summarize: MRS-2 describes a complete condensation of energy to the largest scales, whereas the imposition of the fine-grained quartic Casimir constraint improves agreement with simulations by sharpening the cores and spreading spectral power into smaller-scale modes; MRS-2 is independent of rotation, but descriptions that further conserve higher fine-grained Casimirs are affected by rotation.  Non-conservative simulations with hyperviscosity yield the $\sinh$-like vorticity-streamfunction relationship that agrees with MRS-4, MRS \cite{brands97} and conservative simulations \cite{DF10} but disagrees with MRS-2.  In the absence of hyperviscosity, inviscid spectral simulation fails to conserve the third and higher resolved Casimirs and yields instead the linear relationship of MRS-2 equilibria \cite{Abramov,AM03}, in accord with the idea that hyperviscosity helps to restore the dynamical effects of the conservation laws lost in truncation. 

The outline of the rest of the paper is as follows. \Sref{sec2} introduces the models and methods used in numerical simulation. \Sref{sec3} compares MRS-2 against numerical simulation on the square torus and on the non-rotating sphere, showing that MRS-2 and numerical simulation yield qualitatively similar coherent structures but disagree on quantitative aspects such as the radial vorticity profiles. The question of whether or not MRS-4 can improve agreement with numerical simulation is investigated at first-order in perturbation theory at the end of \sref{sec3}.  \Sref{sec4} extends the comparison between MRS-2 and numerical simulation to rotating spheres where now qualitative discrepancies are found: MRS-2 does not depend on the rotation rate but late-time coherent structures are strongly affected by rotation. Higher-order MRS-N equilibria are altered by rotation, but comparison with numerical simulation is beyond the scope of this paper. Furthermore numerical simulation shows that the late-time structures depend on the initial conditions; the assumption of ergodicity is strongly violated on the rotating sphere.  Some discussion and open questions are presented in \Sref{sec5}. 

\section{Models and methods of numerical simulation \label{sec2}}

The 2D Euler flow with velocity field $\bi{u}(\bi{r})$ is described by the scalar vorticity field $\omega = \bi{n} \cdot (\bi{\nabla} \times \bi{u})$, where $\bi{n}$ is the unit normal vector of the surface. On the torus, the time evolution of the vorticity field is described by the equation of motion (EOM)
\begin{eqnarray}
\frac{\partial \omega}{ \partial t} + J[\psi,~ \omega]= - \nu_2 \bi{\nabla}^4 \omega,
\end{eqnarray} 
where the streamfunction $\psi(\bi{r})$ of the velocity field is determined by $\bi{u} = \bi{n} \times \bi{\nabla}\psi$.  It is related to vorticity by $\omega = \bi{\nabla}^2 \psi$, and the Jacobian or Poisson bracket operator on the torus is given by $J[A,~ B]  \equiv \frac{\partial A}{\partial x} \frac{\partial B}{\partial y} - \frac{\partial A}{\partial y} \frac{\partial B}{\partial x} $.  Subgrid physics is modeled on the right-hand side through the hyperviscosity term with small positive parameter $\nu_2$. 
The eigenmodes of the positive-definite operator $(-\bi{\nabla}^2)$ on the $2\pi \times 2\pi$ torus are plane waves $\{ e_{n_x,n_y}(x,y)\equiv  e^{i (n_x x + n_y y)}/(2\pi)\}$, where wavenumbers $n_x$ and $n_y$ are integers. The corresponding eigenvalue of each plane wave is $k^2 \equiv n_x^2 + n_y^2$.  Numerical integration of the EOM is carried out in real space using a square grid of 200 lattice points in each direction.  The lattice implementation of the operators, including the Jacobian \cite{Arakawa}, ensures conservation of both energy and enstrophy in the absence of hyperviscosity.  
A second-order accurate leapfrog algorithm with a Robert filter parameter $\alpha = 0.001$ \cite{robert66,asselin72} is used to integrate the vorticity field forward in time.  The hyperviscosity is chosen such that the fastest dissipating mode has a decay rate of $46.76$.  The unit of time can be compared to an eddy-turnover time defined as ${L}/{\sqrt{2 E(0)}}$, where $L = 2\pi$ is the length of the domain and $E(0) = \frac{1}{2} \langle \bi{u}^{2} (0)\rangle$ is the average initial energy per unit mass.

Kinetic energy contained in each $e_{n_x,n_y}(x,y)$ plane wave at time $t$ is given by
\begin{eqnarray}
E(n_x, ~n_y;~t) = \frac{1}{8 \pi^2 k^2} |\omega_{n_x,n_y}(t)|^2,
\end{eqnarray}
where $\{ \omega_{n_x,n_y} \}$ are the expansion coefficients of the instantaneous $\omega$ field on the basis of plane waves:
\begin{eqnarray}
\omega(x,y;~ t) = \sum\limits_{n_x=-\infty}^{\infty} \sum\limits_{n_y = -\infty}^{\infty}  \omega_{n_x,n_y}(t) e_{n_x,n_y}(x,y).
\end{eqnarray}
There is no $e_{00}$ component because the torus has no boundary requiring $\langle \omega \rangle = \langle \nabla^2 \psi \rangle = 0$. 
Note that $E(n_x, ~n_y;~t) =E(-n_x, -n_y;~t)$, because the real-valuedness of the $\omega$ field requires that $\omega_{n_x,n_y}^{*} = \omega_{-n_x,-n_y} $. 

On a unit sphere rotating at angular rate $\Omega$, the EOM reads instead
\begin{eqnarray}
\frac{\partial q}{ \partial t} + J[\psi,~ q] = - \nu_2 (\bi{\nabla}^2+2)\bi{\nabla}^2 \zeta
\label{sphereEOM}
\end{eqnarray}
where the Jacobian operator $J[A,~ B] \equiv \frac{1}{\sin \theta}(\frac{\partial A}{\partial \theta} \frac{\partial B}{\partial \phi} - \frac{\partial A}{\partial \phi} \frac{\partial B}{\partial \theta})$. The absolute vorticity field $q = \zeta + f$, where $\zeta$ is the relative vorticity field and $f = 2 \Omega \cos\theta$ is the planetary vorticity; again $\psi$ is the relative streamfunction and $\zeta = \bi{\nabla}^2 \psi$. The angular-momentum-preserving hyperviscosity on the right-hand side of equation \eref{sphereEOM} is a higher-order form of the Newtonian viscous term $\nu (\bi{\nabla}^2+2)\zeta$ \cite{Silberman} and models subgrid physics. For the non-rotating sphere, the absolute vorticity $q$ is the same as the relative vorticity $\zeta$, and we denote both as $\omega$, as on the torus, for simplicity. The eigenmodes of $(-\bi{\nabla}^2)$ on the sphere are spherical harmonics $\{ Y_{\ell m}\}$ with eigenvalues $\ell(\ell+1)$ where $\ell$ is the spherical wavenumber.  The simulation is performed on a spherical geodesic grid \cite{Heikes:1995p113} of $D$ cells as will be specified later; again the lattice operators conserve energy and enstrophy.  The hyperviscosity is chosen such that the most quickly dissipating mode has a decay rate of $4$. To evolve the vorticity forward in time, a second-order accurate leapfrog algorithm and a Robert filter with parameter $\alpha = 0.001$ is used. The eddy-turnover time on the sphere is defined as ${R}/{\sqrt{2 E(0)}}$, where $R=1$ is the radius of the unit sphere.   The program used for the simulations is publicly available \cite{GCM}.

Similar to the torus case, kinetic energy contained in each $Y_{\ell m}$ mode at time $t$ is given by
\begin{eqnarray}
E(\ell, ~m;~t) = \frac{1}{8 \pi \ell(\ell +1)} |\zeta_{\ell m}(t)|^2,
\end{eqnarray}
where $\{ \zeta_{\ell m} \}$ are the expansion coefficients of the instantaneous $\zeta$ field on the basis of spherical harmonics:
\begin{eqnarray}
\zeta(\theta,\phi;~ t) = \sum\limits_{\ell=1}^{\infty} \sum\limits_{m = -\ell}^{\ell} \zeta_{\ell m}(t) Y_{\ell m}(\theta,\phi).
\end{eqnarray}
Again there is no constant component $Y_{00}$ because the sphere also has no boundary.
The kinetic energy in each spherical wavenumber $\ell$ is 
\begin{eqnarray}
E(\ell;~t) = \sum\limits_{m = -\ell}^{\ell} E(\ell, ~m;~t).
\end{eqnarray}
Note that $E(\ell, ~m;~t) =E(\ell,-m;~t )$, because the reality of the $\zeta$ field requires that $\zeta_{\ell m}^{*} = (-1)^{m} \zeta_{\ell,-m} $. The energy spectrum $E(\ell, ~m;~t)$ below combines, for $m > 0$, the energies contained in each pair of $(\ell, m)$ and $(\ell, -m)$ modes. 

\section{Flows on the square torus and the non-rotating sphere\label{sec3}}

We first examine 2D inviscid turbulence on the torus and on the non-rotating sphere.  These surfaces have no boundaries, and flows conserve linear momentum on the torus, and angular momentum on the sphere.  At long times, numerical simulation produces a dipole on the torus \cite{matthaeus91,mat912}, and a quadrupole on the sphere \cite{cho}. MRS-2 qualitatively agrees because MRS-2 captures the physics of inverse energy cascade. The different coherent state has its origin in the different types of momentum that are conserved on the two surfaces. A quadrupole appears on the sphere because the conservation of angular momentum on the sphere forbids the upscale-cascading vorticity field to cascade into the three modes with spherical wavenumber $\ell = 1$ \cite{cho, Majda, Herbert13}, forcing the cascade to stop at $\ell = 2$. On the torus there is no such constraint and the lowest $k = 1$ plane wave is populated instead, yielding a dipole.  Quantitatively, however, the coherent vortices found in numerical simulation show sharper cores than MRS-2. We show below that this is connected to the linear relationship between vorticity and streamfunction that is a consequence of MRS-2; the actual relationship is closer to $\sinh$-like. First-order perturbative MRS-4 that weakly imposes the fine-grained quartic Casimir constraint in MRS-2 shows an either $\sinh$-like or $\tanh$-like vorticity-streamfunction relationship depending on the values of conserved quantities. The $\sinh$-like vorticity-streamfunction relationship is reproduced by perturbative MRS-4 if the fine-grained Casimirs are approximated using the resolved values taken after a short initial period of filament development in numerical simulation.  The relationship between vorticity and streamfunction characterizes the equilibrium state for flows on the torus and for isotropic flows with zero angular momentum on the non-rotating sphere.  We extend the relationship to non-zero angular momentum to account for the anisotropy required by the conservation of angular momentum.  A similar extension was carried out by Herbert \etal by treating the additional conservation of the $\bi{z}$-component of the angular momentum \cite{HDCP12short, HDCP12}.  

\subsection{Numerical simulation}
\begin{figure}
\centerline{\includegraphics[width=12.0 cm]{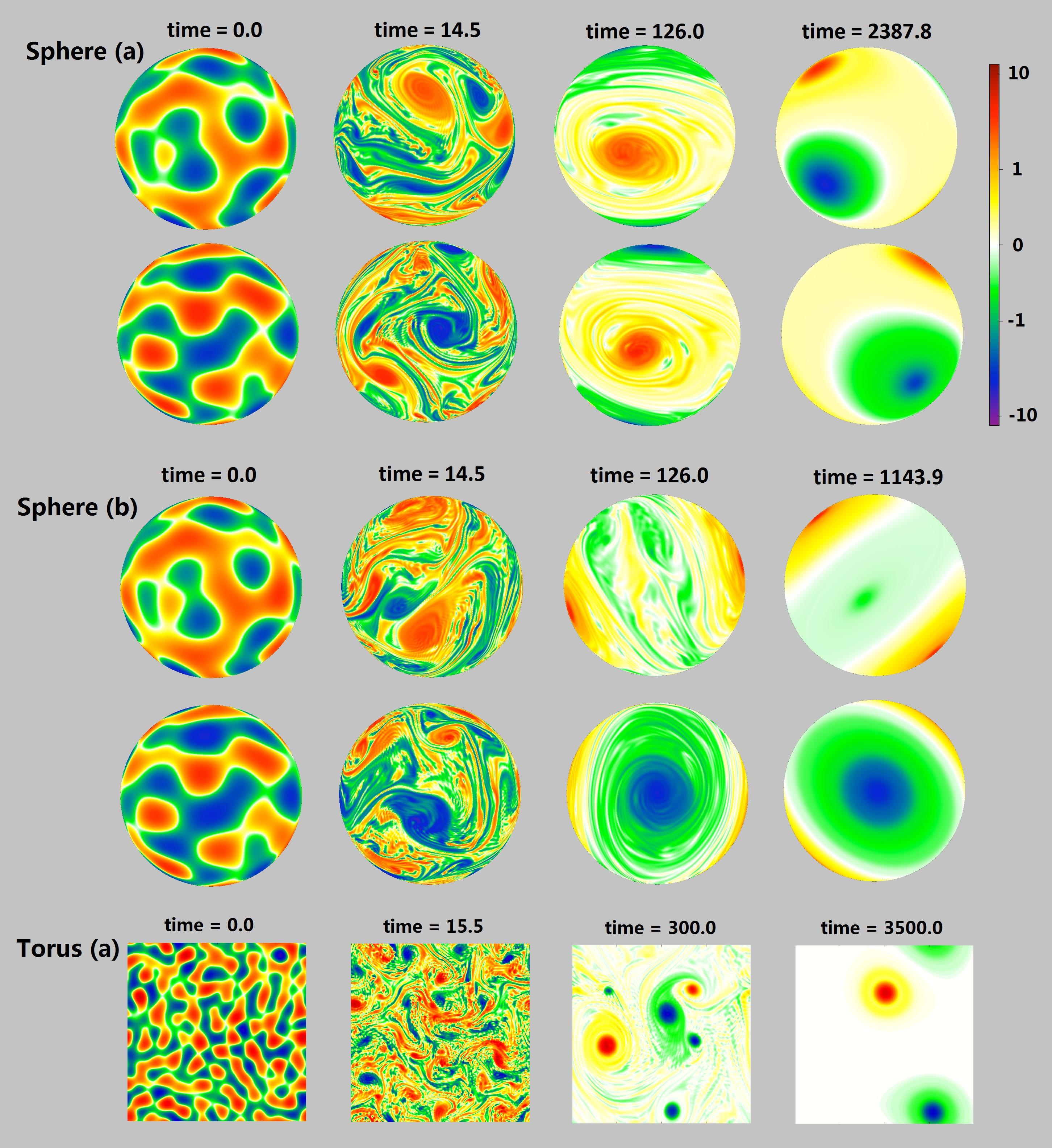}}
\caption{Numerical simulation snapshots of the vorticity field $\omega$ on the non-rotating sphere and the square torus.  Initial states, the development of filaments, and the condensation of kinetic energy into coherent structures and quasi-stable coherent states are shown. Sphere (a) has zero angular momentum, while Sphere (b) has angular momentum in the $\bi{z}$ direction. North-pole (upper row) and south-pole views (lower row) of spheres are shown. The final state of Sphere (a) is a quasi-static quadrupole; Sphere (b) is a quadrupole rotating about the $\bi{z}$ axis.  The final state of Torus (a) is a quasi-static dipole. \label{NS}}
\end{figure}

\Fref{NS} shows the time evolution of the vorticity field in three situations.  The initial vorticity in the case of Sphere (a) is a random superposition of the spherical harmonics with spherical wavenumbers between $\ell = 4$ and $10$.  The absence of $\ell = 1$ components means that total angular momentum is zero (see equation \eref{LPsi} and \ref{app2}).  The complex-valued initial amplitudes of the modes are drawn from a Gaussian distribution with zero mean.  Sphere (b), by contrast, has added net angular momentum, chosen to be along the $\bi{z}$ direction without loss of generality.  Angular momentum about the $\bi{z}$ axis is added through a non-zero amplitude for the mode with $\ell = 1$ and $m = 0$.  The third run, Torus (a), resembles that of Sphere (a) as it has the same initial energy and enstrophy per unit area $\langle \omega^2(0)\rangle$.   The initial vorticity is a superposition of plane waves with random amplitudes for square-wavenumbers, $k^2$, in the same range as $\ell (\ell + 1)$ for Sphere (a).  Simulation parameters are listed in \tref{table1}.  Since negative and positive values of initial vorticity are equally probable, the odd-order Casimirs are initially close to zero and remain small during the evolution with time. 

\Table{\label{table1}Time step $\Delta t$, total number of cells $D$, eddy-turnover time, and order of magnitude of integration time for each run that appears in \Fref{NS}.}
\br
Run&$\Delta t$&D&Eddy-turnover time&Integration time\\
\mr
Sphere (a)&0.01&163842&4.82&$10^3$\\
Sphere (b)&0.01&163842&3.48&$10^3$\\
Torus (a)&0.01&40000&30.37&$10^3$\\
\br
 \endTable

Selective decay of enstrophy at the resolved scales, with energy nearly conserved, is consistent with the ME principle. Non-zero hyperviscosity causes the resolved kinetic energy to decrease by 0.5\% for Sphere (a), 0.24\% for Sphere (b), and 2.7\% for Torus (a) over the course of the time integration. By contrast the resolved enstrophy decreases by a factor of about 6 on the sphere and 13 on the torus. The existence of an inverse energy cascade above the grid scale is readily apparent in \Fref{NS}. Energy spectra confirm that on the spheres energy condensates into the $\ell=2$ modes.  Initial energy in the $\ell = 1$ modes remains constant throughout the time evolution reflecting the conservation of angular-momentum (see \ref{app2}).  On the torus energy condensed into the $k = 1$ modes.


\subsection{MRS-2\label{MRS2nonrot}}

We turn next to MRS-2 equilibria on the torus and on the sphere.  As discussed in \sref{intro}, MRS-2 maximizes the entropy while holding fixed the energy, circulation, and fine-grained enstrophy.  Additionally, on the sphere angular momentum is held constant.  On the torus, linear momentum is instead conserved, and can be set to zero without loss of generality by a boost into an appropriate inertial frame.  The circulation vanishes on both surfaces due to the absence of any boundaries.  Because conservation of angular momentum on the sphere leads to new physics, for the remainder of this section we focus primarily on the sphere, following Herbert \cite{Herbert13}, Majda and Wang \cite{Majda}.  

\subsubsection{On the sphere}

As introduced in \sref{intro}, the MRS theory uses a coarse-grained description of the 2D Euler flows, and each macroscopic state is defined by a local probability distribution $\rho(\bi{r},\sigma)$ of finding the vorticity $\sigma$ inside the small cell of the position $\bi{r}$. What is observed at a finite resolution is the mean field $\bar\omega(\bi{r}) \equiv \int \rho(\bi{r},\sigma) \sigma \rmd \sigma$. The coarse-grained velocity field $\bar\bi{u}$ and the streamfunction $\bar\psi$ of the mean flow are related to the mean vorticity field $\bar\omega$ by $\bar{\bi{u}} = \hat{\bi{n}} \times (\bi{\nabla} \bar\psi)$ and $\bar\omega =  \bi{\nabla}^2 \bar\psi$, where $\hat{\bi{n}}$ is the unit vector normal to the surface.  The mean-field kinetic energy per unit mass is $E[\bar\omega] =- \frac{1}{2} \langle \bar\psi \bar\omega \rangle$.  The angular momentum per unit mass is $\bi{L}[\bar{\omega}] \equiv \int \rmd^2 \bi{r}  (\bi{r} \times \bi{u})$ on the unit sphere.  Expanding the streamfunction in spherical harmonics 
\begin{eqnarray}
\bar\psi = \sum\limits_{\ell,m} \bar\psi_{\ell,m} Y_{\ell,m},
\label{expandPsi}
\end{eqnarray}
is possible if the amplitudes obey
\begin{eqnarray}
\bar\psi_{\ell m} ^{*} = (-1)^m \bar\psi_{\ell, -m},
\label{reality}
\end{eqnarray}
reflecting the real-valuedness of the field.  The angular momentum is determined by the  $\ell = 1$ amplitudes by \cite{Herbert13}
\begin{eqnarray}
 L_x &=& \sqrt{\frac{8 \pi}{3}}(\bar\psi_{11} - \bar\psi_{1,  -1}), 
 \nonumber \\
 L_y &=&  i \sqrt{\frac{8 \pi}{3}}(\bar\psi_{11}+ \bar\psi_{1,  -1}),
 \nonumber \\
 L_z &=& -4 \sqrt{\frac{\pi}{3}}\bar\psi_{10}.
\label{LPsi}
 \end{eqnarray}
MRS-2 on the sphere is thus equivalent to the optimization problem
\begin{eqnarray}
\max\limits_{\rho(\bi{r}, \sigma)} \{    S[\rho]~|~E,  ~ \Gamma_2^{f.g.}, \bar\psi_{1,\pm1} = 0,\bar\psi_{10} \}
\label{problemSphere}
\end{eqnarray}
where we choose the angular momentum to be directed along the $\bi{z}$-axis without loss of generality.  
On the torus the constraints are simpler:
\begin{eqnarray}
\max\limits_{\rho(\bi{r}, \sigma)} \{    S[\rho]~|~E,~ \Gamma_2^{f.g.} \}.
\label{MRS2torus}
\end{eqnarray}
Implicit in the optimization are constraints keeping the vorticity probability distribution function non-negative and normalized at each position $\bi{r}$: 
\begin{eqnarray}
 0 &\leq& \rho(\bi{r}, \sigma),
\nonumber \\
1 &=& \int \rho(\bi{r}, \sigma) \rmd \sigma.
\end{eqnarray}

To solve MRS-2 on the sphere, namely equation \eref{problemSphere}, the angular momentum constraints on $\{ \bar\psi_{1m}, ~m=0,~\pm 1 \}$ are expressed as integrals of the distribution $\rho(\bi{r},\sigma)$ (see \ref{app2}): 
\begin{eqnarray}
-4 \sqrt{\frac{\pi}{3}} \bar\psi_{10} = L_z =  \int \rmd^2 \bi{r}  \rmd\sigma \rho \sigma \cos\theta, \\
0 = L_x = \int  \rmd^2 \bi{r}  \rmd\sigma \rho \sigma \sin\theta \cos\phi, \\
0 = L_y = \int  \rmd^2 \bi{r}  \rmd\sigma \rho \sigma \sin\theta \sin\phi.
\end{eqnarray}
The critical points of the MRS-2 variational problem on the sphere can be found by introducing the Lagrange multipliers $\{ \beta, \gamma_2,  \alpha_1, \alpha_2, \alpha_3, \xi(\bi{r}) \}$ that are real: 
\begin{eqnarray}
 0 &=& \delta S - 4 \pi \beta \delta E - \gamma_2 \delta\Gamma_2^{f.g.} - \alpha_1 \delta L_z - \alpha_2 \delta L_x-\alpha_{3} \delta L_y - \int  \rmd^2 \bi{r}  \rmd\sigma \xi(\bi{r}) \delta \rho \nonumber\\
&=& -\int  \rmd^2 \bi{r}  \rmd\sigma \delta \rho (\ln \rho+1) + \beta \int  \rmd^2 \bi{r}  \rmd\sigma \delta \rho ~\sigma \bar\psi - \gamma_2 \int  \rmd^2 \bi{r}  \rmd\sigma \delta \rho ~ \sigma^2 \nonumber\\
&& -\alpha_1 \int  \rmd^2 \bi{r}  \rmd\sigma \delta \rho ~\sigma \cos\theta - \alpha_2 \int  \rmd^2 \bi{r}  \rmd\sigma \delta\rho ~\sigma \sin\theta \cos\phi \nonumber\\
&&- \alpha_3 \int  \rmd^2 \bi{r}  \rmd\sigma \delta\rho ~\sigma \sin\theta \sin\phi  -  \int  \rmd^2 \bi{r}  \rmd\sigma \xi(\bi{r}) \delta \rho .
\end{eqnarray}
The solution is a Gaussian distribution with the normalization factor $C(\bi{r})$
\begin{eqnarray}
\rho(\bi{r}, \sigma) &=& C(\bi{r}) \times \exp[-\gamma_2 \sigma^2\nonumber\\
&& -\sigma(- \beta \bar\psi +  \alpha_1 \cos\theta + \alpha_2 \sin\theta \cos\phi + \alpha_3 \sin\theta \sin\phi)] \nonumber\\
&=& \sqrt{\frac{\gamma_2}{\pi}} \exp\{-\gamma_2 [\sigma - \frac{1}{2 \gamma_2} (\beta \bar\psi -\alpha_1 \cos\theta - \alpha_2 \sin\theta \cos\phi \nonumber\\
&& -\alpha_3 \sin\theta \sin\phi)]^2\}.
\label{MRS2Distribution}
\end{eqnarray}
The equilibrium distribution is Gaussian because the highest power of the vorticity level $\sigma$ that appears in the constraints is $2$. Generally a statistical mechanical description that conserves a set of fine-grained Casimirs $\{ \Gamma_{i}^{f.g.} \}$ on the non-rotating sphere,
\begin{eqnarray} 
\max\limits_{\rho(\bi{r}, \sigma)} \{    S[\rho]~|~E,  ~\{ \Gamma_i^{f.g.}\},~ \{ \bar\psi_{1,m} \} \},
\end{eqnarray}
yields an equilibrium distribution 
\begin{eqnarray}
\rho(\bi{r}, \sigma) &=& C(\bi{r}) \times \exp[-\sum\limits_i \gamma_i \sigma^i\nonumber\\
&& +\sigma( \beta \bar\psi -  \alpha_1 \cos\theta - \alpha_2 \sin\theta \cos\phi - \alpha_3 \sin\theta \sin\phi)],
\label{genDistri}
\end{eqnarray}
where $\{ \gamma_i \}$ are the Lagrange multipliers enforcing the conservation of $\{ \Gamma_{i}^{f.g.} \}$, and the inverse of the normalization factor
\begin{eqnarray}
 C^{-1}(\bi{r}) &=& \int_{-\infty}^{\infty}  \rmd\sigma \exp[-\sum\limits_i \gamma_i \sigma^i\nonumber\\
&& +\sigma( \beta \bar\psi -  \alpha_1 \cos\theta - \alpha_2 \sin\theta \cos\phi - \alpha_3 \sin\theta \sin\phi)]
\end{eqnarray}
is the partition function $Z(\beta \bar\psi -  \alpha_1 \cos\theta - \alpha_2 \sin\theta \cos\phi - \alpha_3 \sin\theta \sin\phi)$, where
\begin{eqnarray}
Z(u) \equiv \int  \rmd\sigma \exp[-\sum\limits_i \gamma_i \sigma^i + \sigma u].
\end{eqnarray}
The function $Z(u)$ depends on the values of the multipliers $\{ \gamma_i \}$.
Note that in the presence of higher fine-grained Casimirs $\Gamma_n^{f.g.}$ constraints with $n \geq 3$, the equilibrium distribution becomes non-Gaussian due to the $\sigma^n$ term in the exponential. 

The general distribution equation \eref{genDistri} determines a functional relationship between the mean vorticity field $\bar\omega$ and the field $ \beta \bar\psi -  \alpha_1 \cos\theta - \alpha_2 \sin\theta \cos\phi - \alpha_3 \sin\theta \sin\phi$:
\begin{eqnarray}
\bar\omega &=&\frac{ \partial \ln Z}{\partial u}(\beta \bar\psi -  \alpha_1 \cos\theta - \alpha_2 \sin\theta \cos\phi - \alpha_3 \sin\theta \sin\phi)\nonumber\\
&\equiv& \bar\omega [ \beta \bar\psi -  \alpha_1 \cos\theta - \alpha_2 \sin\theta \cos\phi - \alpha_3 \sin\theta \sin\phi].
\end{eqnarray} 
On the torus where the constraints on the angular momentum are absent 
there is a functional relationship between $\bar\omega$ and $\bar\psi$,
\begin{eqnarray}
\bar\omega &=&\frac{ \partial \ln Z}{\partial u}(\beta \bar\psi) \equiv \bar\omega[ \bar\psi],
\end{eqnarray}
where the multiplier $\beta=0$ corresponds to the trivial case with $\bar\omega(\bi{r})=0$ and is not considered. 
This vorticity-streamfunction relationship on the torus characterizes a stationary state of the Euler flows. Here the relationship between $\bar\omega$ and $ \beta \bar\psi -  \alpha_1 \cos\theta - \alpha_2 \sin\theta \cos\phi - \alpha_3 \sin\theta \sin\phi $ on the non-rotating sphere is also referred to as the vorticity-streamfunction relationship for simplicity, though $\bar\omega$ is generally not a function of $\bar\psi$ in the presence of anisotropy represented by nonzero $\{ \alpha_i  \}$. The solution is also a stationary state: it is stationary in the inertial frame of reference for the isotropic case with zero angular momentum, but is stationary in the frame rotating at a constant rate about the direction of the angular momentum for the anisotropic case. Generally in the anisotropic case where nonzero angular momentum requires some $\{ \alpha_i \}$ to be nonzero, the equilibrium solution $\bar\psi(\bi{r})$ is determined by the equation
 \begin{eqnarray}
\nabla^2 \bar\psi = \bar\omega  [ \beta \bar\psi -  \alpha_1 \cos\theta - \alpha_2 \sin\theta \cos\phi - \alpha_3 \sin\theta \sin\phi],
\label{vseqnSphere}
\end{eqnarray}
which can be rewritten as
\begin{eqnarray}
\nabla^2 \bar\psi = \bar\omega  [ \beta \bar\psi -  \bi{a}\cdot \hat\bi{r}(\theta,\phi)],
\end{eqnarray}
where the constant vector $\bi{a} \equiv (\alpha_2, \alpha_3, \alpha_1)$ in Cartesian coordinates and $\hat\bi{r}(\theta,\phi)$ is the unit radial vector at the position $(\theta,\phi)$. We can rotate the coordinate system such that the new unit vector of the $\bi{z}$ axis is $\hat\bi{z}'\equiv\bi{a}/|\bi{a}|$, and using the rotational invariance of inner products, the above equation in the new coordinate system reads
 \begin{eqnarray}
\nabla'^2 \bar\psi = \bar\omega  [ \beta \bar\psi - |\bi{a}|  \cos\theta'],
\end{eqnarray}
where the prime indicates the new $(\theta', \phi')$ spherical coordinates. This is the same type of solution, $\bar\omega=\bar\omega(\bar\psi + \Omega_L \cos\theta)$, studied by Herbert \etal, and they showed that it is stationary in a frame rotating with a constant velocity $\Omega_L$ about $\hat\bi{z}'$ \cite{ HDCP12short,HDCP12}. Thus a general type of solution equation \eref{vseqnSphere} is stationary in a frame rotating at a constant angular velocity $-\bi{a}/\beta$. The solid-body rotation accounts for the nonzero angular momentum $\bi{L}$, and obviously $-\bi{a}/\beta$ should be in the direction of $\bi{L}$; for the general case where the angular momentum is chosen to be directed along the $\bi{z}$-axis, $\alpha_2=\alpha_3=0$ whereas $\alpha_1 \neq 0$. However in the isotropic case with zero angular momentum, $SO(3)$ symmetry of the equilibrium solution $\bar\psi(\bi{r})$ imposed on the vorticity-streamfunction equation \eref{vseqnSphere} requires all $\{ \alpha_i  \}$ to vanish, and the vorticity-streamfunction relationship reduces to the relationship between $\bar\omega$ and $\bar\psi$ as on the torus. The solution of the isotropic case is thus stationary in the inertial frame of reference. The vorticity-streamfunction relationship of MRS-2 is linear as directly read off from the Gaussian distribution equation \eref{MRS2Distribution}: 
\begin{eqnarray}
\bar\omega=  \frac{1}{2 \gamma_2} (\beta \bar\psi  -\alpha_1 \cos\theta - \alpha_2 \sin\theta \cos\phi -\alpha_3 \sin\theta \sin\phi).
\label{linearvs}
\end{eqnarray}
The linear vorticity-streamfunction relationship is closely related to a Gaussian probability distribution.
The nonlinear vorticity-streamfunction relationship arises as a consequence of the non-Gaussian distribution due to higher fine-grained Casimir constraints (see \sref{pert}). As emphasized by Naso \etal, the linear vorticity-streamfunction relationship and the Gaussian distribution are key features of the MRS-2 solution; they further suggested adding more and more fine-grained Casimir constraints as a practical way to go beyond the Gaussian approximation \cite{NCD09}. 

The streamfunction $\bar\psi$ can be solved from the vorticity-streamfunction equation Eq. \eref{vseqnSphere} using the expansion equation \eref{expandPsi}. A method to solve the linear vorticity-streamfunction equation was introduced in the reference \cite{CS96}. The modes with $\ell \geq 2$ satisfy
\begin{eqnarray}
[-\ell(\ell+1)-\frac{\beta}{2\gamma_2}] \bar\psi_{\ell m} &=& 0,
\end{eqnarray} 
and only when $\beta/(2\gamma_2)= -\ell^{*}(\ell^{*}+1)$ for some $\ell^{*} \geq 2$, nonzero modes other than $\ell=1$ exist. Note that the values of $\{ \bar\psi_{1m},~m = 0,~\pm1 \}$ are specified by the angular-momentum constraint, and higher modes must exist to account for the observation that the total energy is larger than the energy contained in the $\ell =1$ modes. Therefore 
\begin{eqnarray}
\frac{\beta}{2 \gamma_2} &=& -\ell^{*} (\ell^{*} +1),
\label{betaMRS2}
\end{eqnarray}
for some $\ell^{*} \geq 2 $.
The solution of the vorticity-streamfunction equation is
\begin{eqnarray}
\bar\psi &=&\bar\psi_{10}Y_{10} + \sum\limits_{m = - \ell^{*}}^{\ell^{*}} \bar\psi_{\ell^{*},m} Y_{\ell^{*},m},
\label{psiMRS2}
\end{eqnarray}
for arbitrary complex amplitudes $\{ \bar\psi_{\ell^{*},m}, ~m =- \ell^{*}, \cdots, ~\ell^{*}  \}$ satisfying equation \eref{reality}. Projecting the vorticity-streamfunction equation to $\ell = 1$ modes determines the other multipliers to be
\begin{eqnarray}
\frac{\alpha_1}{2 \gamma_2} &=& \sqrt{\frac{3}{4 \pi}}[2-\ell^{*} (\ell^{*} +1) ] \bar\psi_{10},\label{a1MRS2}\\
\alpha_2 = \alpha_3 &=& 0.\label{a23MRS2}
\end{eqnarray}
The only undetermined multiplier is $\gamma_2$ and the unused constraints are those on energy and the fine-grained enstrophy.
The energy conservation constrains the overall magnitude of the amplitudes $\{ \bar\psi_{\ell^{*},m} \}$:
\begin{eqnarray}
E &=& \frac{1}{8 \pi} [2 \bar\psi_{10}^2 + \ell^{*}(\ell^{*}+1)\sum\limits_{m = - \ell^{*}}^{\ell^{*}} |\bar\psi_{\ell^{*},m}|^2 ].
\label{energyConstr}
\end{eqnarray}
The $\Gamma_2^{f.g.}$ constraint further requires
\begin{eqnarray}
\Gamma_2^{f.g.} &=& \int  \rmd^2 \bi{r} [ \rmd\sigma \rho (\sigma- \bar\omega)^2 + \bar\omega^2] \nonumber\\
&=& \int  \rmd^2 \bi{r} (\frac{1}{2 \gamma_2}) + \int  \rmd^2\bi{r} \bar\omega^2 \nonumber\\
&=& \frac{2 \pi}{\gamma_2} + 4  \bar\psi_{10}^2 + [\ell^{*}(\ell^{*}+1)]^2 \sum\limits_{m = - \ell^{*}}^{\ell^{*}} |\bar\psi_{\ell^{*},m}|^2 \nonumber\\
&=& \frac{2 \pi}{\gamma_2} + 4  \bar\psi_{10}^2  +  [\ell^{*}(\ell^{*}+1)] \cdot (8 \pi E - 2  \bar\psi_{10}^2 ),
\end{eqnarray}
and that fixes the multiplier $\gamma_2$ as a function of $\{\ell^{*},~ E, ~\Gamma_2^{f.g.},~ \bar\psi_{10}\}$
\begin{eqnarray}
\gamma_2 &=& \frac{2 \pi}{\Gamma_2^{f.g.} -  4 \bar\psi_{10}^2 - [\ell^{*}(\ell^{*}+1)]  (8 \pi E - 2 \bar\psi_{10}^2 )}.
\label{gamma2}
\end{eqnarray}
Note that $\gamma_2 > 0 $ as required by the normalizability of the Gaussian.

Critical points that satisfy all the constraints are described by the equations \eref{MRS2Distribution}, \eref{psiMRS2}-\eref{a23MRS2} and \eref{gamma2}, parameterized by $\ell^{*} \geq 2$ and arbitrary complex amplitudes $\{ \bar\psi_{\ell^{*},m}, ~m =- \ell^{*}, \cdots, ~\ell^{*}  \}$ that satisfy equations \eref{energyConstr} and \eref{reality}. The entropy of the Gaussian distribution equation \eref{MRS2Distribution} depends only on $\gamma_2$:
\begin{eqnarray}
S[\rho] &=& - 2 \pi \ln \gamma_2 + 2 \pi (\ln \pi +1).
\label{entropyMRS2}
\end{eqnarray}
For fixed $\{E,~ \Gamma_2^{f.g.},~\bar\psi_{10}\}$ the entropy is maximized at the smallest $\gamma_2$. Equation \eref{gamma2} combined with $8 \pi E  - 2 \bar\psi_{10}^2 = \ell^{*}(\ell^{*}+1) \sum\limits_{m = - \ell^{*}}^{\ell^{*}} |\bar\psi_{\ell^{*},m}|^2 >0$ shows that the maximum entropy has the smallest $\ell^{*} $, namely,
\begin{eqnarray}
\ell^{*} = 2.
\end{eqnarray}
Note that here for simplicity we only consider the entropy global maximum, assuming that relaxation is complete. If the entropy cannot go to positive infinity for certain distributions $\rho(\bi{r}, \sigma)$ that satisfy the constraints, then the critical point which has higher entropy value than other critical points must be the global maximum and must be locally stable. It is physically reasonable to assume that the entropy global maximum exists, so it is unnecessary to address the stability issue here by showing that the second variations are strictly negative. 
Note that owing to the equivalence of \sref{equivSec}, the above results are the same as those of section 3.2.2 of \cite{HDCP12}. Only the stability issue is different. The
authors of \cite{HDCP12short, HDCP12} argue that the $\ell^{*} = 2$ quadrupole solution  is a saddle point because it can be destabilized by a perturbation with $(l,m) = (1, \pm1)$. However, such a perturbation is forbidden because it does not conserve the $\bi{x}$- and $\bi{y}$-components of the angular momentum (a conservation law not considered in \cite{HDCP12short, HDCP12}). Therefore the quadrupole is actually stable, as also found in \cite{Herbert13}. Also note that other local maxima of entropy can be important if the system gets trapped in one of them, and as argued by Naso \etal, even saddle points can be long-lived because the system may not spontaneously generate the perturbations that destabilize them \cite{NCD09}.

Finally, replacing $\bar\psi_{10}$ with $L_z$ using equation \eref{LPsi}, the MRS-2 equilibrium on the sphere is the Gaussian distribution
\begin{eqnarray}
\rho (\bi{r}, \sigma) &=& \sqrt{\frac{\gamma_2}{\pi}} \exp\{-\gamma_2 [\sigma - \bar\omega(\bi{r})]^2\},
\end{eqnarray}
where 
\begin{eqnarray}
\gamma_2 &=& \frac{2 \pi}{\Gamma_2^{f.g.} +  3 L_z^2/(2\pi) - 48 \pi E }
\label{gamma2MRS2}
\end{eqnarray}
determines the variance of the fine-grained fluctuations, and the coarse-grained mean field is the degenerate $\ell=2$ modes plus the angular-momentum part,
\begin{eqnarray}
\bar\omega(\bi{r}, t) =- 6 \sum\limits_{m = -2}^{2} \bar\psi_{2m}(t) Y_{2m}  + \sqrt{\frac{3}{4\pi}} L_z Y_{10},
\label{MRS2SolutionNonrotSphere}
\end{eqnarray}
where the set of arbitrary complex parameters $\{ \bar\psi_{2m} \}$ satisfy equation \eref{reality} and the energy constraint
\begin{eqnarray}
  E = \frac{3}{64 \pi^2} L_z^2 +  \frac{3}{4 \pi}\sum\limits_{m=-2}^{2} |\bar\psi_{2m}|^2.
\end{eqnarray}
The equilibrium observable $\bar\omega(\bi{r},t)$ depends on the energy and angular momentum, but not on the fine-grained enstrophy $\Gamma_2^{f.g.}$. The value of $\Gamma_2^{f.g.}$ only affects the variance of the unresolved fluctuations. Therefore, the MRS-2 has no initial-value problem. 

The equilibrium coarse-grained vorticity field $\bar\omega(\bi{r},t)$ is static in the absence of angular momentum and is generally a quadrupole, though if it is a pure $Y_{20}$ state or a rotation of it, there are instead two same-signed vortices each covering a hemisphere (see \fref{theoryFig}). Specifically the pure $Y_{20}$ state has a zonal flow pattern where jets move in opposite directions in the northern and southern hemispheres and the velocity vanishes at the equator. A rotation-invariant dimensionless quantity $\eta[\bar\psi]$ defined as
\begin{eqnarray}
\eta[\bar\psi] &\equiv & F^2[\bar\psi]/G^3[\bar\psi],\label{etaPsi}\\
F[\bar\psi] &\equiv& - 2 \bar\psi_{20}^3 +6 \bar\psi_{20} (\bar\psi_{2,-1} \bar\psi_{21}+2\bar\psi_{2,-2} \bar\psi_{22}) \nonumber\\
&&- 3 \sqrt{6} (\bar\psi_{2,-2} \bar\psi_{21}^2 + \bar\psi_{2,-1}^2 \bar\psi_{22}), \label{FPsi}\\
G[\bar\psi] &\equiv& \sum\limits_{m=-2}^{2} |\bar\psi_{2m}|^2, \label{GPsi}
\end{eqnarray}
ranges from $0$ to $4$ and characterizes the shape of the $\ell = 2$ configurations. There are only two vortices in the extreme limit $\eta = 4$.  When $\eta = 0$ there are four vortices of equal magnitude, a ``symmetric quadrupole" state. As $\eta$ increases from $0$ to $4$, one pair of same-signed vortices gradually dominates over the other pair, and the sign of $F[\bar\psi]$ is that of the dominant vortex pair.  For $L_z \neq 0$, a $Y_{10}$ component that corresponds to a solid-body rotation is superposed. The overall configuration of vorticity undergoes a solid-body rotation at angular frequency $L_z/(4 \pi)$ about the $\bi{z}$-axis. The rotation period of the MRS-2 vorticity field for the case of Sphere (b) is about $T = 38.6$, consistent with the estimate of $39$ seen in the simulation coherent state. 

\subsubsection{On the torus}

\begin{figure}
\centerline{\includegraphics[width=10.0 cm]{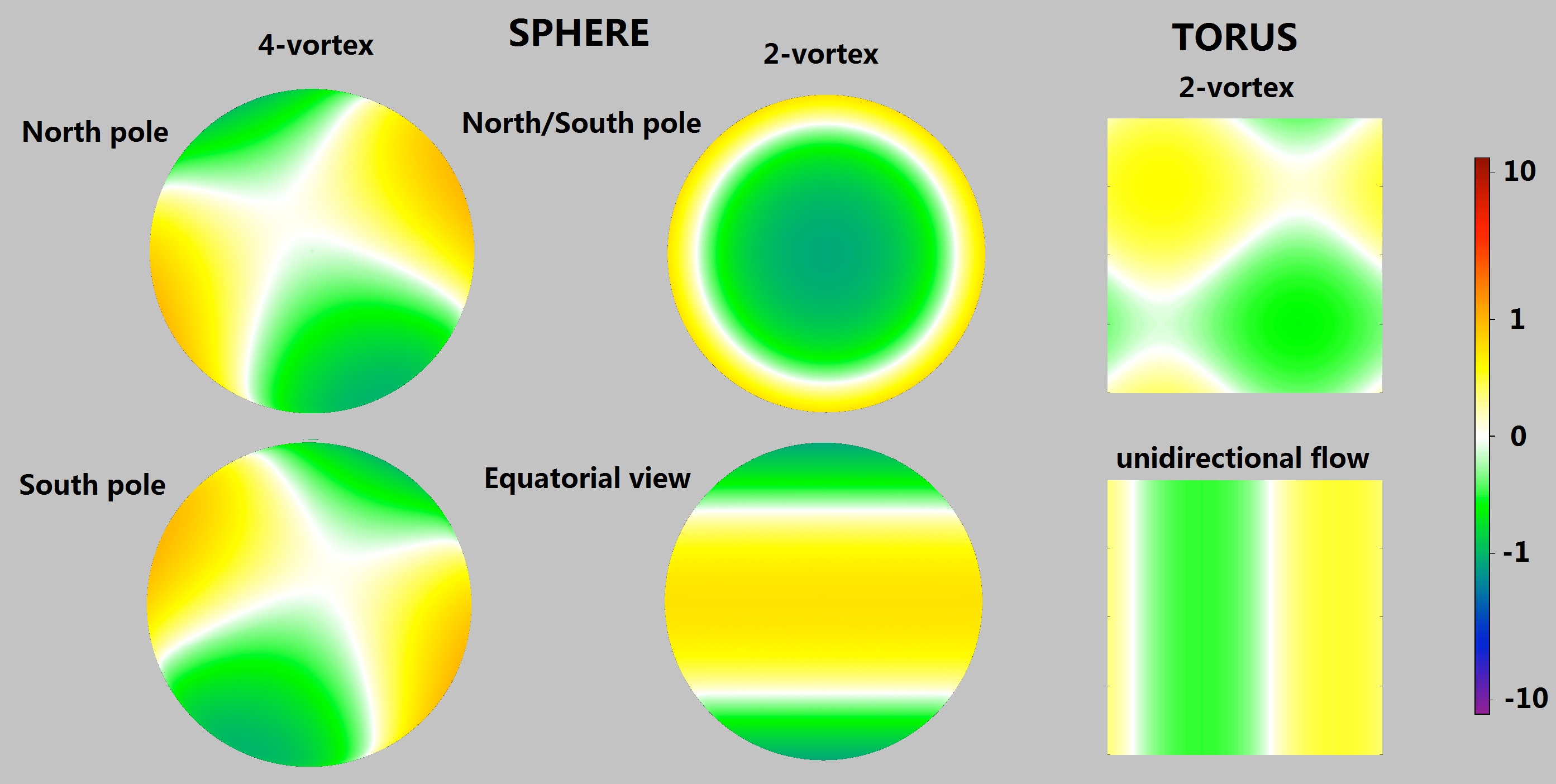}}
\caption{Typical MRS-2 coarse-grained vorticity fields $\bar\omega$ for Sphere (a) and Torus (a) based upon the initial energies of the corresponding numerical simulation. The equilibrium configuration for Sphere (a) is generally a static quadrupole but as a special case a two-vortex state can occur.  On Torus (a) the coarse-grained vorticity is generally a static two-vortex state but unidirectional flow is also a possibility. \label{theoryFig}}
\end{figure}

On the torus the streamfunction $\bar\psi$ may be expanded in the basis of plane waves
\begin{eqnarray}
 \{ e_{n_x, n_y} (x,y)\equiv \frac{1}{2\pi}e^{i(n_x x +n_y y)}, ~n_x, ~n_y =0, \pm 1, \pm2, \cdots  \}.
\label{torusBasis}
\end{eqnarray}
 The equilibrium state is degenerate in the four lowest modes with $(n_x, n_y)=(0, \pm 1), (\pm 1, 0)$,
\begin{eqnarray}
\bar\omega(x,y) &=& -\bar\psi_{10} \cdot e_{10} (x) - \bar\psi_{01} \cdot e_{01} (y) + c.c.,
\label{torusMRS2sol}
\end{eqnarray}
where $c.c.$ represents the complex conjugate terms and $\{ \bar\psi_{10}, \bar\psi_{01} \}$ are arbitrary complex amplitudes constrained in the overall magnitude by energy. The solution is static and generally is a dipole, though for the special cases of amplitudes with $\bar\psi_{10} = 0$ or $\bar\psi_{01} = 0$, the flow is unidirectional with two jets of opposite directions (see \fref{theoryFig}). The special dipole case with symmetry between $\bi{x}$ and $\bi{y}$ directions, $|\bar\psi_{10}| = |\bar\psi_{01}|$, is denoted as ``symmetric dipole". The possibility of either dipoles or unidirectional flows was noted in a generalized-entropy description \cite{BS09}. 

\subsubsection{Equivalence with the ME principle \label{equivSec}}

All the equilibrium mean-field solutions show the condensation of energy at the largest possible scales, similar to the phenomenological ME principle.  Reference \cite{NCD09} showed that the MRS-2 optimization problem of equation \eref{MRS2torus} is equivalent to the ME principle at the coarse-grained level: 
\begin{eqnarray}
\min\limits_{\bar\omega(\bi{r})} \{  \Gamma_2^{c.g.} ~[\bar\omega]~| ~E, ~\Gamma \}.
\label{ME-torus}
\end{eqnarray}
The equivalence can be generalized to the sphere if the constraint on the angular momentum is included:
\begin{eqnarray}
\min\limits_{\bar\omega(\bi{r})} \{  \Gamma_2^{c.g.} ~[\bar\omega]~| ~E, ~\Gamma, ~\bi{L}   \}.
\label{ME-sphere}
\end{eqnarray}
Again the constraints of circulation $\Gamma$ in equations \eref{ME-torus} and \eref{ME-sphere} are trivial. To show the equivalence on the sphere, following the approach in reference \cite{NCD09}, the mean field of the solution $\rho(\bi{r},\sigma)$ to the MRS-2 variational problem equation \eref{problemSphere} is found in two steps. First, impose an additional constraint that requires the local vorticity distribution $\rho(\bi{r},\sigma)$ to have a specific mean $\bar\omega(\bi{r})$ and the variational problem becomes
\begin{eqnarray}
S^{*}[\bar\omega] = \max\limits_{\rho(\bi{r}, \sigma)} \{    S[\rho]~|~E,  ~ \Gamma_2^{f.g.}, \bi{L}, ~ \bar\omega(\bi{r})\}.
\label{newVar}
\end{eqnarray}
The extremal entropy $S^{*} [\bar\omega]$ for the new problem is found.  As the second step, vary over all possible mean vorticity fields $\bar\omega(\bi{r})$ to maximize $S^{*} [\bar\omega]$
\begin{eqnarray}
\max\limits_{\bar\omega(\bi{r})} \{  S^{*} [\bar\omega]~| ~E, ~ \Gamma_2^{f.g.}, \bi{L}\},
\label{newVar2}
\end{eqnarray}
and find $\bar\omega(\bi{r})$ that is the mean field of the distribution $\rho(\bi{r},\sigma)$ that maximizes the entropy thus solving the MRS-2 problem. The constraints on $E$ and $\bi{L}$ in equation \eref{newVar} can be absorbed into the $\bar\omega(\bi{r})$ constraint because energy and angular momentum only depend on the mean field $\bar{\omega}(\bi{r})$, not the fine-grained fluctuations. The equation \eref{newVar} also appears in reference \cite{NCD09}
\begin{eqnarray}
S^{*}[\bar\omega] =  \max\limits_{\rho(\bi{r}, \sigma)} \{    S[\rho]~|~ \Gamma_2^{f.g.},\bar\omega (\bi{r}) \}.
\end{eqnarray} 
These authors showed that the extremal entropy $S^{*}[\bar\omega]$ is a monotonically increasing function of $(\Gamma_2^{f.g.}-\Gamma_2^{c.g.})$.  Equation \eref{newVar2} can be rewritten as
\begin{eqnarray}
\max\limits_{\bar\omega(\bi{r})} \{ \Gamma_2^{f.g.}-\Gamma_2^{c.g.}~| ~E, ~ \Gamma_2^{f.g.}, \bi{L}\},
\end{eqnarray}
or equivalently,
\begin{eqnarray}
\min\limits_{\bar\omega(\bi{r})} \{ \Gamma_2^{c.g.}~| ~E, ~ \Gamma_2^{f.g.}, \bi{L}\}.
\label{newVar3}
\end{eqnarray}
The fine-grained enstrophy constraint $\Gamma_2^{f.g}$ in equation \eref{newVar3} is irrelevant because it only serves to constrain the fine-grained fluctuations for any mean field $\bar\omega(\bi{r})$. Thus equation \eref{newVar3} coincides with the ME problem, equation \eref{ME-sphere}, completing the proof. That the same vorticity-streamfunction relationship obtains from MRS-2 and from ME under the additional constraints of just the $\bi{z}$-component and the norm of the angular momentum, instead of all the three components as considered here, was mentioned by Herbert \cite{Herbert13}.  

\subsection{Quantitative difference between MRS-2 and numerical simulation \label{MRS2NSNonrot}}

\begin{figure}
\centerline{\includegraphics[width=14.0 cm]{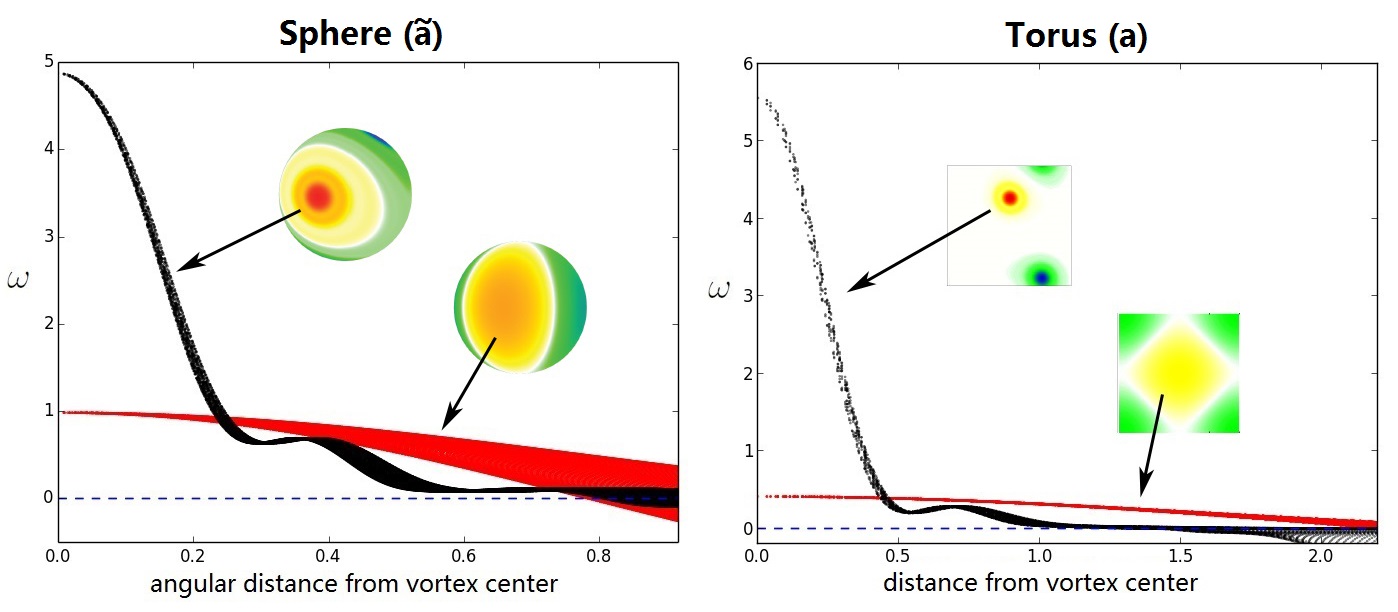}}
\caption{Scatter plots of the radial vorticity profiles $\omega(r)$ of a typical positive vortex obtained from numerical simulation and in MRS-2.  Black: numerical simulation of Sphere ($\tilde{a}$) (similar to Sphere (a)) and Torus (a).  Red:  MRS-2 symmetric quadrupole and symmetric dipole solutions based upon the initial energies of the corresponding numerical simulations.
\label{vortexFig}}
\end{figure}

\begin{figure}
\centerline{\includegraphics[width=14.0 cm]{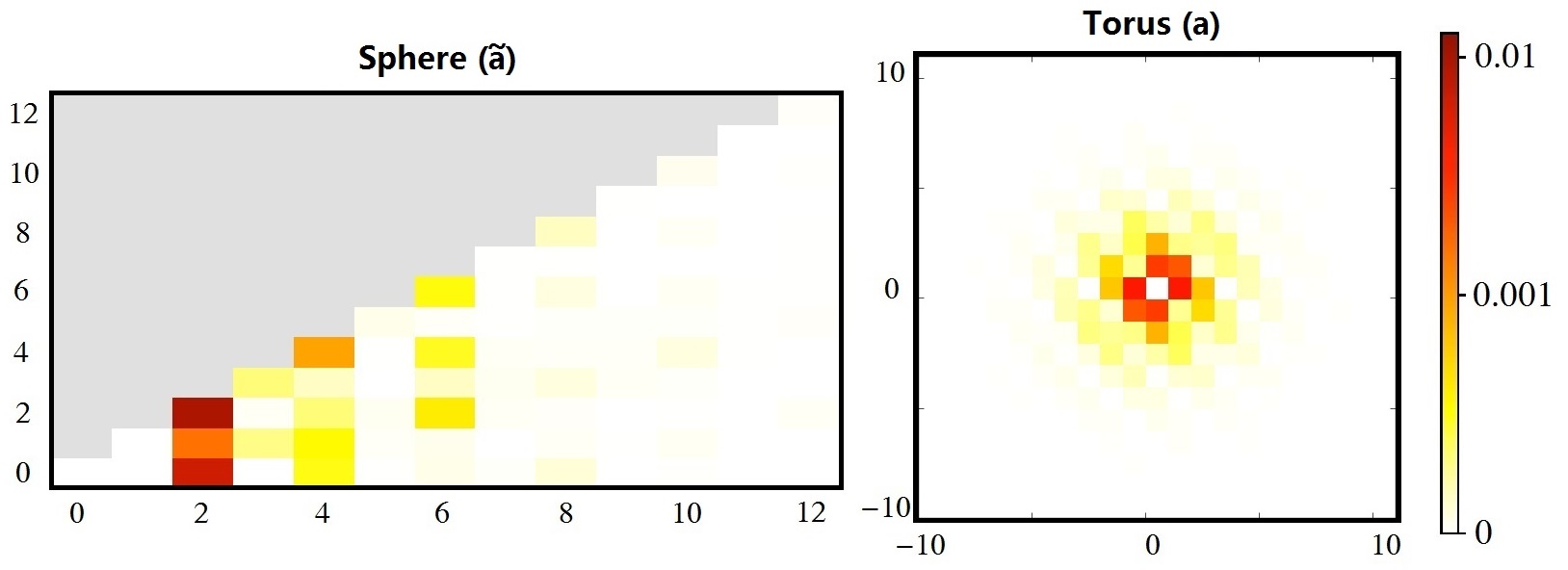}}
\caption{Energy spectrum $E(\ell,m)$ for the coherent state of Sphere ($\tilde{a}$) and $E(n_x, n_y)$ for that of Torus (a). For Sphere ($\tilde{a}$), spherical wavenumber $\ell$ is plotted along the horizontal axis and the azimuthal wavenumber $m$ is along the vertical. For Torus (a), the $\bi{x}$-direction wavenumber $n_x$ is along the horizontal and the $\bi{y}$-direction wavenumber $n_y$ is along the vertical.  The presence of $\ell>2$ modes on the sphere and $k>1$ modes on the torus is readily apparent.
\label{spectraNonrotFig}}
\end{figure}

\begin{figure}
\centerline{\includegraphics[width=14.0 cm]{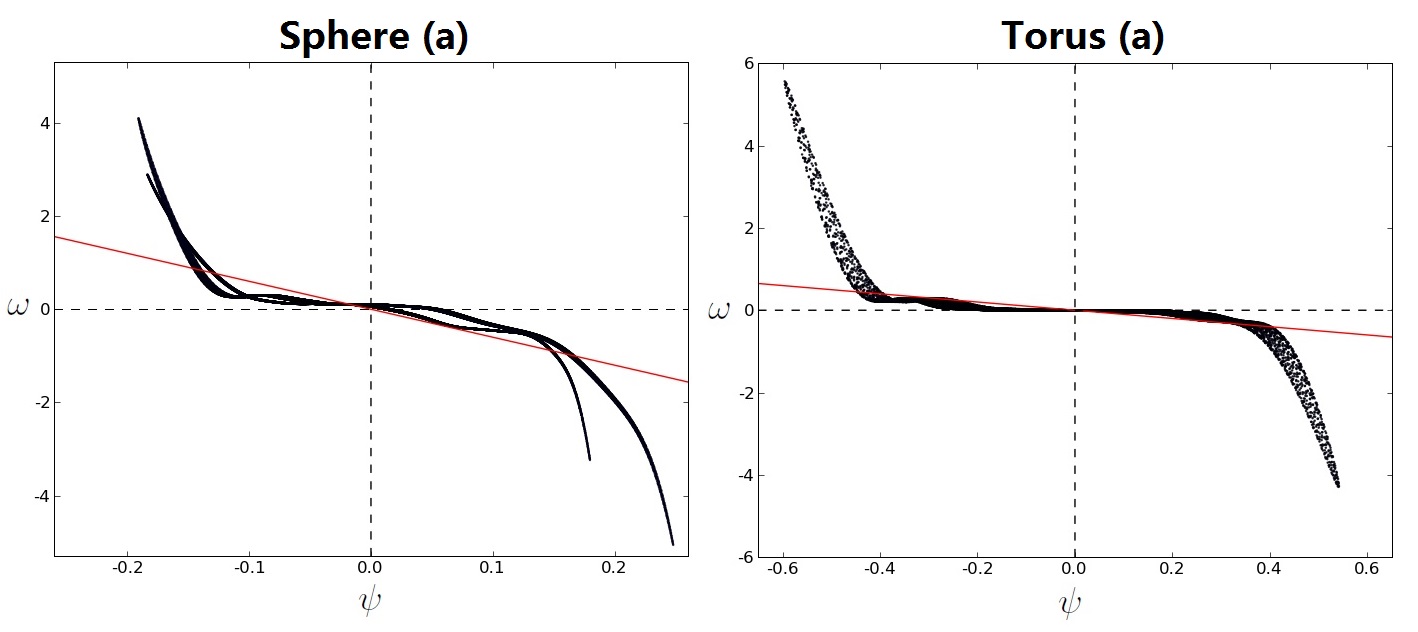}}
\caption{Scatter plots of vorticity $\omega$ versus streamfunction $\psi$ for the coherent states of Sphere (a) and Torus (a) found by numerical simulation.  Corresponding contour plots are shown in \fref{NS}. The red straight line represents MRS-2 equilibria.
\label{omegaPsiFig}}
\end{figure}

\begin{figure}
\centerline{\includegraphics[width=14.0 cm]{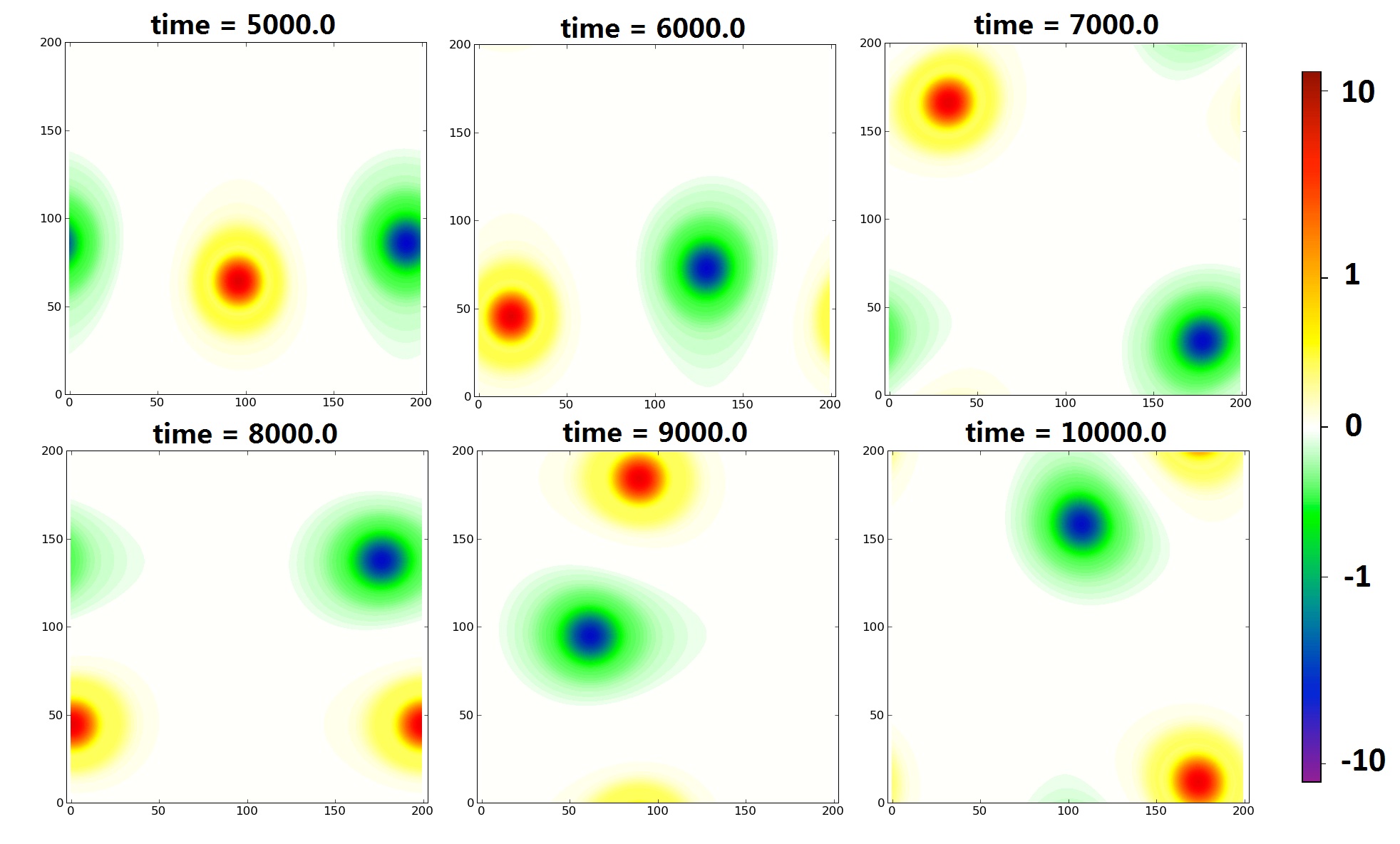}}
\caption{Numerical simulation snapshots of vorticity field $\omega(\bi{r})$ for Torus (a) show that the radial vorticity profile $\omega(r)$ of the coherent vortices peaks sharply at the core in contrast to what is found at MRS-2 equilibria. The coherent structures are stable over long periods of time.\label{long-time}}
\end{figure}

Coherent structures found in numerical simulations on the non-rotating sphere and on the square torus agree qualitatively with MRS-2 equilibria. Quantitatively numerical simulation at long times leads to symmetric quadrupole and symmetric dipole configurations rather than other degenerate states permitted by MRS-2. Moreover the vortex cores of simulation coherent states are much sharper than those of MRS-2 symmetric quadrupole and symmetric dipole equilibria. \Fref{vortexFig} shows scatter plots of vorticity versus the distance from a positive vortex center for both the simulation coherent states and MRS-2 symmetric equilibria. The black markers correspond to the numerical simulations Sphere ($\tilde{a})$ and Torus (a). Sphere ($\tilde{a}$) is a similar run to Sphere (a): its initial state is a different random superposition of the same wavenumbers constrained by the same total energy. The red markers are the MRS-2 symmetric equilibrium configurations based on the same initial energies of the corresponding numerical simulations. The flat peaks of MRS-2 vorticity profiles contrast with the sharp peaks of numerical simulations. The sharp vortex cores have modes with wavenumbers higher than those at MRS-2 equilibria. The corresponding energy spectra (\fref{spectraNonrotFig}) clearly show the existence of $\ell>2$ modes on the sphere and $k>1$ modes on the torus. The feature of sharp cores is also related to the shape of the vorticity-streamfunction relationship. The vorticity-streamfunction relationship for the coherent state in the absence of angular momentum has a $\sinh$-like shape that disagrees with the linear relationship of MRS-2. \Fref{omegaPsiFig} shows scatter plots of vorticity versus streamfunction for the coherent states of Sphere (a) and Torus (a). The corresponding contour plots are shown in \Fref{NS}.  The $\sinh$-like shape \footnote{The curve can be $\pm \sinh$-like due to the sign convention adopted for streamfunction $\psi$, and both are denoted as $\sinh$-like for simplicity. Likewise the ``$\tanh$-like" vorticity-streamfunction relationship as will be mentioned later can refer to $\pm \tanh$-like.} observed in numerical simulation contrasts with the straight line of MRS-2 equilibria. That the scatter plot for Sphere (a) shows two branches is related to the dynamically-trapped asymmetry between the two same-signed coherent vortices: for example, one of the negative vortex is much weaker than the other one. The asymmetry indicates that the coherent structures still retain some memory of the details of the initial states; it cannot be related to equilibrium features. The scatter plots also show that the vorticity along each streamline of the fluid is approximately single-valued. Upon reaching such a state, the nonlinear advection term in the EOM becomes small, energy redistribution among different scales due to nonlinear interaction has almost stopped, and the structure decays linearly under hyperviscosity. Thus the energy in the higher-wavenumber modes will never completely go to the lowest modes to agree with MRS-2. This is confirmed by extending the integration time of Torus (a): \fref{long-time} shows that during the long time period from $t = 3500$ to $t = 10~000$, the coherent vortices drift slowly around but the shape of the radial vorticity profile $\omega(r)$ maintains the same sharp peak.

\begin{figure}
\centerline{\includegraphics[width=15.0 cm]{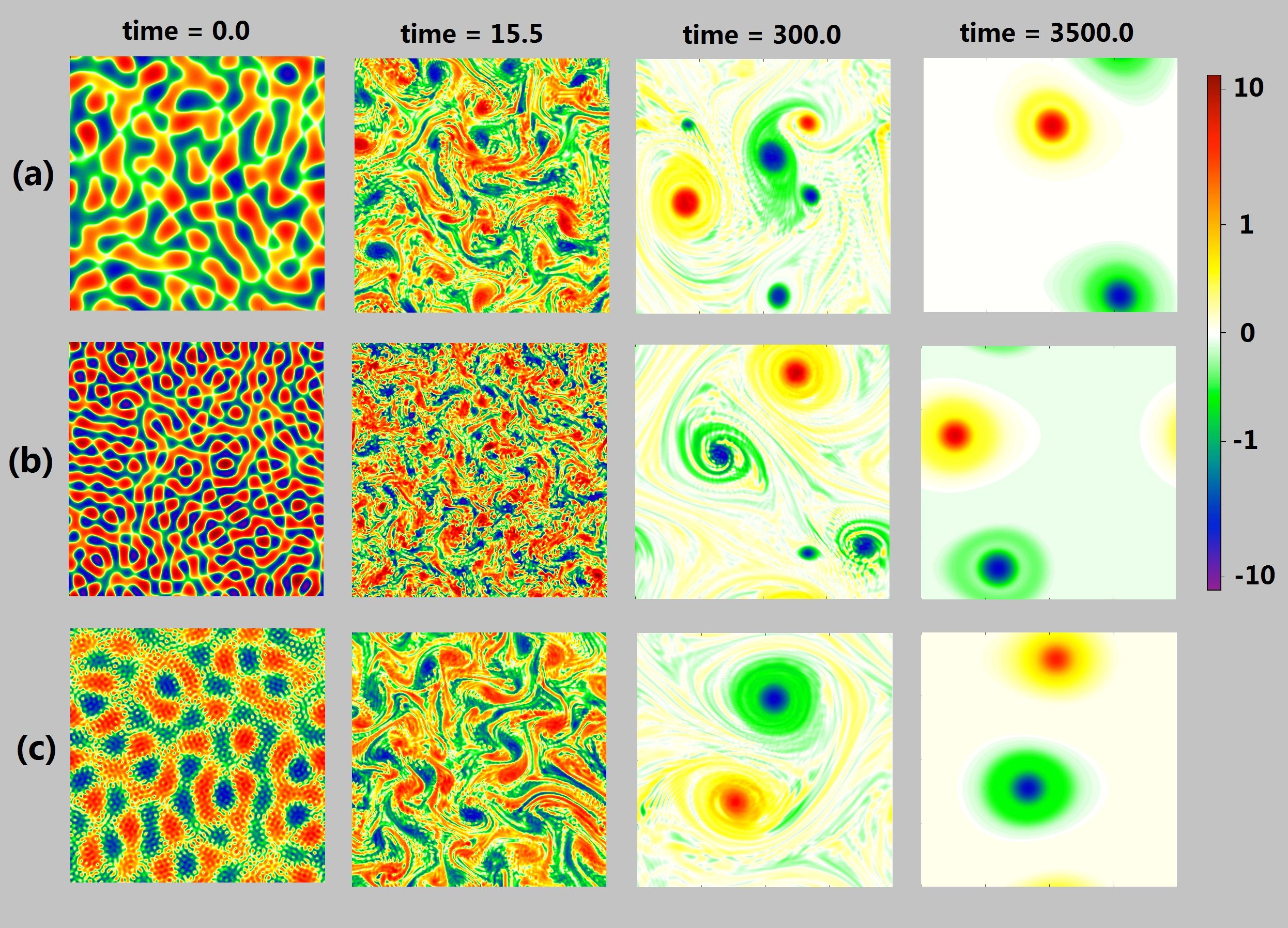}}
\caption{Same as \fref{NS} but for three different initial states on the square torus, namely Torus (a), (b) and (c). The initial states have random amplitudes over the different ranges of wavenumbers  listed in \tref{table2}, with distinct initial values for the resolved Casimirs.  When the resolved odd-order Casimirs are initially close to zero, the dipole coherent structures and radial vorticity profiles $\omega(r)$ are qualitatively insensitive to changes in the initial even resolved Casimirs. \label{highK}}
\end{figure}

\begin{figure}
\centerline{\includegraphics[width=10.0 cm]{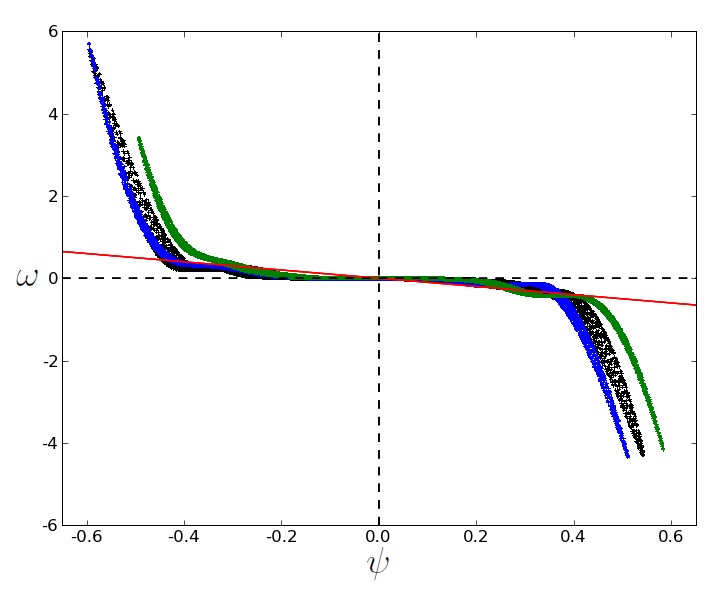}}
\caption{Scatter plots of the vorticity $\omega$ versus the streamfunction $\psi$ for the coherent states of Torus (a), (b) and (c).  Black points are for Torus (a), blue for Torus (b) and green are for Torus (c).  Corresponding contour plots are shown in \fref{highK}. The red straight line represents MRS-2 equilibrium.  Note that the three vorticity-streamfunction relationships are similar.\label{omegaPsiFigTorus}}
\end{figure}

The shape of the radial vorticity profile $\omega(r)$ and the vorticity-streamfunction relationship for the coherent structure are insensitive to changes in the initial resolved Casimirs, when the odd-order resolved Casimirs are close to zero initially. This contrasts with the findings in conservative simulations that $\omega(r)$ and $\omega$-$\psi$ relationship vary with initial Casimirs \cite{AM03, DF10}. There is no contradiction because the resolved Casimirs in non-conservative simulations are not the exact Casimirs of the underlying Euler flows.  Here numerical simulation for the torus is performed again but with two different random initial states. The initial states of the three runs have the same energy but different ranges of wavenumbers that lead to distinct resolved Casimirs. The initial state of Torus (b) contains plane waves whose square-wavenumbers $k^2$ are about ten times those of Torus (a). The effective wavenumber, estimated as $\sqrt{\langle \omega^2 \rangle/(2 E)}$ using energy per unit mass and the normalized enstrophy, is almost doubled compared to that of Torus (a). The enstrophy is increased by a factor of $3.8$, the quartic Casimir by a factor of about $13$, sixth by a factor of about $44$, and higher even-order Casimirs are even more drastically increased. The initial state of Torus (c) is chosen such that lower Casimirs remain close to Torus (a) whereas the higher Casimirs are significantly changed, by adding a tiny component with very high wavenumbers to a low-wavenumber background. Compared with Torus (a), the initial enstrophy of Torus (c) is unchanged, the quartic Casimir is only changed by $9.3 \%$ and sixth by $23\%$, but the high-wavenumber component largely increases the high even-order Casimirs. Details of the initial states are listed in \tref{table2}. \Fref{highK} shows the snapshots of the vorticity field at different times. Torus (a) is included for comparison. The vorticity profiles $\omega(r)$ for coherent vortices among the three systems have a similar sharp peak at the core different from MRS-2 equilibria. \Fref{omegaPsiFigTorus} shows that the three corresponding vorticity-streamfunction curves are almost the same. The vorticity-streamfunction curve can be translated along the $\psi$-axis without changing the physics, because any constant can be added to the definition of $\psi$ without affecting the vorticity field. The arbitrary constant in $\psi$ is chosen such that the spatial mean of $\psi$ vanishes. The small relative translation among the three curves is thus a result of the asymmetry between positive and negative vortices.  Apart from an irrelevant translation, the three curves are almost the same.  Initially distinct Casimirs also become closer in value at later times as shown in \tref{table2}. Similar phenomena are also observed on the non-rotating sphere with zero initial angular momentum.

\Table{Range of square-wavenumber $k^2$ in the initial state, energy per unit mass $E$ and normalized enstrophy $\langle \omega^2 \rangle$ at the initial ($t = 0$) 
and final ($t = 3500$) times for each run.}
\br
Run&Range of $k^2$ & $E(0)$ & $\langle \omega^2 \rangle (0)$ & $E(3500)$ & $\langle \omega^2 \rangle (3500)$\\
\mr
Torus (a) & $[20,110]$ & 0.0214 & 2.4869 & 0.0208 & 0.19\\
Torus (b) & $[200,250]$ & 0.0214 & 9.4124 & 0.0197 & 0.17\\
Torus (c) & $[30,50]$ and $[2460,2500]$ & 0.0214 & 2.4869 & 0.0209 & 0.13\\
\br
\label{table2} 
\endTable

The $\sinh$-like vorticity-streamfunction relationship obtained from numerical simulation differs from the linear relationship found at MRS-2 equilibria. The result supports the idea that hyperviscosity as a subgrid model helps to restore the conservative properties of 2D Euler flows that are lost in spatial discretization. Without hyperviscosity, inviscid truncated spectral simulation on the torus conserves energy and enstrophy, but not the third and higher Casimirs. It robustly approaches the energy-enstrophy equilibrium (see \cite{Shepherd87} and references therein) and produces the linear vorticity-streamfunction relationship of MRS-2 equilibria \cite{AM03,Abramov}. By contrast, the $\sinh$-like vorticity-streamfunction relationship is a consequence of the third and higher Casimir constraints. Conservative simulation of Euler flows found the $\sinh$-like vorticity-streamfunction relationship if the conserved global vorticity distribution has zero skewness but nonzero kurtosis \cite{DF10}; statistical mechanical descriptions that respect all conservation laws of the 2D Euler flows confirm this \cite{DF10, brands97}. The non-conservative simulation with hyperviscosity starts from random initial conditions with approximate symmetry between positive and negative vorticity, so the modeled underlying Euler flow is likely to have approximately zero skewness, but nothing constrains its kurtosis to be zero. Thus the observed equilibrium $\sinh$-like vorticity-streamfunction relationship is likely to be in agreement with inviscid Euler flows that conserve all the infinite Casimirs, suggesting that the use of hyperviscosity as subgrid model helps remedy the problem of non-conservation of higher Casimirs in the non-conservative simulations. 

\subsection{Perturbative MRS-4\label{pert}}

To investigate whether or not further imposing the fine-grained quartic Casimir constraint in MRS-2 improves agreement with simulation coherent structures by reproducing the observed quantitative features, we consider the case of Sphere (a),
\begin{eqnarray}
\max\limits_{\rho(\bi{r}, \sigma)} \{    S[\rho]~|~E,  ~ \Gamma_2^{f.g.}, ~ \Gamma_4^{f.g.},\bar\psi_{1,\pm1} = \bar\psi_{10}=0 \},
\label{MRS4}
\end{eqnarray}
where the vorticity distribution $\rho(\bi{r}, \sigma)$ is non-negative and normalized at each position $\bi{r}$. The calculation is similar to that in \sref{MRS2nonrot}.
The equilibrium solution is a non-Gaussian local vorticity distribution
\begin{eqnarray}
\rho (\bi{r}, \sigma) =&& C(\bi{r}) \times \exp[-\gamma_4 \sigma^4  - \gamma_2 \sigma^2\nonumber\\
&& - \sigma (-\beta \bar{\psi} + \alpha_1 \cos\theta + \alpha_2 \sin \theta \cos \phi + \alpha_3 \sin \theta \sin \phi)], 
\label{nonGaussian}
\end{eqnarray}  
where $C(\bi{r})$ is the normalization factor and $\gamma_4$ is the Lagrange multiplier that enforces the $\Gamma_4^{f.g.}$ constraint. The vorticity-streamfunction equation
\begin{eqnarray}
\bar{\omega}(\bi{r})= \int  \rmd\sigma \rho \sigma = \bi{\nabla}^2 \bar\psi
\label{eqn}
\end{eqnarray}
has to be solved to express $\bar{\psi}$ in terms of multipliers. However the mean of a non-Gaussian distribution cannot be found analytically and the problem is thus impossible to solve exactly. Instead, we investigate a perturbative approach by making the assumption that the distribution is close to the Gaussian distribution $\rho_0 (\bi{r},\sigma)$ of MRS-2 as given in equation \eref{MRS2Distribution}.
The non-Gaussian equilibrium solution \Eref{nonGaussian} can be rewritten as
\begin{eqnarray}
\rho (\bi{r}, \sigma) = C_1(\bi{r}) \cdot \rho_0(\bi{r}, \sigma) \times \exp(-\gamma_4 \sigma^4),
\label{nonGaussian2}
\end{eqnarray}
where $C_1(\bi{r})$ is a normalization factor.  If $|\gamma_4|$ is so small that $\gamma_4 \sigma^4 $ is small in a region around origin where the Gaussian has significant size, then the condition
\begin{eqnarray}
\int _{-\infty}^{\infty}  \rmd\sigma \rho_0(\bi{r},\sigma)\gamma_4 \sigma^4  \ll 1
\label{condition}
\end{eqnarray}
 is satisfied. In that case the factor $\exp(-\gamma_4 \sigma^4 )$ in equation \eref{nonGaussian2} can be expanded:
\begin{eqnarray}
\rho (\bi{r}, \sigma) &=& C_1(\bi{r}) \cdot \rho_0(\bi{r}, \sigma) \times (1-\gamma_4 \sigma^4) + {\cal O}(\gamma_4^2),
\end{eqnarray}
and equation \eref{eqn} and constraints can be expressed in terms of the moments of the Gaussian. The normalization of $\rho (\bi{r}, \sigma)$ up to the first order gives $C_1(\bi{r}) = 1+\gamma_4 \cdot m_4 (\bi{r})$, where $m_4(\bi{r})\equiv \int  \rmd\sigma \rho_0 (\bi{r}, \sigma) \cdot\sigma^4$ is the fourth-order moment of the Gaussian. Thus the perturbation expansion of the local vorticity distribution is
\begin{eqnarray}
\rho (\bi{r}, \sigma) &=& \rho_0(\bi{r}, \sigma) \times \{1+\gamma_4 [m_4(\bi{r}) - \sigma^4]\}+ {\cal O}(\gamma_4^2). \label{expand}
\end{eqnarray}
Note that the perturbation assumption equation \eref{condition} may not be satisfied in many situations, but is useful in studying how adding the $\Gamma_4^{f.g.}$ constraint in MRS-2 changes the equilibrium features.  If $\gamma_4 = 0$ it reduces to MRS-2.  Perturbative MRS-4 investigated here imposes the fine-grained enstrophy and fine-grained quartic Casimir constraints microcanonically and thus differs from the generalized-entropy problem studied perturbatively by Bouchet and Simonnet \cite{BS09} where these two constraints are treated canonically. 

In first-order perturbation theory, we assume the following perturbation expansion of the field and the multipliers:
\begin{eqnarray}
\bar\psi (\bi{r}) &=& \bar\psi^{(0)} (\bi{r})+ \gamma_4 \bar\psi^{(1)}(\bi{r}) + {\cal O}(\gamma_4^2), \label{expand1}\\
\alpha_i &=& \alpha_i^{(0)} +  \gamma_4 \alpha_i^{(1)} + {\cal O}(\gamma_4^2), ~i = 1,2,3, \label{expand2}\\
\beta &=& \beta^{(0)} +  \gamma_4 \beta^{(1)} + {\cal O}(\gamma_4^2), \label{expand3}\\
\gamma_2 &=& \gamma_2^{(0)} \cdot (1-\gamma_4 \cdot s) + {\cal O}(\gamma_4^2), \label{expand4}
\end{eqnarray}
where $\{ \bar\psi^{(j)},~ \alpha_i^{(j)}, ~\beta^{(j)},~ \gamma_2^{(0)},~ s\}$ are all of  ${\cal O}(1)$. After some non-trivial calculation the detail of which is presented in \ref{app1sphere}, the vorticity-streamfunction relationship for the globally maximized entropy is found and it is now nonlinear,
\begin{eqnarray}
\bar\omega = -6 \bar\psi + \gamma_4 \cdot \left(\frac{\beta^{(1)}}{2 \gamma_2^{(0)}}-6 s + \frac{18}{[\gamma_2^{(0)}]^2} \right) \bar\psi+\frac{432 \gamma_4 {\bar\psi}^3}{\gamma_2^{(0)}} + {\cal O}(\gamma_4^2).
\label{vs-MRS4}
\end{eqnarray}
Calculation further reveals that the first-order correction does not lift the zeroth-order degeneracy in $\ell=2$ modes but only sharpens or weakens the cores of the zeroth-order equilibrium vortices. The sign of the coefficient of the cubic term in \eref{vs-MRS4} is crucial.  Since $\gamma_2^{(0)}>0$ as required by the normalizability of Gaussian, for $\gamma_4>0$ the vorticity-streamfunction curve bends downward for large negative streamfunction at the cores of positive vortices and bends upward for large positive streamfunction at the cores of negative vortices. The vorticity-streamfunction curve is $\tanh$-like and the cores of vortices are weakened by the first-order correction. If $\gamma_4<0$, the vorticity-streamfunction relationship is $\sinh$-like and the cores are sharpened.  The multiplier $\gamma_4$ is determined by $\{ E, ~\Gamma_2^{f.g.}, ~\Gamma_4^{f.g.}\}$ through equation \eref{gamma4solution} and can take positive or negative values depending on the conserved quantities. Thus MRS-4 suffers from the problem of initializing the fluctuation-dependent Casimirs. Details of the calculation can be found in \ref{app1sphere}.

\begin{figure}
\centerline{\includegraphics[width=12.0 cm]{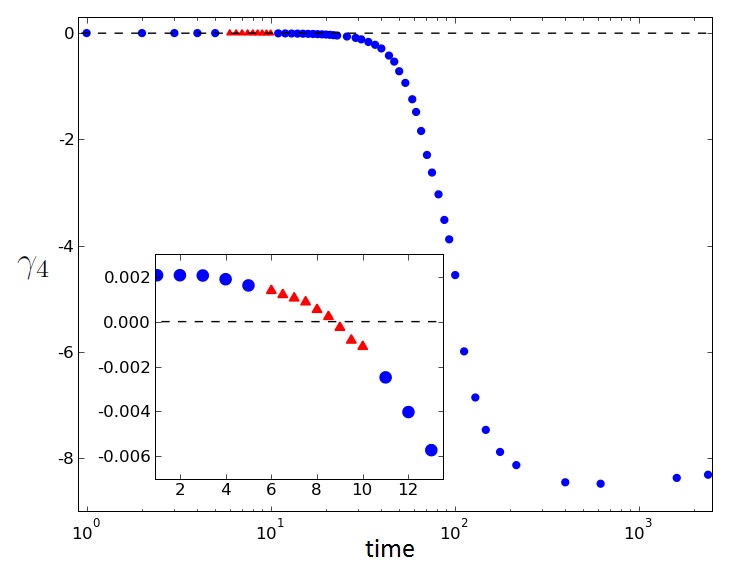}}
\caption{The Lagrange multiplier $\gamma_4$ versus time. Here the resolved energy and Casimirs at different times in simulation Sphere (a) are used to calculate $\gamma_4$ in first-order perturbative MRS-4. The multiplier $\gamma_4$ is initially positive but becomes negative during the short initial period of filament development. Only the red triangles are in the perturbative regime where the perturbative MRS-4 is self-consistent. Positive $\gamma_4$ corresponds to a $\tanh$-like vorticity-streamfunction relationship, whereas negative $\gamma_4$ yields a $\sinh$-like one.  The tendency for $\gamma_4$ to become negative within the perturbative regime shows that MRS-4 equilibria produces the sharpened cores and $\sinh$-like vorticity-streamfunction relationship seen in simulation coherent structures. Inset: magnified view of the region where $\gamma_4$ changes sign.\label{gamma4Fig}}
\end{figure}

\begin{figure}
\centering
\subfloat[][]{\includegraphics[width=8.0 cm]{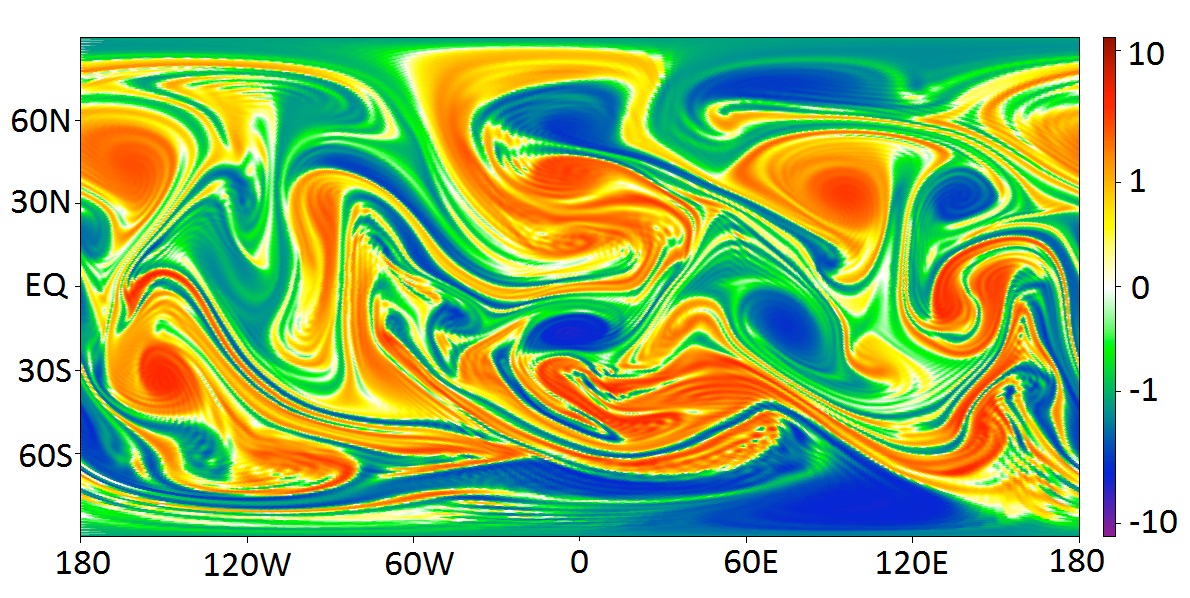}}\\
\subfloat[][]{\includegraphics[width=8.0 cm]{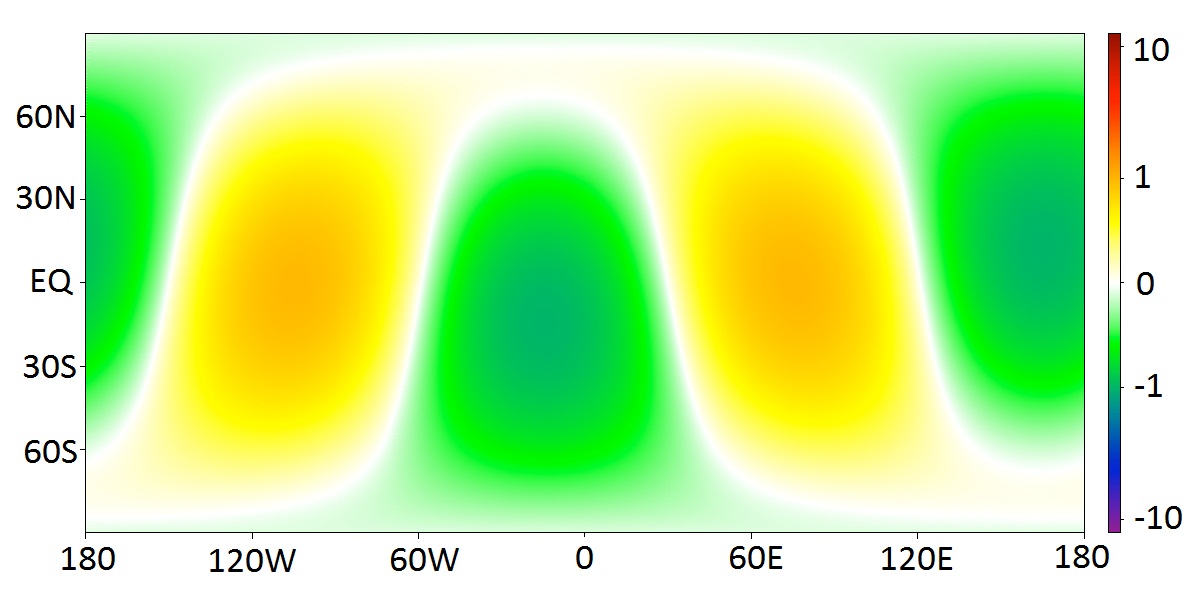}}\\
\subfloat[][]{\includegraphics[width=8.0 cm]{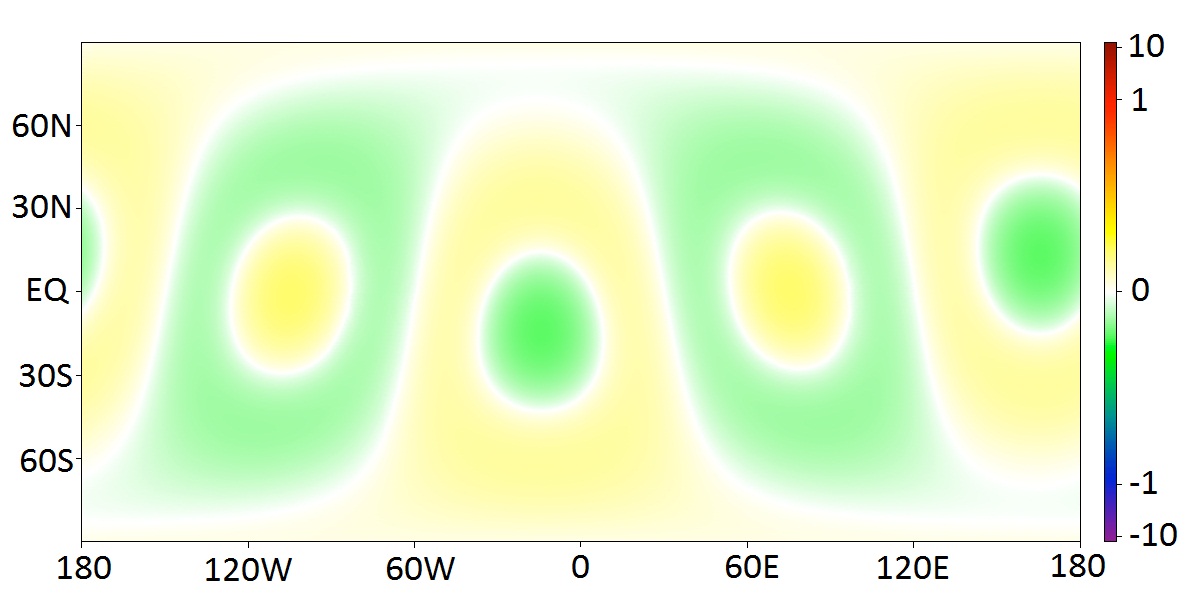}}
\caption{(a) Numerical simulation snapshot of the vorticity field of Sphere (a) at time $t = 9.0$.  (b) and (c) The equilibrium mean vorticity field obtained from perturbative MRS-4 using the resolved energy and Casimirs of Sphere (a) at $t = 9.0$.  The zeroth-order vorticity field of MRS-2 equilibria (b) and its first-order correction (c) when summed together yield the first order perturbative MRS-4 result.  The corrections sharpen the cores of the zeroth-order vortices and thus improve agreement with simulation coherent structures seen in \fref{NS}.   
\label{pertfig}}
\end{figure}

To deal with the initial value problem in first-order perturbative MRS-4, we adopt the intuitive argument described in \sref{intro} that the coarse-grained Casimirs become better approximations of the fine-grained ones at later times. The resolved energy and Casimirs taken at different time $t$ of simulation Sphere (a) are used to calculate the first-order perturbative MRS-4 equilibrium, and the features of the MRS-4 equilibrium of the underlying Euler flow are revealed in the tendency as $t$ increases.  The resolved quantities taken at different times are substituted into equation \eref{gamma4solution} to compute the multiplier $\gamma_4$ and the plot of $\gamma_4$ versus time is shown in \fref{gamma4Fig}. The multiplier $\gamma_4$ is initially small and positive, crosses zero during the short initial period of filament development, decreases to large negative values during further vortex stretching and merging, and becomes stable after coherent structure is formed. The validity of perturbation expansion equations \eref{expand1} -  \eref{expand4} needs be checked. The cases when all the first-order corrections are within one-tenth of the zeroth orders are shown with red triangles in the plot, and those are when the initial fine filaments are developing. The perturbative MRS-4 is only self-consistent inside the small early-time perturbative regime. Indeed from the strongly nonlinear nature of the $\sinh$-like vorticity-streamfunction relationship for simulation coherent structures, one would expect a perturbation theory that assumes small departure from a linear vorticity-streamfunction relationship to break down at late times.  Nevertheless as the energy spectrum shows an inverse cascade during the early-time perturbative regime, perturbative MRS-4 may still reveal, in its tendency as $t$ increases, the statistical effect of the fine-grained quartic Casimir constraint.  The perturbative MRS-4 equilibrium mean field depends especially on the sign of $\gamma_4$ that varies with time; only negative $\gamma_4$ agres with the observed sharp cores and $\sinh$-like vorticity-streamfunction relationship, and that appears after the initial transient period.
The tendency for $\gamma_4 $ to go to negative values as $t$ increases within the perturbative regime shows that MRS-4 equilibrium agrees with the quantitative features of sharp cores in the simulation coherent structures. \Fref{pertfig} shows a zeroth-order vorticity field and its first-order correction using the resolved values of energy and Casimirs taken at $t = 9.0$ when $\gamma_4$ is small negative. The numerical simulation snapshot is also included to show that the fluid is undergoing initial filament development at $t = 9.0$. The correction sharpens the cores of the zeroth-order equilibrium vortices and agrees with the features of simulation coherent vortices.

If the fine-grained cubic Casimir constraint is also weakly imposed, the vorticity-streamfunction relationship obtained from maximizing the entropy becomes
\begin{eqnarray}
\bar\omega &=&-6\bar\psi  -\frac{3 \gamma_3}{4 [\gamma_2^{(0)}]^2}  + [\frac{\gamma_3 \tilde\beta^{(1)}+\gamma_4 \beta^{(1)}}{2\gamma_2^{(0)}} -6( \gamma_3 \tilde{s} + \gamma_4 s) +\frac{18 \gamma_4}{[\gamma_2^{(0)}]^2} ] \bar\psi - \frac{54 \gamma_3 {\bar\psi}^2}{\gamma_2^{(0)}}  \nonumber\\
&&+\frac{432 \gamma_4 {\bar\psi}^3}{\gamma_2^{(0)}}+{\cal O}(\gamma_3^2,~ \gamma_3 \gamma_4, ~\gamma_4^2). 
\end{eqnarray}
Here the first-order terms in the perturbation expansion are extended from $\gamma_4 \beta^{(1)} $ to $\gamma_3 \tilde\beta^{(1)}+\gamma_4 \beta^{(1)}$, from $\gamma_4 s$ to $\gamma_3 \tilde{s} + \gamma_4 s$ and so on. Nonzero $\gamma_3$ breaks the odd symmetry of $\bar\omega(\bar\psi)$, and that corresponds to asymmetry between positive and negative vortices: the $\gamma_3$ -contribution of first-order correction either sharpens the positive vortices and weakens the negative vortices, or weakens the negative and sharpens the positive. Such asymmetry is almost absent for systems such as Sphere (a) with initial approximate symmetry between positive and negative resolved vorticity, so the constraint on the fine-grained cubic Casimir is not needed here. First-order perturbative MRS-4 can be extended to the square torus. By contrast the first-order correction now partly lifts the degeneracy of the zeroth-order field. Instead of any dipoles and unidirectional flows described by equation \eref{torusMRS2sol}, only the unidirectional flows and symmetric dipole are allowed. The first-order correction sharpens or weakens the cores of the zeroth-order field \footnote{The `cores' of the unidirectional flows refer to where the vorticity takes large absolute values and that is at the boundary between jets.} depending on $\gamma_4$ that is determined by $\{ E,~ \Gamma_2^{f.g.}, ~\Gamma_4^{f.g.} \}$, similar to the sphere. Details of the calculation on the torus are presented in \ref{app1torus}.

The first-order perturbation theory $\max\limits_{\rho(\bi{r}, \sigma)} \{    S[\rho]~|~E,  ~ \Gamma_2^{f.g.}, ~ \Gamma_4^{f.g.} \}$ is the microcanonical version of a first-order perturbative generalized-entropy problem \cite{BS09, BouchetReview}. The generalized-entropy problem involves only the mean field $\omega \equiv \bar{\omega}$ rather than the whole local probability distribution $\rho (\bi{r}, \sigma)$, and is given by
\begin{eqnarray}
\max\limits_{\omega(\bi{r})} \left\{ C_s[\omega] \equiv- \int \rmd^2 \bi{r} s(\omega) ~|~E   \right\},
\label{simplerProblem}
\end{eqnarray}
where assuming the small-energy limit when the vorticity field $\omega$ is close to zero and assuming the symmetry between negative and positive values of vorticity, the convex function $s(\omega)$ is approximated as
\begin{eqnarray}
s(\omega) = \frac{\omega^2}{2} - \frac{a_4 \omega^4}{4}.
\end{eqnarray}
The sign of $a_4$ is crucial: if $a_4>0$, the $\omega-\psi$ relationship is $\sinh$-like, while if $a_4<0$, it is $\tanh$-like \cite{BS09}. The perturbative approach to the generalized-entropy problem further assumes that $a_4$ is small, and is applied to turbulence on the square torus. The zeroth-order perturbative generalized-entropy problem has $s(\omega) = \omega^2/2$ and is exactly the ME principle, equivalent to MRS-2. A degeneracy in the solution is found at the zeroth order as in MRS-2, but the degeneracy is lifted at the first order by nonzero $a_4$. The first-order perturbative generalized-entropy description can be related, by using the general result by Bouchet, to the first-order perturbative MRS-4. Bouchet showed that the grand-canonical variational problem of MRS,
\begin{eqnarray}
\max\limits_{\rho(\bi{r}, \sigma)} \left\{    S[\rho]-\int \rmd^2 \bi{r}  \rmd \sigma \alpha(\sigma) \rho (\bi{r},\sigma)~|~E \right\},
\label{generalGrandCanonical}
\end{eqnarray}
where $\alpha(\sigma)$ is the Lagrange multiplier enforcing the conservation of global vorticity distribution $\int  \rmd^2 \bi{r} \rho (\bi{r}, \sigma)$, is equivalent to a generalized-entropy problem equation \eref{simplerProblem} with a specific choice of $s(\omega)$ determined by $\alpha(\sigma)$ \cite{bouchet08}.  Note that the grand-canonical variational problem was proposed by Ellis \etal \cite{eht02} with the prior distribution $\chi(\sigma) \equiv \exp[-\alpha(\sigma)]$, and the relation between the prior $\chi(\sigma)$ and the generalized entropy $ C_s[\omega]  $ in equation \eref{simplerProblem} was discussed in detail in \cite{chavanis05, chavanis08}. The microcanonical variational problem $\max\limits_{\rho(\bi{r}, \sigma)} \{    S[\rho]~|~E, ~ \Gamma_2^{f.g.}, ~ \Gamma_4^{f.g.}\}$ is related to the grand-canonical variational problem $\max\limits_{\rho(\bi{r}, \sigma)} \{    S[\rho]-\gamma_2 \Gamma_2^{f.g.} - \gamma_4 \Gamma_4^{f.g.}~|~E \}$, an example of equation \eref{generalGrandCanonical} with $\alpha(\sigma) = \gamma_2 \sigma^2 + \gamma_4 \sigma^4 $. Applying the general result in the reference \cite{bouchet08} and assuming that $\gamma_4$ is small, the equivalent generalized-entropy problem has $s(\omega) = \omega^2/2 + \gamma_4 \omega^4/(2 \gamma_2^{(0)}) +{\cal O}(\gamma_4^2)$.  First-order perturbative MRS-4 and the first-order perturbative generalized-entropy problem studied in references \cite{BS09, BouchetReview} are related if
\begin{eqnarray} 
a_4 &\equiv& -2\gamma_4/\gamma_2^{(0)}.
\end{eqnarray}
 The limit of small $a_4$ coincides with that of small $\gamma_4$ and the two descriptions can be compared. The shape of vorticity-streamfunction curve is related to the sign of $a_4$ in the generalized-entropy problem the same way as it is related to the sign of $\gamma_4$ here, because the microcanonical and grand-canonical descriptions share the same critical points. However conclusions about degeneracy lift in the zeroth-order solution differ: the first-order correction selects either symmetric dipole (if $a_4>0$) or unidirectional flows (if $a_4 <0$) in the perturbative generalized-entropy problem, whereas it selects both here. Thus in the perturbative generalized-entropy problem, only unidirectional flows with $\tanh$-like vorticity-streamfunction relationship or symmetric dipole with $\sinh$-like relationship can occur, while unidirectional flows with $\sinh$-like relationship and symmetric dipole with $\tanh$-like relationship are also allowed in the first-order perturbative MRS-4. The disagreement on degeneracy lift results from the fundamental difference between microcanonical and grand-canonical ensembles: conserved quantities $\{ E, ~\Gamma_2^{f.g.}, ~\Gamma_4^{f.g.}\}$ are fixed when varying over all zeroth-order solutions in the microcanonical description, while energy $E$ and the multiplier ratio $ a_4 $ are fixed in the perturbative generalized-entropy problem. The multiplier ratio $a_4$ for symmetric dipole differs from that for unidirectional flows in the microcanonical approach (compare equations \eref{gamma4Solution1} and \eref{gamma4Solution3}), and thus difference in degeneracy lift is understandable. 

Chavanis and Sommeria proposed a different microcanonical perturbative approach to go beyond MRS-2 \cite{CS96}, and their results have many similarities with perturbative MRS-4. They studied the strong mixing limit where the energy constraint is weakly imposed in the MRS theory and expanded in terms of $\beta \bar\psi \sigma \ll 1$. The zeroth order yields the uniform vorticity field, corresponding to complete mixing. MRS-2 is recovered at the first order where the vorticity-streamfunction relationship is linear and the equilibrium is independent of the third and higher fine-grained Casimirs. The fine-grained cubic and quartic Casimirs enter at the second order and the $\bar\omega$-$\bar\psi$ relationship is nonlinear with $\bar\psi^2$ and $\bar\psi^3$ terms. For a symmetric distribution in positive and negative vorticity on a boundaryless surface, the $\bar\omega$-$\bar\psi$ relationship is $\sinh$-like if $Ku-3>0$ and $\tanh$-like if $Ku-3<0$, where $Ku\equiv \Gamma_4^{f.g.}/(\Gamma_2^{f.g.})^2$ is the kurtosis of the fine-grained vorticity distribution \cite{CS96}. This is similar to the result of the perturbative generalized-entrophy problem based on the sign of $a_4$ and perturbative MRS-4 based on the sign of $\gamma_4$. Obviously the sign of $\gamma_4$ (see equation \eref{gamma4solution}) is not given by the sign of $Ku-3$. The dependence of the nonlinear $\bar\omega$-$\bar\psi$ relationship on the conserved quantities differ in these two microcanonical perturbative approaches, because they address entirely different limiting cases: the approach studied by Chavanis and Sommeria takes the small $\beta$ limit based on the full MRS, whereas perturbative MRS-4 takes the small $\gamma_4$ limit based on the truncated MRS-4 approach.

\section{Rotating sphere\label{sec4}}

We now turn to the problem of 2D fluid motion on the surface of the rotating sphere.  MRS-2 on a rotating sphere was solved by Majda and Wang \cite{Majda} and by Herbert \cite{Herbert13}.  We review the solution below.  MRS-2 equilibria describe a complete condensation of energy to the largest possible scale.  The structure of the MRS-2 mean field is rotation-independent, in contrast to simulation coherent structures whose most energetic wavenumbers and anisotropy tend to increase with rotation rate.  The reason why MRS-2 shows no structural changes due to rotation has its origin in the linear vorticity-streamfunction relationship. Here the vorticity-streamfunction relationship refers to the relationship between the mean absolute vorticity and a combined field of the mean relative streamfunction and the $\ell=1$ spherical harmonics. The structure of higher-order MRS-N equilibrium depends on the rotation rate, but whether the dependence agrees with numerical simulation is beyond the scope of this paper. Numerical simulation also shows that the assumption of ergodicity is significantly violated on the rotating sphere, due to the rotation-induced failure to develop broad-band turbulence and the dynamical trapping effect of the anisotropic structures. The breakdown of ergodicity poses a serious difficulty in relating simulation coherent structures to statistical equilibrium, and it will be further addressed in \sref{sec5}. 

\subsection{MRS-2 on the rotating sphere}
MRS-N is extended to the rotating sphere by redefining the distribution $\rho(\bi{r}, \sigma)$ as that of the absolute vorticity $q$ \cite{Herbert13, Majda}. MRS for the rotating sphere is specified by 
\begin{eqnarray}
\max\limits_{\rho(\bi{r}, \sigma)} \{ S[\rho]~|~E;  ~\{ \Gamma_n^{f.g.} \};~\{\bar\psi_{1m}(t)\} \},
\label{MRSTheoryRot}
\end{eqnarray}   
where the kinetic energy
\begin{eqnarray}
E = -\frac{1}{2} \langle \bar\psi \bar\zeta \rangle,
\end{eqnarray} 
and the infinite Casimirs of the fine-grained absolute vorticity
\begin{eqnarray}
\Gamma_n^{f.g.} \equiv \int \rmd^2 \bi{r} \overline{q^n}  = \int \rho(\bi{r}, \sigma)~ \sigma^n~ \rmd\sigma \rmd^2 \bi{r},\ n=1, ~2, ~3, \cdots, 
\end{eqnarray}
  are again conserved.  Rotation partly breaks $SO(3)$ symmetry down to axial symmetry about the $\bi{z}$-axis.  The total angular momentum precesses about the $\bi{z}$-axis (see \cite{Herbert13, Majda} and \ref{app2}).  Consequently the three $\ell = 1$ amplitudes time evolve as $\bar\psi_{1 0}(t) = \bar\psi_{1 0}(0)$, $\bar\psi_{1 1}(t)  = \bar\psi_{1 1}(0) e^{i \Omega t}$ and $\bar\psi_{1 -1}(t)  = \bar\psi_{1 -1}(0) e^{-i \Omega t}$.  Thus the dynamics of the $\ell =1$ modes decouples from that of the $\ell>1$ amplitudes \cite{Herbert13, Majda}. Previous work has treated the angular momentum in different ways.  Herbert \etal studied a generalized-entropy description which is the same as the ME principle taking into account only the conservation of the $\bi{z}$-component of angular momentum $L_z$, rather than all three components, and found that the equilibrium relative vorticity field is a dipole \cite{HDCP12short, HDCP12}. Lim and his collaborators have studied a different model on the sphere for which the 2D fluid is coupled to the solid sphere and none of the three components of angular momentum is conserved. They studied an energy-relative-enstrophy approach (a form of ME) \cite{DL07, Lim07} and an energy-enstrophy-circulation statistical mechanical description (a form of MRS-2) \cite{lim12, lim13}, and found in certain parameter regimes sub- or super-rotating flows. They further studied the effect of adding the relative quartic Casimir constraint into the energy-relative-enstrophy description \cite{LS09}, analogous to imposing the fine-grained quartic Casimir constraint in MRS-2 as investigated here.  Finally a previous study also treated the  $\ell=1$ modes as random variables statistically independent from the $\ell>1$ modes \cite{Herbert13} rather than deterministic time-dependent constraints as we do here. 

Applying the approach in \sref{MRS2nonrot} to rotating sphere, it is straightforward to show that MRS-2 on the rotating sphere
\begin{eqnarray}
\max\limits_{\rho(\bi{r}, \sigma)} \{ S[\rho]~|~E,~  \Gamma_2^{f.g.} ,~\{\bar\psi_{1m}(t)\} \}
\label{MRS2TheoryRot}
\end{eqnarray}
is equivalent to the ME principle
\begin{eqnarray}
\min\limits_{\bar{q}(\bi{r})} \left\{  \int \rmd^2 \bi{r} \bar{q}^2 ~| ~E,  ~\{ \bar\psi_{1m} (t) \}   \right\}.
\end{eqnarray}
Now
\begin{eqnarray}
\int \rmd^2 \bi{r} \bar{q}^2&=& \int \rmd^2 \bi{r} (\bar{\zeta}+ 2 \Omega \cos\theta)^2 \nonumber\\
&=&  \int \rmd^2 \bi{r} \bar{\zeta}^2+8 \Omega \sqrt{4\pi/3} \bar\psi_{10}+ 4 \Omega^2 \int \rmd^2 r \cos^2 \theta,
\label{enstrophyRelation}
\end{eqnarray}
where in the last expression, the second term is held constant by the angular momentum constraint and the third term is a constant. Therefore, the ME principle is equivalent to
\begin{eqnarray}
\min\limits_{\bar{\zeta}(\bi{r})}  \left\{  \int \rmd^2 \bi{r} \bar{\zeta}^2 ~| ~E,  ~\{ \bar\psi_{1m} (t) \}   \right\}.
\end{eqnarray}
The rotation rate $\Omega$ drops out and the description is no different from that of the non-rotating sphere. The solution is: 
\begin{eqnarray}
\bar\zeta(\theta,\phi, t) = -2 \sum\limits_{m=-1}^{1} \bar\psi_{1m}(t)~ Y_{1m}(\theta,\phi) - 6 \sum\limits_{m=-2}^{2} \bar\psi_{2m}(t)~ Y_{2m}(\theta,\phi),
\end{eqnarray}
set by any complex parameters $\{ \bar\psi_{2m} \} $ that satisfy the reality condition equation \eref{reality} and the energy constraint
\begin{eqnarray}
 E =\frac{1}{4 \pi} (\sum\limits_{m = -1}^{1} |\bar\psi_{1m}|^2 + 3 \sum\limits_{m=-2}^{2} |\bar\psi_{2m}|^2).
\end{eqnarray}
The form of the solution is preserved by the EOM, and is generally quasi-periodic with several frequencies (the $\ell = 1$ modes precesses at one frequency, and the $\ell = 2$ modes oscillate at one or two frequencies).  For zero net angular momentum, the solution is a pure quadrupole that undergoes solid-body rotation about the $\bi{z}$-axis at angular rate $-\Omega/3$.  

\subsection{Numerical simulation on the rotating sphere \label{NSrot}}

The physics of rotation and 2D turbulence in combination has been the subject of a great deal of study. The dynamics of 2D inviscid flows on the rotating sphere can be classified into two regimes, turbulent regime and wave regime, by comparing the relative strength of the nonlinear advection term to the linear Coriolis term \cite{rhines}. The turbulent regime is where the dominant spherical wavenumbers are much larger than the Rhines wavenumber \cite{ny97} (the symbol $\ell$ for spherical wavenumber is not to be confused with the conventional symbol for length) 
\begin{eqnarray}
\ell_\beta = \sqrt{\frac{\pi \Omega}{4 \sqrt{2 E(0)}}}, 
\end{eqnarray}
and the nonlinear advection dominates, whereas the dominant wavenumbers of the wave regime are much smaller than the Rhines wavenumber and the fluid follows linear wave-like dynamics. If the system starts in the turbulent regime, the canonical picture has upscale-cascading energy reach the Rhines wavenumber at which point the inverse cascade is suppressed by rotation, the nonlinear turbulent behavior is replaced by the linear Rossby wave motion and the system enters the wave regime. Triad interaction of Rossby waves thus transfers energy into modes at small zonal wavenumbers giving rise to anisotropy. However if the system is initially in the wave regime, it stays there during the whole evolution. Here we investigate coherent structures in the presence of rotation by numerically integrating initial states with zero total angular momentum forward in time on rotating spheres. Numerical simulation shows that for fixed rotation rate, the coherent structure characterized by its degree of anisotropy and its most energetic wavenumber $\ell_\mathrm{m}$ is robust against changes in initial states, as long as the system starts in the turbulent regime so that the broad-band turbulence is generated in the evolution process. That suggests the existence of equilibrium-like features in coherent structures that only depend on rotation rates. That both anisotropy and the most energetic wavenumber $\ell_\mathrm{m}$ tend to increase with rotation rate agrees with the qualitative features of Rhines theory, though quantitatively $\ell_\mathrm{m}$ is not found to scale precisely as $\sqrt{\Omega}$ as it does in the classical Rhines picture \cite{cho}.
The detailed behavior of coherent structures as rotation rate increases can be further classified into three regimes with small, intermediate and large rotation rates respectively.
However despite the robust equilibrium-like features, numerical simulation also shows many non-equilibrium-like features of coherent structures, indicating the breakdown of ergodicity. Asymmetry in strengths and signs of vorticity is dynamically trapped by anisotropic structures and thus details of coherent structures are sensitive to initial states. Failure to sufficiently develop turbulence also keeps the system far away from equilibrium. If the system is in the wave regime from the onset, turbulence is never fully developed and at late times the fluid shows a less zonal structure.

\begin{figure}
\centerline{\includegraphics[width=15.0 cm]{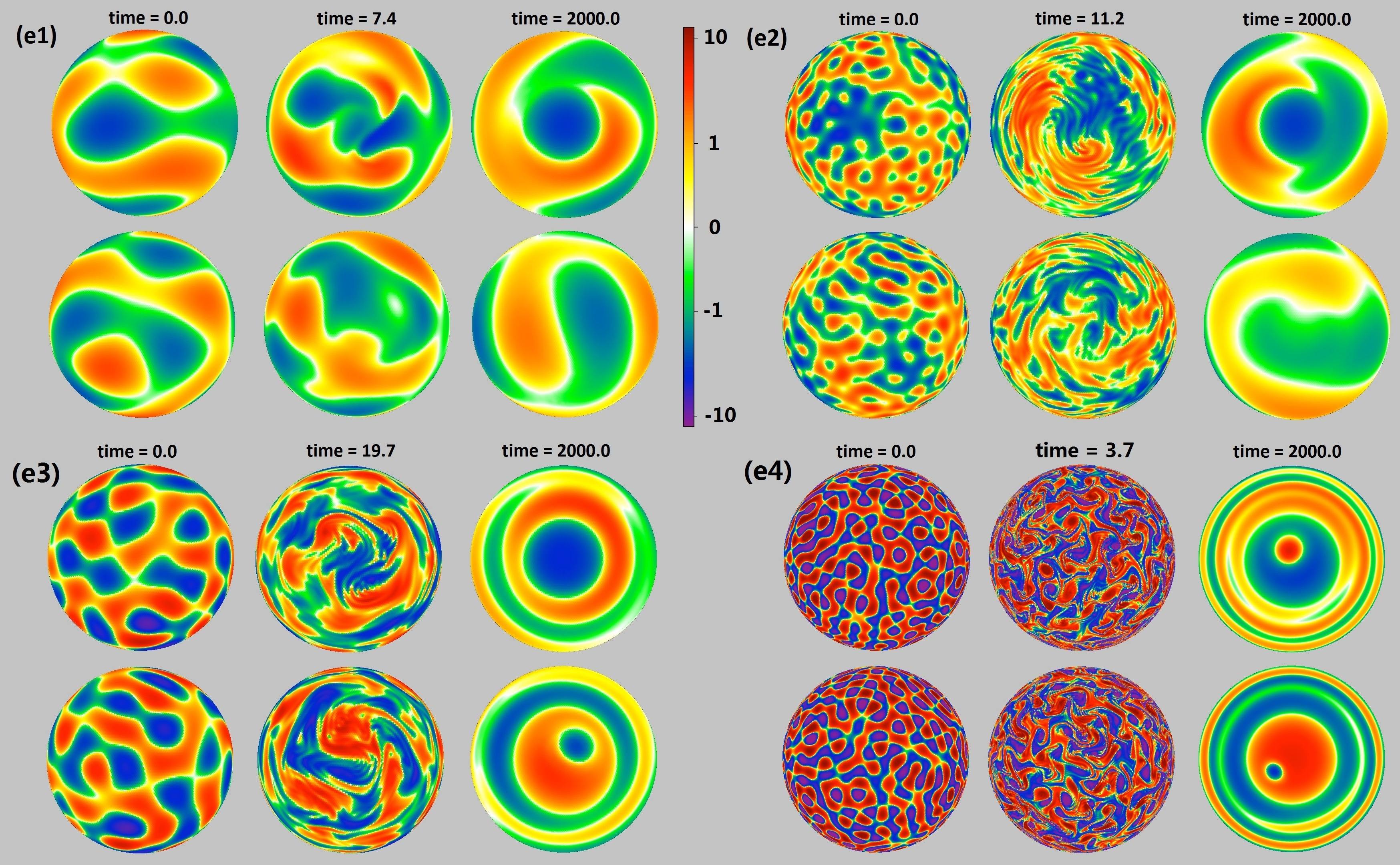}}
\caption{Numerical simulation snapshots of the relative vorticity field $\zeta(\bi{r})$ for four runs at the same rotation rate. The initial state, the state at about the time when the energy spectrum is broadest, and the coherent state, with a north-pole view (upper row) and south-pole view (lower row), are shown. Details of the four runs are listed in \tref{table4}.  Systems initially in the turbulent regime (Spheres (e3) and (e4)) and in the wave regime (Spheres (e1) and (e2)) evolve into qualitatively different coherent structures. The difference suggests that a failure to develop broad-band turbulence keeps the system far away from equilibrium.  \label{rotatee1-e4}}
\end{figure}

\begin{figure}
\centerline{\includegraphics[width=15.0 cm]{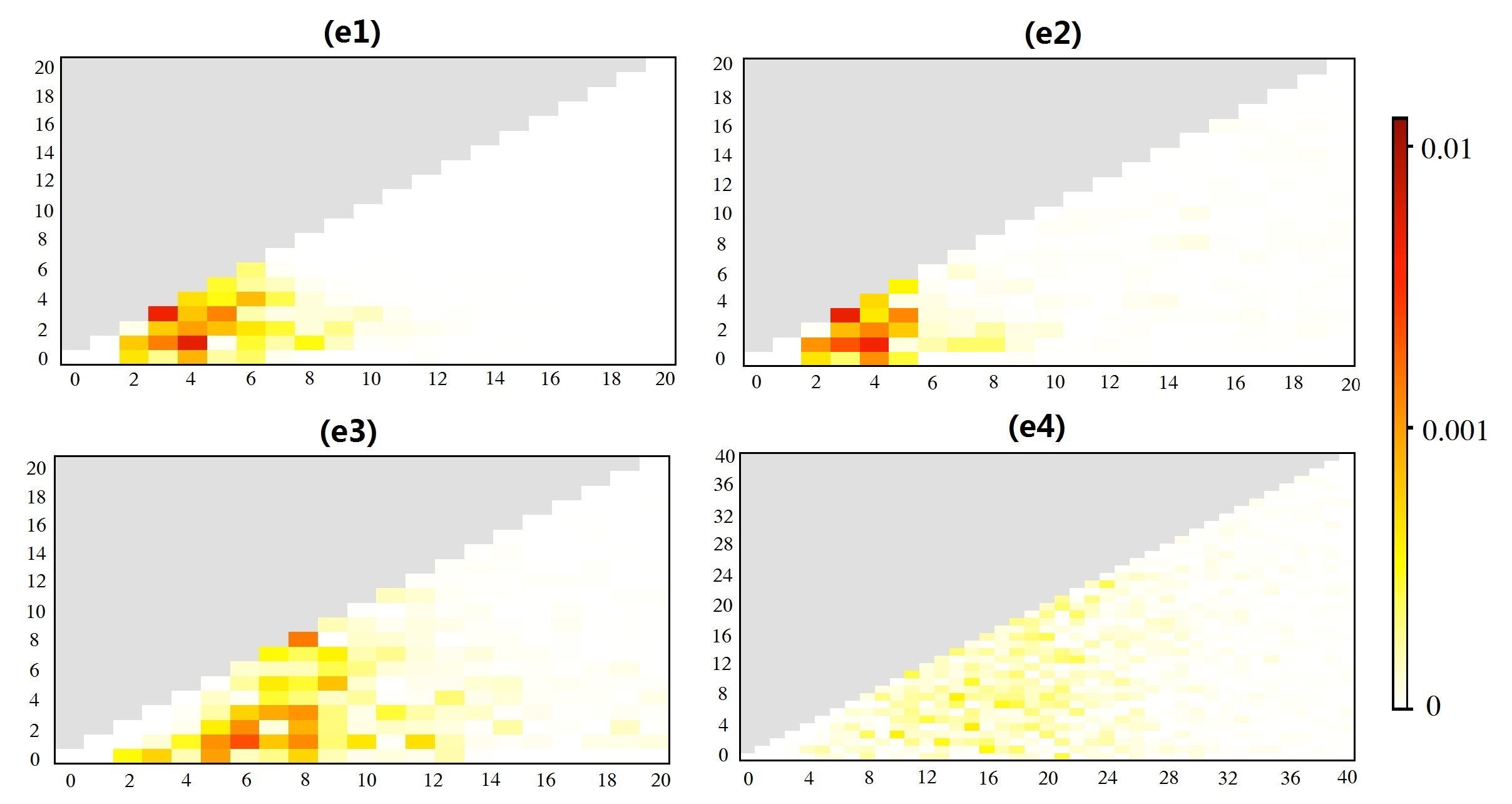}}
\caption{Energy spectrum $E(\ell, m)$ when spread most broadly for each of the four runs shown in \fref{rotatee1-e4}.  Spherical wavenumber $\ell$ is plotted along the horizontal axis and the azimuthal wavenumber $m$ is along the vertical. Broad-band turbulence develops for systems initially in the turbulent regime (Spheres (e3) and (e4)) in contrast to those initially in the wave regime (Spheres (e1) and (e2)). \label{e1-e4spectrumBroad}}
\end{figure}

\begin{figure}
\centerline{\includegraphics[width=15.0 cm]{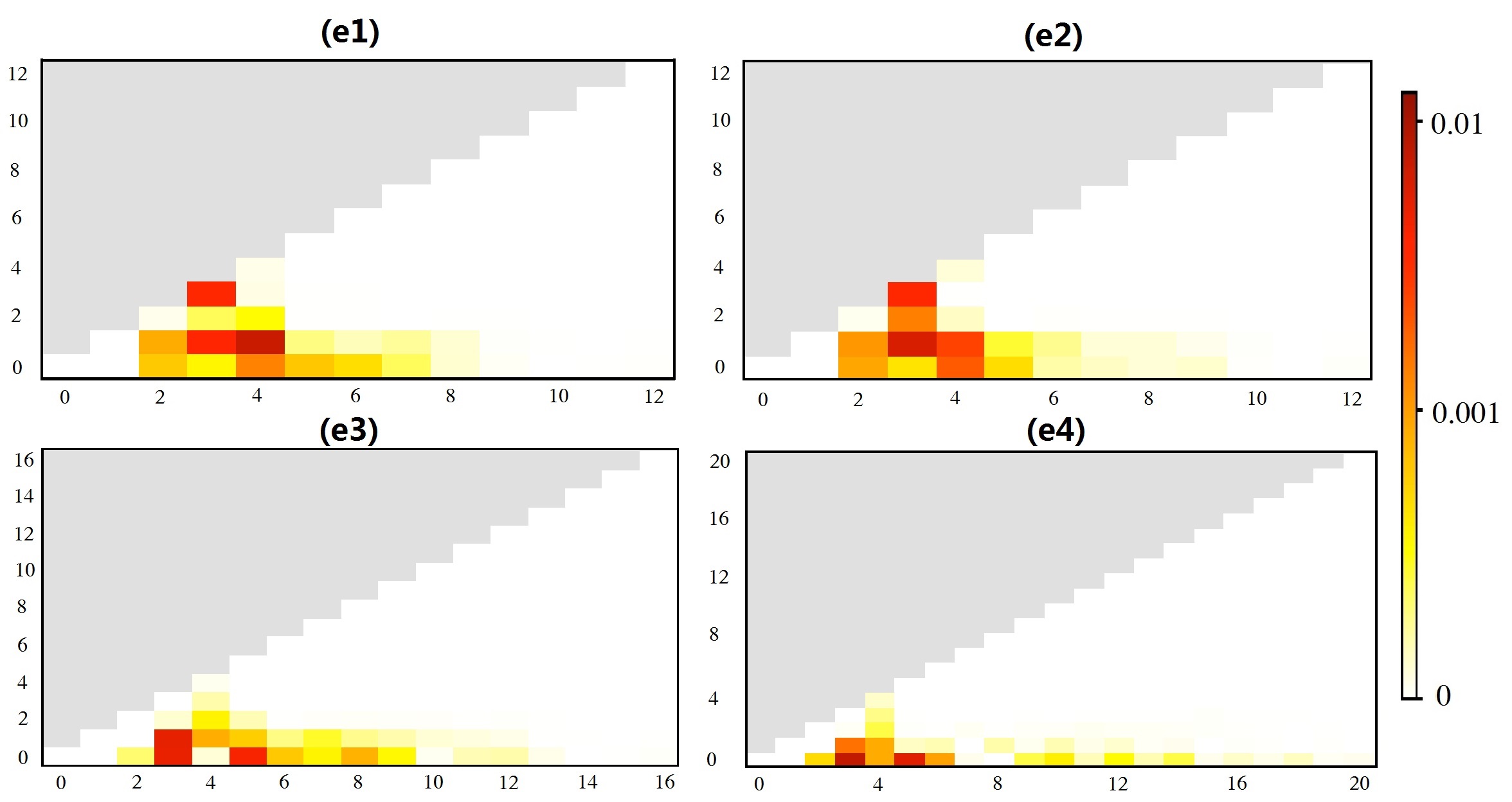}}
\caption{Same as \fref{e1-e4spectrumBroad} but for the coherent states shown in \fref{rotatee1-e4}. Energies are concentrated in the zonal modes with azimuthal wavenumber $m =0$ for Spheres (e3) and (e4), but not for Spheres (e1) and (e2).   \label{e1-e4spectrumFinal}}
\end{figure}

First we show that systems starting in the turbulent regime display equilibrium-like coherent structures whereas those starting in the wave regime appear to stay far away from equilibrium. The result implies that in order to compare coherent structures with statistical equilibrium, only the systems starting in the turbulent regime are useful. Here we integrate four different initial states with the same energy forward in time on the sphere rotating at the same rate. Simulation parameters and measures of the four runs are summarized in \tref{table4} and snapshots of the relative vorticity fields at different times are shown in \fref{rotatee1-e4}. The Rhines wavenumber $\ell_\beta $ is the same for all runs but the initial wavenumbers are different. A rough estimate of the initial wavenumber can be provided by the quantity $\ell_0 \equiv\sqrt{\langle \zeta^2 (0)\rangle/[2 E(0)]}$, henceforth denoted as the initial effective wavenumber, where $E(0)$ is the initial energy and $\langle \zeta^2(0) \rangle$ is the initial relative enstrophy per unit area. The initial state of Sphere (e1) contains only spherical harmonics with wavenumbers smaller than $\ell_\beta $, and its $\ell_0$ is only about a half of $\ell_\beta $. Sphere (e2) has its initial energy mostly concentrated at the wavenumbers smaller than $\ell_\beta $ but it also has a tiny component with wavenumbers much larger than $\ell_\beta $. The method of using $\ell_0$ to estimate the wavenumber that characterizes the dynamics fails in this case because there are two separate bands of wavenumbers. The range of the initial wavenumbers for Sphere (e3) encompasses $\ell_\beta $, but the wavenumbers larger than $\ell_\beta $ dominate and $\ell_0>\ell_\beta $. The initial state of Sphere (e4) only has wavenumbers much larger than $\ell_\beta$. \Fref{rotatee1-e4} shows that the motion of Sphere (e1) is wave-like, whereas that of Spheres (e3) and (e4) is turbulent. The high-wavenumber component of Sphere (e2) undergoes turbulent mixing, but the energy in these modes is always at least one order of magnitude smaller than that of the low-wavenumber background, and thus the linear wave-like dynamics still dominates. The energy spectrum $E(\ell, m)$ at the time when it is broadest for each run as shown in \fref{e1-e4spectrumBroad} reveals that broad-band turbulence where energy is shared widely among many wavenumbers is better developed during the initial evolution for Spheres (e3) and (e4) than for Spheres (e1) and (e2). \Fref{rotatee1-e4} further shows that systems starting in turbulent regime and in wave regime evolve into qualitatively different coherent states. The coherent structures are banded zonal structures for Spheres (e3) and (e4) and are less zonal for Spheres (e1) and (e2), but the most energetic wavenumbers for all runs are almost the same (see \tref{table4}). Here for simplicity only the most energetic wavenumber $\ell_\mathrm{m}$ is discussed, despite the fact that a few other dominant wavenumbers may contain less but comparable energy as $\ell_\mathrm{m}$ and these wavenumbers may also be the robust feature of the coherent state. Energies at the coherent state are concentrated in the zonal modes for Spheres (e3) and (e4), but not for Spheres (e1) and (e2) (see \fref{e1-e4spectrumFinal}). The phenomenon that coherent structures can show different configurations of the same lowest modes depending on whether broad-band turbulence is developed has also been observed in simulations of 2D evolving turbulence on the square torus \cite{YMC03,Yin04}. These authors time evolved the 2D Euler equation with added viscosity and found that although the broad-band initial condition generally produces symmetric dipoles with $\sinh$-like vorticity-streamfunction relationship, initial states where most energies are concentrated in few narrow-band low wavenumbers can lead to other asymmetric configurations of the degenerate $k=1$ modes with various shapes of the vorticity-streamfunction relationship. The latter systems appear less turbulent and fail to generate broad-band turbulence, and thus may be considered as too far from equilibrium to produce the equilibrium features. 

\Table{\label{table4} Range of initial wavenumber $\ell$, initial effective wavenumber $\ell_0\equiv \sqrt{\langle \zeta^2 \rangle /(2E)}$, the most energetic wavenumber $\ell_\mathrm{m}$ at the coherent state, and total number of cells $D$ for each run. Rotation rate $\Omega = 18.85$, Rhines wavenumber $\ell_\beta = 8.45$, initial eddy-turnover time equals 4.8 and time step $\Delta t = 0.005$ for all runs.  For the coherent state found on Sphere (e1), the energy spectrum peak oscillates between $\ell=3$ and $\ell = 4$ at short time scales. }
\br
Run                     & Range of initial $\ell$              &   $\ell_0$ &   $\ell_\mathrm{m}$       & D\\ 
\mr
Sphere (e1)           &   $[2,6]$                               &   4.83 &  $3$ or $4$   & 40962 \\
Sphere (e2)          &    $[2,5]$ and $[20,25]$   &      -- &   $3$  & 40962 \\
Sphere (e3)            &   $[8,12]$                       &    10.21  &  $ 3 $    & 40962\\
Sphere (e4)            &   $[27,30]$                       &    28.73  &  $ 3 $   & 163842\\
\br
\endTable

Spheres (e1) - (e4) are in the intermediate-rotation regime as will be discussed later. Similar qualitative difference between systems starting in the turbulent regime and the wave regime is also found on rapidly rotating spheres. Yoden \etal performed time integration from initial states with different energy spectra on spheres with slow, intermediate and fast rotation rates \cite{yihy99}, but they did not point out that the behavior of systems starting in the wave regime is qualitatively different from those starting in the turbulent regime. The initial state with the lowest-wavenumber band in their paper, with the run denoted as cpn010, is in the turbulent region for slow rotation and in the wave regime for the fastest rotation. The non-equilibrium-like feature of systems starting in the wave regime can be inferred from figure 6(a) in \cite{yihy99} where unlike other systems with initial spectra of high-wavenumber bands, cpn010 does not show strongest circumpolar jets for the fastest rotation. Their integration time corresponds to only $t \sim {\cal{O}}(10)$ here, and seems insufficient to develop the flow to coherent state because the vorticity fields are still developing at that time \cite{tyh07}. Of course the choice of integration time depends on the evolution of the specific physical quantities under investigation. This paper focuses on the relative vorticity field which is a reasonable choice for the purpose of comparing with the equilibrium solution $\rho(\bi{r}, \sigma)$ and revealing the structure unobscured by the $2\Omega \cos\theta$ term, whereas the physical-space quantity that previous studies have mostly focused on is the velocity field \cite{YY93, cho, yihy99, tyh071, tyh07}.

\begin{figure}
\centerline{\includegraphics[width=15.0 cm]{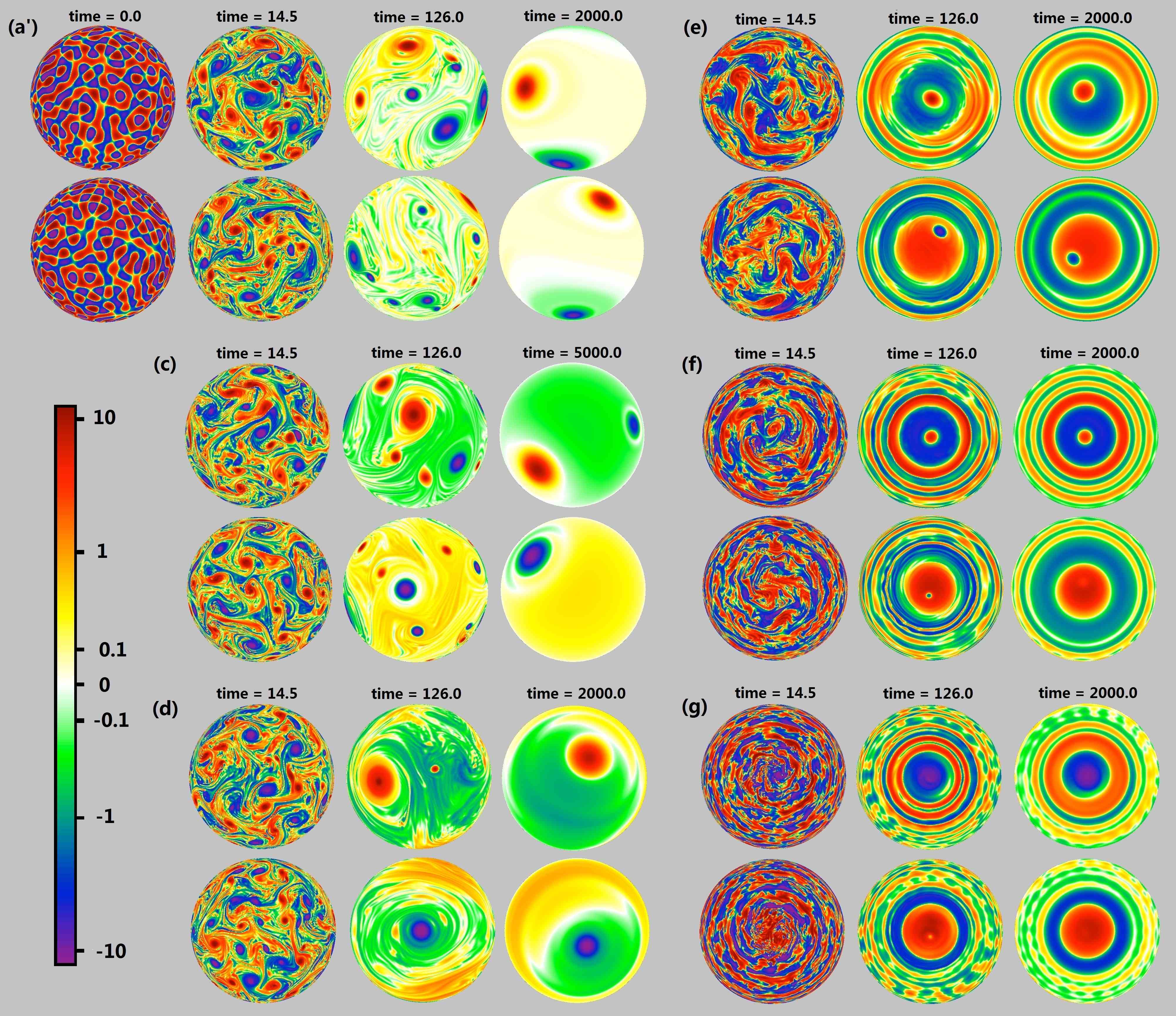}}
\caption{Numerical simulation snapshots of the relative vorticity field $\zeta(\bi{r})$ show that the same initial state evolves differently on spheres rotating at different rates. Details of the six runs are listed in \tref{table3}. For each run, the north-pole (upper row) and south-pole (lower row) views are shown. High rotation rates tend to arrest the energy inverse cascade at larger wavenumbers and create more anisotropic zonal flows. \label{rotate}}
\end{figure}

We investigate how the equilibrium-like features of coherent structures depend on rotation, by time evolving the same initial state on spheres with different rotation rates. Although details of coherent structures depend sensitively on initial conditions because dynamical trapping by anisotropic structures forbids the system to sample the whole phase space (compare Spheres (e3) and (e4) in \fref{rotatee1-e4}), the degree of anisotropy and the most energetic wavenumber are robust for fixed rotation rate. \Tref{table3} summarizes the simulation parameters and measures, and \fref{rotate} shows the snapshots of the evolution. Sphere (e4) is included and renamed as Sphere (e). The detailed behavior can be classified into three regimes with small, intermediate and large rotation rates respectively \cite{YY93,cho, tyh071}. Spheres (c) and (d) are at small rotation rates. Their coherent states are quasi-static quadrupoles with the most energetic wavenumber $\ell_\mathrm{m} =2$ undergoing overall rotation about the $\bi{z}$-axis, similar to MRS-2. However the rotation period of the quadrupole configuration is slightly different from that of the MRS-2 equilibria due to the presence of $\ell>2$ modes. The quadrupole of Sphere (c) rotates with a period of roughly $108$, larger than the MRS-2 period $100$ by $8\%$, and Sphere (d) roughly $12$, smaller than the MRS-2 period $14.5$ by $17\%$. As rotation rate increases in the slow-rotation regime, anisotropy of coherent structures increases as the vortex pairs appear more confined to the poles. Sphere (e) is at intermediate rotation rate and its coherent state has stable zonal banded structures with no apparent difference between high and low latitudes. Its most energetic wavenumber is larger than the slow-rotation cases. When the sphere rotates rapidly as in Spheres (f) and (g), the zonal banded coherent structures develop extreme latitudinal inhomogeneities where most of the kinetic energy is accumulated near the poles. As rotation rate increases in this regime, stronger and narrower circumpolar vorticity structures that correspond to an easterly circumpolar jet appear in high latitudes. Takehiro \etal performed quantitative numerical studies and proposed a scaling theory that the averaged speed and width of the circumpolar jet are proportional to $\Omega^{1/4}$ and $\Omega^{-1/4}$ respectively, in the asymptotic limit of large rotation rate $\Omega$ \cite{tyh071, tyh07}. The study of how the width of the circumpolar jet decreases as $\Omega$ increases is similar but maybe not exactly the same as the investigation here about how the most energetic wavenumber $\ell_\mathrm{m}$ increases with $\Omega$ at large but finite $\Omega$, even though the circumpolar jets contain most of the kinetic energy. The relative vorticity field of the flow pattern with strong polar easterly jets shows an approximate anti-symmetry under the reflection across equator. Thus the low odd-$\ell$ modes that show north-south anti-symmetry dominate over the low even-$\ell$ modes that show north-south symmetry, and the most energetic wavenumber is odd at rapid rotation. For intermediate and fast rotation, the most energetic wavenumber $\ell_\mathrm{m} $ at the coherent state increases with the rotation rate but is always smaller than the Rhines wavenumber (see \tref{table3}) in accord with the findings of Cho and Polvani \cite{cho}. 

\Table{\label{table3} Rotation rate $\Omega$, rotation period $T\equiv 2\pi/\Omega$, Rhines wavenumber $\ell_\beta$, the most energetic wavenumber of the coherent state $\ell_\mathrm{m}$, and the time step $\Delta t$ for each run. The initial effective wavenumber $\ell_0 = 28.75$, initial eddy-turnover time equals 4.8, and the total number of cells $D= 163842$ for all runs.}
\br
Run & $\Omega$ & $T$ & $\ell_\beta$ &  $\ell_\mathrm{m}$ &  $\Delta t$\\
\mr
Sphere (a') & 0 & $\infty$ & 0 &     2    & 0.005\\
Sphere (c) & 0.19 & 33.33 & 0.84     & 2  & 0.005\\
Sphere (d) & 1.30 & 4.83 & 2.22 &     2  & 0.005\\
Sphere (e) & 18.85 & 0.33 & 8.45    &  3  & 0.005\\
Sphere (f) & 62.83 & 0.10 & 15.43     & 5 &0.002\\
Sphere (g) & 188.50 & 0.03 & 26.72    &  5  &0.002\\
\br
\endTable

\subsection{Qualitative difference between MRS-2 and numerical simulation}

Simulation coherent structures and MRS-2 equilibria are qualitatively different, because the coherent structures depend on rotation rate whereas MRS-2 describes a rotation-independent complete condensation of energy to $\ell = 2$.  That MRS-2 equilibria are structurally independent of rotation has its origin in the linear vorticity-streamfunction relationship. Here the vorticity-streamfunction relationship refers to the relationship between the absolute vorticity field $\bar{q}$ and the field $\beta \bar\psi- \alpha_1 \cos\theta - \alpha_2 \sin\theta \cos \phi - \alpha_3 \sin\theta \sin\phi$, as a direct extension of the non-rotating case described in \sref{MRS2nonrot}. The general statistical mechanical description that conserves a set of fine-grained Casimirs $\{ \Gamma_{i}^{f.g.} \}$ on the rotating sphere,
\begin{eqnarray} 
\max\limits_{\rho(\bi{r}, \sigma)} \{    S[\rho]~|~E,  ~\{ \Gamma_i^{f.g.}\},~ \{ \bar\psi_{1,m} (t) \} \},
\end{eqnarray}
yields a functional relationship between $\bar{q}$ and $\beta \bar\psi- \alpha_1 \cos\theta - \alpha_2 \sin\theta \cos \phi - \alpha_3 \sin\theta \sin\phi$, and the function is determined by the values of the multipliers for $\{ \Gamma_i^{f.g.} \}$.
When the transverse angular momentum constraints $\{ \bar{\psi}_{1,\pm1}(t) \}$ are removed by setting their Lagrange multipliers $\alpha_2=\alpha_3=0$, we recover the type of solution $\bar{q} = F(\bar\psi +\Omega_L \cos\theta)$ obtained by Herbert \etal \cite{HDCP12short, HDCP12}, where $F$ is an arbitrary function and the state is stationary in a frame rotating at angular rate $\Omega_L$ about the $\bi{z}$ axis with respect to the co-rotating frame of the sphere. Anisotropy due to rotation of the sphere leads to nonzero $\{ \alpha_i \}$, and thus $\bar{q}$ is generally not a function of $\bar\psi$ and the equilibrium field is non-static under advection in the co-rotating frame of the sphere.  A functional relationship between $\bar{q}$ and $\bar{\psi}$ is a sufficient but unnecessary condition of static configuration, and a static configuration only means the single-valuedness of $\bar{q}$ along each streamline $\bar\psi = const.$. In special cases, the statistical equilibrium can be static and can even have functional relationship between $\bar{q}$ and $\bar\psi$: if the system has axial symmetry about the $\bi{z}$-axis and $\alpha_2 = \alpha_3= 0$, the zonal equilibrium solution $\bar{q}(\theta)$ with the corresponding $\bar{\psi}(\theta)$ is static; if $\bar\psi (\theta)$ is further invertible, there is a functional relationship between $\bar{q}$ and $\bar{\psi}$ despite the presence of the nonzero $\alpha_1 \cos\theta$ term. 

MRS-2 yields the linear vorticity-streamfunction relationship
\begin{eqnarray}
\bar{q} &=& \frac{1}{2\gamma_2} (\beta \bar\psi- \alpha_1 \cos\theta - \alpha_2 \sin\theta \cos \phi - \alpha_3 \sin\theta \sin\phi).
\end{eqnarray}
The equilibrium field is solved from the following vorticity-streamfunction equation
\begin{eqnarray}
 \bi{\nabla}^2 \bar\psi + 2 \Omega \cos \theta 
&=& \frac{1}{2\gamma_2} (\beta \bar\psi- \alpha_1 \cos\theta - \alpha_2 \sin\theta \cos \phi - \alpha_3 \sin\theta \sin\phi).
\label{MRS-2rotating}
\end{eqnarray}
The Coriolis term $2 \Omega \cos \theta$ in equation \eref{MRS-2rotating} is decoupled from modes with spherical wavenumber $\ell>1$, and as the $\ell =1$ modes are fixed by angular momentum, the problem reduces to the non-rotating case. That MRS-2 shows no structural changes due to rotation has its origin in the linear vorticity-streamfunction relationship.  This conclusion is based on the conservation of the $\bi{z}$-component angular momentum $L_z$. The equilibrium solution $\bar\psi$ of the generalized-entropy description with a linear vorticity-streamfunction relationship (or equivalently an ME principle) in the absence of any angular momentum constraint is rotation-dependent: it can be any configuration of $\ell=1$ modes that are constrained by energy on the non-rotating sphere, whereas $\bar\psi = \sqrt{3E} \cos\theta$ for any nonzero value of rotation rate $\Omega$ and represents a westward zonal flow \cite{HDCP12}. Further imposing the $L_z$ constraint makes the solution rotation-independent: the solution has a fixed $Y_{10}$ part, plus any configuration of $\{ Y_{1,\pm1}\}$ that contains the rest of the energy, regardless of rotation rate $\Omega$ \cite{HDCP12short, HDCP12}. Here as the angular momentum $\bi{L}(t)$ is imposed as a constraint, higher-order MRS-N with the addition of higher fine-grained Casimir constraints would yield a nonlinear vorticity-streamfunction relationship and thus the equilibrium solution would structurally depend on $\Omega$. However that calculation lies beyond the scope of this work. Shepherd's rigorous analytical work on 2D inviscid flows \cite{Shepherd87} provides an alternative perspective to see the conclusion that the effect of rotation on the structure of the equilibrium solution cannot be seen in MRS-2 but can only be seen by further imposing the higher fine-grained Casimir constraints. Shepherd argued that the quadratic invariants, energy and enstrophy, do not distinguish between the rotating and non-rotating cases, because the absolute enstrophy $\int  \rmd^2 \bi{r} q^2$ is only a linear combination of the relative enstrophy $\int  \rmd^2 \bi{r} \zeta^2$, the separately conserved $\bi{z}$-component angular momentum and a constant (see equation \eref{enstrophyRelation}). Thus the effect of rotation is expressed only in the third and higher Casimir invariants. Shepherd further mathematically proved that for a sufficiently large rotation rate, 2D Euler flows evolving from a non-trivial set of initial states remain anisotropic at all times, and thus the dynamics cannot be ergodic on the phase-space of constant energy and enstrophy that yields isotropic statistics. Shepherd's proof of anisotropy uses the conservation of the higher Casimirs, showing that the effect of rotation is expressed in the higher Casimir constraints. 

\begin{figure}
\centerline{\includegraphics[width=14.0 cm]{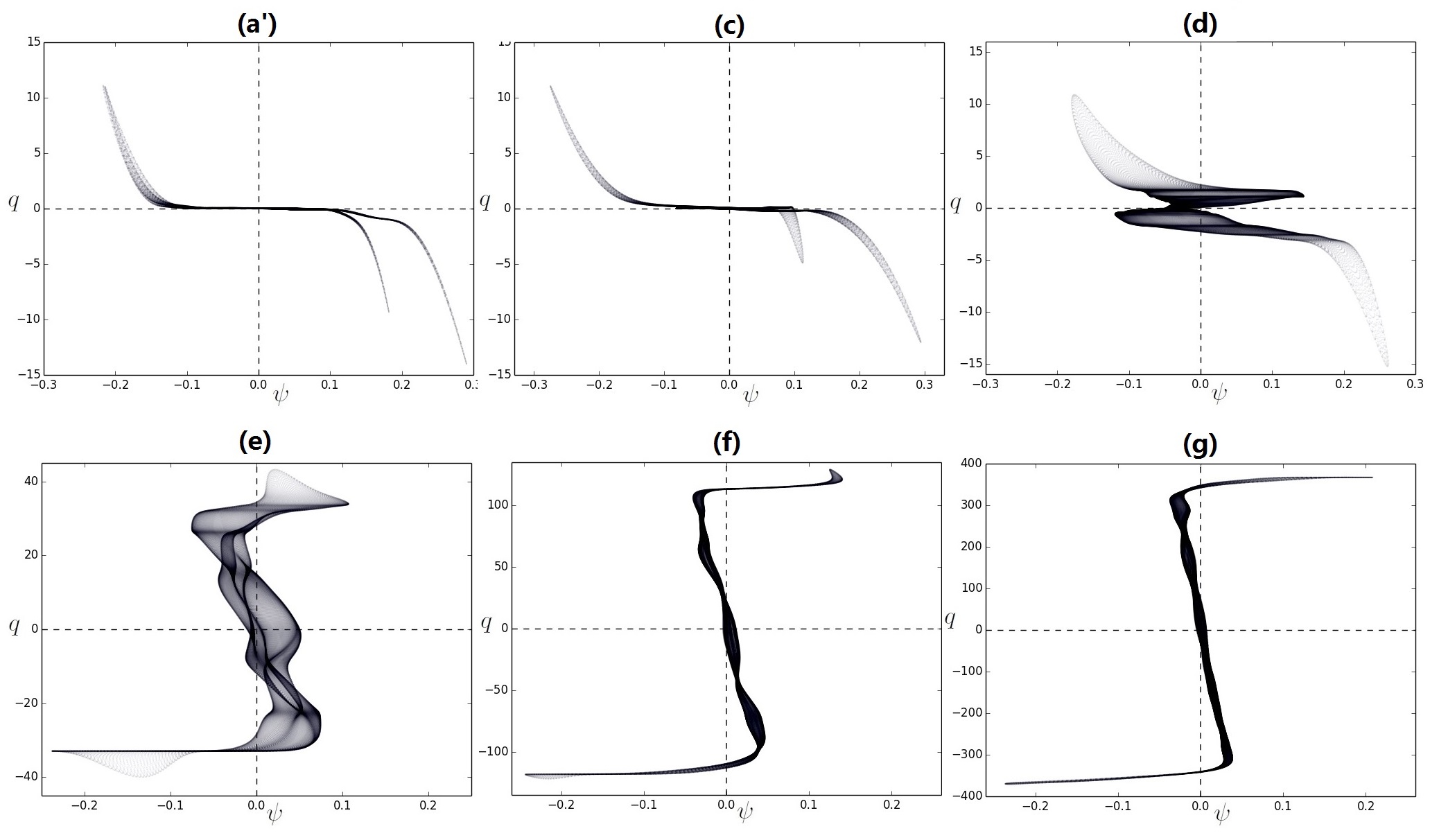}}
\caption{Scatter plots of the absolute vorticity $q$ versus streamfunction $\psi$ for the coherent states on spheres with different rotation rates as shown in \fref{rotate}. The figures are arranged in order of increasing rotation rate.  The relationship between $q$ and $\psi$ cannot be described with one or even several functions. 
\label{qPsiFig}}
\end{figure}

\begin{figure}
\centerline{\includegraphics[width=14.0 cm]{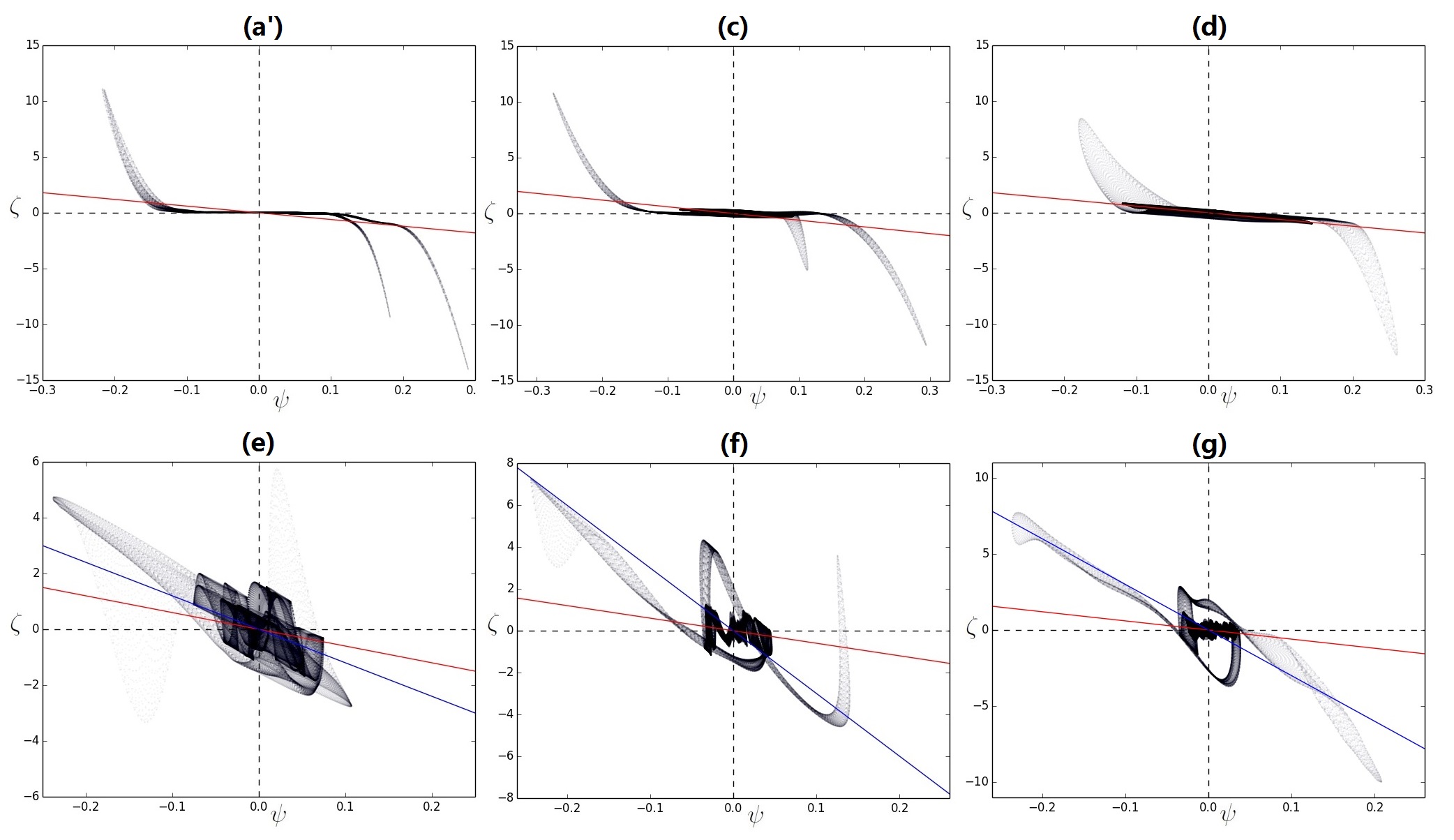}}
\caption{Same as \fref{qPsiFig} but for the relative vorticity $\zeta$ versus the relative streamfunction $\psi$. MRS-2 corresponds to the same red straight line for all rotation rates. Blue lines are the lines $\zeta = -\ell_\mathrm{m}(\ell_\mathrm{m}+1) \psi$ for the most energetic spherical wavenumber $\ell_\mathrm{m}$.  As the sphere rotates faster, the $\zeta$-$\psi$ relationship in simulation coherent structures deviates further from MRS-2.  Steeper slopes with rotation rates support the Rhines picture of an inverse energy cascade that is arrested at increasingly large spherical wavenumbers. 
\label{zetaPsiFig}}
\end{figure}

The vorticity-streamfunction relationship deserves further discussion.  When MRS-2 is applied to systems with zero angular momentum on the rotating sphere such as Spheres (c) - (g) as shown in \sref{NSrot}, the $\ell=1$ modes of equation \eref{MRS-2rotating} fixes the multipliers to be  
\begin{eqnarray}
\alpha_1 &=& -4 \gamma_2 \Omega,\\
\alpha_2 &=& \alpha_3=0.
\end{eqnarray}  
The anisotropy due to rotation is indicated by nonzero $\alpha_1$ and the axial symmetry requires $\alpha_2$ and $\alpha_3$ to vanish.
The MRS-2 relationship between the absolute vorticity field $\bar{q}$ and the relative streamfunction $\bar\psi$ is generally not functional due to the extra spatial dependence from the nonzero $\alpha_1$ term,
\begin{eqnarray}
\bar{q} (\bi{r}) = -6 \bar\psi (\bi{r}) + 2 \Omega \cos \theta, 
\end{eqnarray}
but the relationship between the relative vorticity field $\bar\zeta$ and the relative streamfunction $\bar\psi$ is the same straight line for all rotation rates,
\begin{eqnarray}
\bar{\zeta} (\bi{r}) = -6 \bar\psi (\bi{r}) .
\label{MRS2zetaPsi}
\end{eqnarray}
Scatter plots of the absolute vorticity $q$ versus the streamfunction $\psi$ for the coherent states of simulations Spheres (a') - (g) are shown in \fref{qPsiFig}, and those of the relative vorticity $\zeta$ versus the streamfunction $\psi$ are shown in \fref{zetaPsiFig}. The corresponding contour plots are given in \fref{rotate}. Again different behaviors are found in slow-, intermediate- and fast-rotation regimes. The $q$-$\psi$ plots of Spheres (c) and (d) show that the relationship between $q$ and $\psi$ becomes broad bands that cannot be described by one or several functions in the presence of small rotation. That agrees with the general statistical mechanical equilibrium for an anisotropic system where functional relationship exists between $\bar{q}$ and $\beta \bar\psi- \alpha_1 \cos\theta - \alpha_2 \sin\theta \cos \phi - \alpha_3 \sin\theta \sin\phi$, but generally not between $\bar{q}$ and $\bar\psi$. For Sphere (e) with an intermediate rotation rate, the $q$-$\psi$ plot has a wave-like vertical broad band with horizontal lines extending from it, again showing no functional relationship. For Spheres (f) and (g) under rapid rotation, the formation of quasi-static zonal structures makes the $q$-$\psi$ relationship almost a curve, though not a function.
The effect of rotation is readily seen in the $\zeta$-$\psi$ plots. The MRS-2 equilibria correspond to the same red straight line given in equation \eref{MRS2zetaPsi} for all rotation rates, and the relationship between $\zeta$ and $\psi$ in simulation coherent structures deviates from it as the sphere rotates faster. The scatter plots also show broad bands under slow rotation. 

Rapid rotation, however, separates the relationship between $\zeta$ and $\psi$ into two parts: the weak banded structures in the low- and mid-latitudes correspond to the part near the origin where $\zeta$ fluctuates about zero with no apparent correlation with $\psi$, while the two strong circumpolar jets are the two branches reaching out from those small fluctuations. The asymmetry between the north and south polar vortices is readily seen in the different slopes and lengths of the two circumpolar-jet branches, showing the breakdown of ergodicity due to confinement by the zonal flows.
The slope of the circumpolar-jet sector seen in the $\zeta$-$\psi$ plot is related to the most energetic wavenumber and the width of the circumpolar jet. The slope of the $\zeta$-$\psi$ relationship tends to increase with the rotation rate in accord with the Rhines picture where the inverse energy cascade is arrested at a larger wavenumber as the sphere rotates faster. 
Moreover a comparison between the $q$-$\psi$ plots and the $\zeta$-$\psi$ plots reveals information about structures at different latitudes encoded in the $2 \Omega \cos\theta$ difference between $q$ and $\zeta$. Compared to the $\zeta$-$\psi$ plots, the $q$-$\psi$ plots for the slowly-rotating systems Spheres (c) and (d) show a relative shift along the $q$-axis between the two branches. The two vortex pairs of the quadrupole structure are confined near different poles as seen in \fref{rotate}. For Spheres (e), (f) and (g) under intermediate and rapid rotation, the horizontal lines at extremal values of $q$ in $q$-$\psi$ plots are related to the circumpolar jets. They contrast the lines with slopes roughly close to $-\ell_\mathrm{m} (\ell_\mathrm{m}+1)$ in $\zeta$-$\psi$ plots because $2 \Omega $ is much larger than the extremal $\zeta$ values. The wave-like bands between the $q$ extrema in the $q$-$\psi$ plots contrast the random fluctuation near the $\zeta=0$ line in the $\zeta$-$\psi$ plots; banded structures trap random relative vorticity at various latitudes in the low- and mid-latitudes. That the wave-like band in the $q$-$\psi$ plot of Sphere (e) spans a comparable range of $\psi$ values as the horizontal lines shows that in the intermediate-rotation regime the alternating jets at mid- and low-latitudes are of comparable strength to the circumpolar jet. This wave-like band shrinks to a $\psi$-range much narrower than the horizontal lines, as the circumpolar jets dominate the banded structure at large rotation rates in Spheres (f) and (g). 

\section{Discussion and Conclusions\label{sec5}}

Coherent structures that appear at late times in numerical simulations of 2D inviscid fluids with hyperviscosity as the subgrid model were compared against statistical equilibria obtained from a series of MRS-N statistical mechanical descriptions. MRS-2 conserves up to the fine-grained enstrophy and completely condenses energy at the largest possible length-scale; in the absence of rotation it agrees qualitatively with the simulation coherent structures. Perturbative imposition of conservation of $\Gamma_4^{f.g.}$ up to the first order was shown to enhance the vorticity in the inner cores, improving agreement with the simulation coherent structures, and realizing a nonlinear vorticity-streamfunction relationship. Effects of rotation can only be captured by imposing higher fine-grained Casimir constraints beyond MRS-2.   The result lends support to the (somewhat paradoxical) idea that hyperviscosity as a subgrid model helps to restore the dynamical effects of the conservation laws lost in the spatial discretization.  An interesting extension would be to solve the first-order perturbative MRS-4 in the rotating case and make a comparison with the numerical simulations, but that work lies beyond the scope of the present paper.

Statistical descriptions are limited by the breakdown of ergodicity. In the absence of stochastic forcing the 2D Euler equations are non-ergodic \cite{Bouchet,bkl01}:  Time averages along an arbitrary trajectory of the dynamical system yields a different result than ensemble averages.  The breakdown of ergodicity means that details of the final coherent structures retain some memory of the initial conditions. There are two sources of such non-ergodicity in numerical simulation as discussed in \sref{MRS2NSNonrot} and \sref{NSrot}: the failure to generate broad-band turbulence with energy shared broadly among many wavenumbers, and the dynamical confinement of vorticity. If broad-band turbulence is generated at early times, near equilibrium can be reached at later times; otherwise the system remains away from equilibrium and coherent structures that arise are sensitive to the initial condition.  Initial states with energy concentrated at a few low wavenumbers \cite{YMC03, Yin04} and initial states in the wave regime on a rotating sphere both fail to generate broad-band turbulence. Dynamical confinement of vorticity occurs in all systems. It can block the system from sampling the available phase-space and lead to asymmetry in coherent structures. Even in the absence of rotation, for systems where the broad-band turbulence is generated during evolution, there remain small differences between the final coherent vortices as vorticity can be dynamically trapped inside vortices. The double branches in the plot of vorticity versus streamfunction seen in \fref{omegaPsiFig} reflect such asymmetry; this cannot be ascribed to globally conserved quantities. An idea for overcoming the non-ergodicity due to vortex self-confinement within the framework of the MRS theory was discussed in the references \cite{CS98, BCPS99}.

Breakdown of ergodicity due to confinement is more severe on the rotating sphere because of strong anisotropy. Initial asymmetry in the strengths and widths of jets and polar vortices is maintained by confinement by the zonal flows in the presence of rotation, but paradoxically rapid rotation seems to restore some of the symmetry \cite{cho}.  Comparison of the two circumpolar-jet branches of each $\zeta$-$\psi$ plot for Spheres (e), (f) and (g) in \fref{zetaPsiFig} illustrates this feature.
Sensitivity of the final state to initial conditions is also observed in the more complicated general circulation models. For instance, in a primitive equation simulation of hot Jupiter exoplanets, Thrastarson and Cho found that different initial conditions lead to markedly different coherent structures \cite{TC10}.  Breakdowns in ergodicity for deterministic dynamical systems such as these pose serious challenges to approaches based on equilibrium statistical mechanics.  The addition of small stochastic forcing to restore ergodicity may not help because the forcing may also alter the long-time behavior. For example in the presence of small-scale random forcing, the inverse cascade on the rotating sphere need not be arrested at the Rhines wavenumber.  At late times, jets can undergo sudden merging and disappearance, settling into a quasi-steady state of only two or three jets regardless of the rate of rotation \cite{OTY10}. Nevertheless non-ergodicity does not completely rule out the use of equilibrium statistical mechanics, but only limits its usefulness. Equilibrium statistical mechanics can make qualitatively accurate predictions about whether the coherent states on the torus are unidirectional or dipolar, even though broad-band turbulence is not well generated for systems that evolve into unidirectional flows \cite{YMC03, LN13}. Whether equilibrium statistical mechanics works depends not only on the breakdown of ergodicity in the system, but also on the type of coherent structure. 

Note that equilibrium statistical mechanical descriptions introduced in \sref{intro} all fail to consider the topological constraint that vortex contours of 2D Euler flows cannot cross \cite{frisch}. Conservation of Casimirs is a necessary but incomplete constraint on the inviscid flow.  Even conservative simulations that conserve many Casimir-like quantities beyond the first two \cite{AM03} may not respect the topological constraint. Nevertheless the topological constraint may be irrelevant to coarse-grained 2D Euler flows above a finite resolution scale, even as it plays an important role in studies of contour dynamics (see for example \cite{Dritschel88}). 

Finally we point out a possible application of equilibrium descriptions of fluids on rotating spheres to understanding gravity around spinning black holes through the AdS/CFT correspondence \cite{CLMRS12}. 

\ack
We are grateful for helpful discussions with Freddy Bouchet, James Cho, Kiori Obuse, Peter Weichman and Allan Adams. We also thank the two anonymous referees for thoughtful reports that significantly improved the manuscript. WQ thanks the Global COE program at Kyoto University for the opportunity to visit M. Yamada's research group.  JBM thanks the Aspen Center for Physics (supported in part by NSF Grant No. 1066293) for its hospitality during the summer workshop on  ``Stochastic Flows and Climate Modeling.'' WQ and JBM thank the Kavli Institute for Theoretical Physics for hosting their visits (supported in part by the NSF Grant No. NSF PHY11-25915). This work is supported in part by the NSF under grant Nos. DMR-0605619 and CCF-1048701.

\appendix

\section{First-order perturbative MRS-4 \label{app1}}
\subsection{Non-rotating sphere \label{app1sphere}}
The variational problem equation \eref{MRS4} on the non-rotating sphere can be solved by assuming that the fine-grained quartic Casimir constraint $\Gamma_4^{f.g.}$ is weakly imposed. As a first step, we expand the vorticity-streamfunction equation $\bar\omega = \int  \rmd\sigma ~\rho(\bi{r},\sigma)\cdot \sigma =  \nabla^2 \bar\psi$ order by order. The vorticity-streamfunction equation
\begin{eqnarray}
 \nabla^2 \bar\psi &=&  \int  \rmd\sigma ~\rho (\bi{r},\sigma)\cdot \sigma \nonumber\\
&=& \int  \rmd\sigma~ \rho_0(\bi{r}, \sigma) \times \{1+\gamma_4 [m_4(\bi{r}) - \sigma^4]\}\cdot \sigma+ {\cal O}(\gamma_4^2) \nonumber\\
&=& m_1(\bi{r}) + \gamma_4 [m_1(\bi{r})\cdot m_4(\bi{r}) - m_5(\bi{r})  ]+ {\cal O}(\gamma_4^2),
\label{expandOmegaPsiEqn}
\end{eqnarray}
where the perturbation expansion equation \eref{expand} is used, and $m_i(\bi{r})$ is the $i$-th-order moment of the Gaussian distribution $\rho_0(\bi{r},\sigma)$, $m_i(\bi{r}) \equiv \int  \rmd\sigma \rho_0(\bi{r},\sigma) \sigma^i$. Each moment $m_i(\bi{r})$ is a polynomial function of the mean $[\beta \bar\psi(\bi{r})- \alpha_1 \cos\theta - \alpha_2 \sin\theta \cos\phi -  \alpha_3 \sin\theta \sin\phi ]/(2 \gamma_2)$ and the variance $1/(2\gamma_2)$, so the right-hand side of equation \eref{expandOmegaPsiEqn} can be expressed as a function of $\{ \gamma_4, \bar\psi(\bi{r}), \{\alpha_i\},\beta, \gamma_2, \theta,\phi\}$. Then we plug the perturbation expansion equations \eref{expand1} - \eref{expand4} for $\{\bar\psi(\bi{r}), \{\alpha_i\}, \beta, \gamma_2 \}$ into equation \eref{expandOmegaPsiEqn}, and arrange terms by orders of $\gamma_4$. The bars on $\psi$ and $\omega$ will be omitted in the following for simplicity. 
At 
${\cal O}(1)$, it reduces to that of MRS-2: 
\begin{eqnarray}
\nabla^2 \psi^{(0)} 
&=& \frac{1}{2 \gamma_2^{(0)}} (- \alpha_1^{(0)} \cos\theta - \alpha_2^{(0)}  \sin\theta \cos\phi - \alpha_3^{(0)}  \sin\theta \sin\phi )\nonumber\\
&& + \frac{\beta^{(0)} }{2\gamma_2^{(0)}} \psi^{(0)}.
\end{eqnarray}
Projecting the equation to $\ell = 1$ modes and using the zero-angular-momentum constraint at ${\cal O}(1)$, $\psi_{10}^{(0)} = \psi_{1,\pm1}^{(0)} = 0$, yields
\begin{eqnarray}
\alpha_1 ^{(0)}= \alpha_2^{(0)} = \alpha_3^{(0)} = 0. \label{a123MRS4Sphere}
\end{eqnarray}
The nontrivial zeroth-order solution is a degenerate state with spherical harmonics $\{ Y_{\ell^{*}, m}, ~m = -\ell^{*}, \cdots, \ell^{*} \}$ for some $\ell^{*} \geq 2$. Calculation of MRS-2 in \sref{MRS2nonrot} has already shown that $\ell^{*} = 2$ maximizes the entropy at the zeroth order in perturbation theory. Thus at zeroth-order the entropy is maximized for
\begin{eqnarray}
\psi^{(0)} &=& \sum\limits_{m = -2}^{2} \psi_{2 m}^{(0)} Y_{2 m},\label{zerothSphere}\\
\beta^{(0)} &=& -12 \gamma_2^{(0)}, \label{betaMRS4Sphere}
\end{eqnarray}
where the arbitrary amplitudes $\{  \psi_{2m}^{(0)} \}$ satisfy equation \eref{reality}.
The $SO(3)$ symmetry of the optimization problem requires that any rotation of the maximum entropy fields, $\psi^{(0)} = \sum\limits_{m=-2}^{2} \psi_{2m}^{(0)} Y_{2m}$ with its corresponding correction, also maximizes the entropy. After extracting the three arbitrary rotation angles, the five real degrees of freedom in $\{ \psi_{2m}^{(0)} \}$ reduce to two rotation-invariant variables $F[\psi^{(0)}]$ and $G[\psi^{(0)}]$ as defined in equations \eref{FPsi} and \eref{GPsi} using the zeroth-order solution. The dimensionless quantity $\eta[\psi^{(0)}]$ defined by equation \eref{etaPsi} characterizes the shape of the configuration. The optimization problem is also invariant under a change of sign of the field. Therefore the shape and overall magnitude $\{ \eta[\psi^{(0)}],~ G[\psi^{(0)}]\}$ suffice to describe the zeroth-order solution of the optimization problem. The definition of $\eta$, $F$ and $G$ can be extended to any $g(\theta,\phi) = \sum\limits_{m=-2}^{2} g_{2m}Y_{2m}$ by replacing $\{\bar\psi_{2m}\}$ with $\{  g_{2m} \}$. 

At ${\cal O}(\gamma_4)$, further using equations (\ref{a123MRS4Sphere}) and (\ref{betaMRS4Sphere}) to simplify the equation, the vorticity-streamfunction equation reads
\begin{eqnarray}
(\nabla^2  +6)\psi^{(1)} &=&  \frac{1}{2\gamma_2^{(0)}}( - \alpha_{1}^{(1)} \cos\theta- \alpha_{2}^{(1)} \sin\theta \cos\phi - \alpha_{3}^{(1)} \sin\theta \sin\phi )\nonumber\\
&&  + \left( \frac{\beta^{(1)}}{2 \gamma_2^{(0)}}-6 s + \frac{18}{[\gamma_2^{(0)}]^2} \right) \psi^{(0)}+\frac{432 [{\psi}^{(0)}]^3}{\gamma_2^{(0)}}.
\label{vsEqnLinearOrder}
\end{eqnarray}
The zero-angular-momentum constraint $\psi_{10}^{(1)} =\psi_{1,\pm1}^{(1)} = 0 $ again requires that the multipliers $\{ \alpha_i \}$ vanish,
\begin{eqnarray}
\alpha_1^{(1)} = \alpha_2^{(1)}=\alpha_3^{(1)}=0.
\label{alphaLinearOrder}
\end{eqnarray}
Simplifying equation \eref{vsEqnLinearOrder} using equation \eref{alphaLinearOrder}  and projecting it to $Y_{\ell,m}$ yields
\begin{eqnarray}
[-\ell(\ell + 1) +  6] \psi_{\ell, m}^{(1)}&=&  \left( \frac{\beta^{(1)}}{2 \gamma_2^{(0)}}-6 s + \frac{18}{[\gamma_2^{(0)}]^2} \right) \psi_{2 m}^{(0)} \delta_{\ell, 2}\nonumber\\
+\frac{432}{\gamma_2^{(0)}}\sum\limits_{ m_1=-2}^{2} \sum\limits_{ m_2=-2}^{2} && B_{\ell,m; m_1, m_2} \psi_{2, m_1}^{(0)} {\psi}_{2, m_2}^{(0)} {\psi}_{2, m-m_1 - m_2}^{(0)} ,
\label{vsEqnLinearOrderExpand}
\end{eqnarray}
where the constant coefficients
\begin{eqnarray}
 B_{\ell,m; m_1, m_2} &\equiv& \int  \rmd^2\bi{r}~ Y_{\ell,m}^{*} Y_{2, m_1} Y_{2, m_2} Y_{2, m-m_1 - m_2}.
\end{eqnarray}
Note that spherical harmonics have well-defined parity: spherical harmonics with odd $\ell$'s are odd with respect to the reflection about the origin, whereas those with even $\ell$'s are even. The $\ell=2$ spherical harmonics have even parity and so do the product of three of them, so the expansion of $Y_{2, m_1} Y_{2, m_2} Y_{2, m-m_1 - m_2}$ on the basis of spherical harmonics cannot have terms with odd $\ell$'s. Furthermore the expansion of $Y_{2, m_1} Y_{2, m_2} Y_{2, m-m_1 - m_2}$ can only have modes with $0 \leq \ell \leq 2+2+2$, because the expansion of any product $Y_{\ell_1 m_1} Y_{\ell_2 m_2}$ on the basis of spherical harmonics can only have modes satisfying $|\ell_1-\ell_2|\leq \ell \leq \ell_1+\ell_2$. This is a property of the Clebsch-Gordan coefficients of the $SO(3)$ group. Therefore the coefficients $B_{\ell,m; m_1, m_2}$ are nonzero only for $\ell = 0,~2,~4,~6$. Taking $\ell = 0,~ 4,~ 6$ and $m = -\ell, \cdots, ~\ell$, the nonzero components of $\psi^{(1)}$ are given as functions of the zeroth-order field and multipliers: 
\begin{eqnarray}
\psi_{\ell m }^{(1)} &=& \frac{432  }{[6-\ell(\ell + 1) ]\cdot \gamma_2^{(0)}}\sum\limits_{ m_1=-2}^{2} \sum\limits_{ m_2=-2}^{2}  B_{\ell,m; m_1, m_2}  \psi_{2, m_1}^{(0)} {\psi}_{2, m_2}^{(0)} {\psi}_{2, m-m_1 - m_2}^{(0)}.\nonumber\\
\label{cor}
\end{eqnarray}
Taking $\ell = 2$ of equation \eref{vsEqnLinearOrderExpand}, the left-hand side vanishes, and the equation becomes a nonlinear eigenvalue problem with eigenvalue $ \beta^{(1)}/(2 \gamma_2^{(0)})-6 s + {18}/{[\gamma_2^{(0)}]^2} $ and eigenvector $(\psi_{2,-2}^{(0)},~\psi_{2,-1}^{(0)},~\psi_{20}^{(0)},~\psi_{21}^{(0)},~\psi_{22}^{(0)})$: 
\begin{eqnarray}
0=f_{2m} &\equiv&  \left( \frac{\beta^{(1)}}{2 \gamma_2^{(0)}}-6 s + \frac{18}{[\gamma_2^{(0)}]^2} \right) \psi_{2 m}^{(0)}  \nonumber\\
+\frac{432}{\gamma_2^{(0)}}  \sum\limits_{ m_1=-2}^{2} \sum\limits_{ m_2=-2}^{2} && B_{2 m; m_1, m_2} \psi_{2, m_1}^{(0)} {\psi}_{2, m_2}^{(0)} {\psi}_{2, m-m_1 - m_2}^{(0)} .
\label{nonlinearEigen}
\end{eqnarray}
That result is equivalent to requiring $G[f]\equiv \sum\limits_{m = -2}^{2} |f_{2m}|^2=0$:
\begin{eqnarray}
0 &=& G[f]\nonumber\\
& =&\frac{G[\psi^{(0)}]}{196 \pi^2 [\gamma_2^{(0)}]^4} \left(3240  \gamma_2^{(0)}\cdot G[\psi^{(0)}] + 7 \pi \{\beta^{(1)}  \gamma_2^{(0)}-12 s  [\gamma_2^{(0)}]^2+ 36 \}\right)^2 \nonumber
\end{eqnarray}
fixing $\beta^{(1)}$ without placing any constraint on the zeroth-order field:
\begin{eqnarray}
\beta^{(1)}  = -12  \left(\frac{270 \cdot G[\psi^{(0)}]}{7 \pi}  +\frac{3}{\gamma_2^{(0)}} - s \gamma_2^{(0)}\right).
 \label{b1}
\end{eqnarray}
The $\ell = 2$ components of $\psi^{(1)}$ are left arbitrary at this order. We can set them to be zero for simplicity.

The above solution, substituted into the constraints of energy, fine-grained enstrophy and fine-grained quartic Casimir, determines $\{ \eta[\psi^{(0)}],~ G[\psi^{(0)}],~ s, ~\gamma_2^{(0)}, ~\gamma_4\}$. The constraints become
\begin{eqnarray}
E  &=&   \frac{3}{4 \pi} G[\psi^{(0)}]  + {\cal O}(\gamma_4^2),\label{energyEqn}\\
\Gamma_2^{f.g.} &=& 36  \cdot G[\psi^{(0)}]+ \frac{2 \pi}{\gamma_2^{(0)}}- \frac{6 \pi \gamma_4}{[\gamma_2^{(0)}]^3} - \frac{108 \gamma_4 \cdot G[\psi^{(0)}] }{[\gamma_2^{(0)}]^2}\nonumber\\
&& + \frac{2 \pi s \gamma_4}{\gamma_2^{(0)}}+ {\cal O}(\gamma_4^2), \label{Gamma2Eqn}\\
\Gamma_4^{f.g.} &=& \frac{4860\cdot G[\psi^{(0)}]^2}{7 \pi} + \frac{3 \pi (1+ 2 s \gamma_4)}{[\gamma_2^{(0)}]^2} + \frac{108  (1+ s \gamma_4)\cdot G[\psi^{(0)}]}{\gamma_2^{(0)}} \nonumber\\
&&-\frac{ 24 \pi \gamma_4}{[\gamma_2^{(0)}]^4} - \frac{1080 \gamma_4 \cdot  G[\psi^{(0)}] }{[\gamma_2^{(0)}]^3} - \frac{87480  \gamma_4  \cdot G[\psi^{(0)}]^2}{7 \pi [\gamma_2^{(0)}]^2 }\nonumber\\
&&-\frac{ 1508155200 \gamma_4 \cdot  G[\psi^{(0)}]^3}{49049 \pi^2 \gamma_2^{(0)}} - \frac{ 34992000  \gamma_4 \cdot G[\psi^{(0)}]^3 \cdot\eta[\psi^{(0)}]}{7007 \pi^2 \gamma_2^{(0)}}\nonumber\\
&& + {\cal O}( \gamma_4^2).
\label{Gamma4Eqn}
\end{eqnarray}
Equation \eref{energyEqn} determines the overall magnitude of the zeroth-order field
\begin{eqnarray}
G[\psi^{(0)}] &=& {4 \pi E}/{3},
\label{Gsolution}
\end{eqnarray}
and it is substituted into the other two constraints to eliminate $G[\psi^{(0)}]$. The enstrophy constraint equation \eref{Gamma2Eqn} at ${\cal O}(1)$ gives 
\begin{eqnarray}
\gamma_2^{(0)} &=&  \frac{ 2\pi}{\Gamma_2^{f.g.} - 48 \pi E},
\label{gamma2solution}
\end{eqnarray} 
the same as the MRS-2 solution equation \eref{gamma2MRS2}. At ${\cal O}(\gamma_4)$ the enstrophy constraint fixes the first-order correction to the multiplier $\gamma_2$:
\begin{eqnarray}
s &=& \frac{ 3 \Gamma_2^{f.g.} (\Gamma_2^{f.g.}-48 \pi E)}{4 \pi^2}.
\label{ssolution}
\end{eqnarray}
Using the solution of $\gamma_2^{(0)}$ and $s$, the quartic Casimir constraint equation \eref{Gamma4Eqn} is re-expressed in the form
\begin{eqnarray}
\Gamma_4^{f.g.} &=& P_1(E, ~\Gamma_2^{f.g.}) +\gamma_4 \cdot P_2(E, ~\Gamma_2^{f.g.}, ~\eta[\psi^{(0)}])+{\cal O}( \gamma_4^2),
\end{eqnarray}
where $P_1$ and $P_2$ are polynomial functions. Thus the small multiplier
\begin{eqnarray}
\gamma_4 &=& \frac{\Gamma_4^{f.g.}-P_1}{P_2} +{\cal O}( \gamma_4^2)\nonumber\\
&=&\left( \frac{3456 E^2 \pi}{7} - \frac{3 (\Gamma_2^{f.g.})^2}{ 4 \pi} + \Gamma_4^{f.g.}\right) / (  \frac{43951693824 E^4 \pi}{49049} - \frac{4112197632 E^3 \Gamma_2^{f.g.}}{49049} \nonumber\\
&&+ \frac{ 15552 E^2 (\Gamma_2^{f.g.})^2}{7 \pi} - \frac{ 3  (\Gamma_2^{f.g.})^4}{8 \pi^3}+ \frac{ 1990656000 E^4 \pi \eta}{7007 }- \frac{41472000 E^3  \Gamma_2^{f.g.} \eta}{7007}  )\nonumber\\
&&+{\cal O}( \gamma_4^2).
\label{gamma4solution}
\end{eqnarray}
To sum up, the equilibrium vorticity field is 
\begin{eqnarray}
\psi(\theta, \phi) &=&  \sum\limits_{m=-2}^{2} \psi_{2,m}^{(0)} Y_{2m}(\theta,\phi) +  \gamma_4 \sum\limits_{\ell=0,4,6} \sum\limits_{m=-\ell}^{\ell} \psi_{\ell,m}^{(1)} (\{ \psi_{2m}^{(0)}\}) \cdot Y_{\ell m}(\theta,\phi)\nonumber\\
&& + {\cal O}(\gamma_4^2) ,
\end{eqnarray}
where $\psi_{\ell,m}^{(1)} (\{ \psi_{2m}^{(0)}\})$ is given in equation (\ref{cor}).
The solution has eight parameters  $\{  \psi_{2m}^{(0)},~ \gamma_2^{(0)},~ s, ~\gamma_4 \}$.  The three $SO(3)$ degrees of freedom and the sign in $\psi^{(0)}$ are arbitrary due to symmetry, and only five variables $\{\eta[\psi^{(0)}],~ G[\psi^{(0)}],~  \gamma_2^{(0)},~s, ~\gamma_4\}$ parameterize the equilibrium solution. The solution has a free parameter $\eta[\psi^{(0)}]$ and other parameters $\{ G[\psi^{(0)}],~  \gamma_2^{(0)},~s, ~\gamma_4\}$ are determined by the conserved quantities and $\eta[\psi^{(0)}]$ through equations \eref{Gsolution}, \eref{gamma2solution}, \eref{ssolution} and \eref{gamma4solution}. That $\eta[\psi^{(0)}]$ is undetermined shows that degeneracy in the zeroth-order solution is not lifted by first-order perturbation. Furthermore, maximizing entropy up to first order does not select any particular zeroth-order configuration because substituting the solution into the entropy yields
\begin{eqnarray}
S[\rho]&=& -2 \pi\ln\frac{2 \pi}{\Gamma_2^{f.g.}-48 \pi E}+2 \pi( \ln\pi +1 )  +  {\cal O}(\gamma_4^2),
\end{eqnarray}
where the zeroth-order is the MRS-2 entropy equation \eref{entropyMRS2} and the first-order correction vanishes for all $\eta[\psi^{(0)}]$.

The vorticity-streamfunction relationship is 
\begin{eqnarray}
\omega  = -6 \psi + \gamma_4 \cdot \left( \frac{\beta^{(1)}}{2 \gamma_2^{(0)}}-6 s + \frac{18}{[\gamma_2^{(0)}]^2} \right) \psi+\frac{432 \gamma_4 {\psi}^3}{\gamma_2^{(0)}} + {\cal O}(\gamma_4^2).
\end{eqnarray}
As discussed in \sref{pert}, the sign of $\gamma_4$ determines whether the first-order correction sharpens or weakens the cores of zeroth-order vortices. Another way to see this is by noting that for any given zeroth-order field with specific energy, the first-order correction is proportional to the same field:
\begin{eqnarray}
\gamma_4 \psi^{(1)}(\theta, \phi) 
&=& \frac{\gamma_4}{\gamma_2^{(0)}} \sum\limits_{\ell=0,4,6} \sum\limits_{m=-\ell}^{\ell} \frac{432  }{6-\ell(\ell + 1)}\sum\limits_{ m_1=-2}^{2} \sum\limits_{ m_2=-2}^{2}  B_{\ell,m; m_1 , m_2} \nonumber\\
&& \psi_{2, m_1}^{(0)} \psi_{2, m_2}^{(0)} {\psi}_{2, m-m_1 - m_2}^{(0)}  Y_{\ell m}(\theta,\phi)\nonumber\\
&\propto&\sum\limits_{\ell=0,4,6} \sum\limits_{m=-\ell}^{\ell} \frac{432 }{6-\ell(\ell + 1)}\sum\limits_{ m_1=-2}^{2} \sum\limits_{ m_2=-2}^{2}  B_{\ell,m;m_1, m_2} \nonumber\\
&&\psi_{2, m_1}^{(0)} \psi_{2, m_2}^{(0)} {\psi}_{2, m-m_1 - m_2}^{(0)}  Y_{\ell m}(\theta,\phi).
\end{eqnarray}
The proportionality factor is $\gamma_4/\gamma_2^{(0)}$ and changing the sign of $\gamma_4$ turns a core-sharpening correction to a core-weakening one or the opposite. This picture is useful in understanding the behavior of statistical equilibrium when the resolved values of energy, enstrophy and quartic Casimir at different times in a simulation such as Sphere (a) are used. Energy changes little above the grid scale during the relaxation but enstrophy and quartic Casimir change significantly. The calculated zeroth-order equilibrium is almost unchanged over time, whereas the effect of first-order correction changes because $\gamma_4$ changes with enstrophy and the quartic Casimir as shown in \fref{gamma4Fig}.

\subsection{Torus \label{app1torus}}
The calculation can be extended to $2\pi \times 2\pi $ torus. The exact problem reads
\begin{eqnarray}
\max\limits_{\rho(\bi{r}, \sigma)} \{  S[\rho]~|~E, ~\Gamma_2^{f.g.},~\Gamma_4^{f.g.}\}.
\end{eqnarray}
The calculation is similar to that on the sphere but without the constraint of vanishing angular-momentum.
At ${\cal O}(1)$, the vorticity-streamfunction equation reduces to that of MRS-2: 
\begin{eqnarray}
\nabla^2 \psi^{(0)} &=&  \frac{\beta^{(0)} }{2 \gamma_2^{(0)}} \psi^{(0)}.
\end{eqnarray}
At zeroth order the entropy is maximized at the degenerate equilibrium of MRS-2:
\begin{eqnarray}
\psi^{(0)} &=& A \cdot  e_{10} + B \cdot e_{01} +c.c. ,\label{zerothTorus}\\
\beta^{(0)} &=& -2 \gamma_2^{(0)}, \label{beta}
\end{eqnarray}
where $\{ A,~ B\}$ are arbitrary complex amplitudes, $\{ e_{jk}\}$ is the eigenbasis of the Laplacian operator as defined in equation \eref{torusBasis} and $c.c.$ represents complex conjugates. The arbitrary phases of $\{ A,B\}$ represent the translational degrees of freedom along $\bi{x}$ and $\bi{y}$ axes, and the translation-invariant quantities $|A|$ and $|B|$ describe the zeroth-order solution.
The vorticity-streamfunction equation at ${\cal O}(\gamma_4)$ is
\begin{eqnarray}
(\nabla^2  +1)\psi^{(1)} =  \left\{ \frac{\beta^{(1)}}{2 \gamma_2^{(0)}} + \frac{3}{[\gamma_2^{(0)}]^2}-s \right\} \psi^{(0)}+ \frac{2 [{\psi}^{(0)}]^3}{\gamma_2^{(0)}}.
\label{vsEqnLinearOrderTorus}
\end{eqnarray}
Projecting it to $\{ e_{\pm1,0}, ~e_{0,\pm1}\}$ yields the nonlinear eigenvalue problem with the two equations
\begin{eqnarray}
0&=&  \left\{ \frac{\beta^{(1)}}{2 \gamma_2^{(0)}} + \frac{3}{[\gamma_2^{(0)}]^2}-s \right\} \cdot A + \frac{3 }{ 2 \pi^2 \gamma_2^{(0)}}\cdot A \cdot (|A|^2 + 2 |B|^2) ,\label{eigentorus1}\\
0&=& \left\{ \frac{\beta^{(1)}}{2 \gamma_2^{(0)}} + \frac{3}{[\gamma_2^{(0)}]^2}-s  \right\}  \cdot B + \frac{3}{ 2 \pi^2 \gamma_2^{(0)}}\cdot B \cdot (2 |A|^2 +  |B|^2) 
\label{eigentorus2}
\end{eqnarray}
and their complex conjugates. Nontrivial solution is one of the three cases. First, if the flow is unidirectional along the $\bi{x}$-direction with $|A| = 0$ but $|B| \neq 0$, equation \eref{eigentorus1} is trivial, and equation \eref{eigentorus2} determines $\beta^{(1)}$:
\begin{eqnarray}
\beta^{(1)} &=& -\frac{6}{\gamma_2^{(0)}}+ 2\cdot s\cdot \gamma_2^{(0)} - \frac{3}{\pi^2} \cdot |B|^2.
\end{eqnarray} 
Likewise if the flow is unidirectional along the $\bi{y}$-direction with $|B| = 0$ but $|A| \neq 0$, 
\begin{eqnarray}
\beta^{(1)} &=&-\frac{6}{\gamma_2^{(0)}}+ 2\cdot s\cdot \gamma_2^{(0)} - \frac{3}{\pi^2} \cdot |A|^2.
\end{eqnarray} 
If it is a dipole with $|A|\neq 0$ and $|B| \neq 0$, the eigenvalue problem only allows symmetric dipoles with
\begin{eqnarray}
|A| &=& |B|, \\
\beta^{(1)} &=&-\frac{6}{\gamma_2^{(0)}}+ 2\cdot s\cdot \gamma_2^{(0)} - \frac{9}{\pi^2} \cdot |A|^2.
\end{eqnarray}
In contrast to the sphere, the nonlinear eigenvalue problem on the torus not only determines $\beta^{(1)}$, but also partly lifts the degeneracy in the zeroth-order field. Projecting equation \eref{vsEqnLinearOrderTorus} to other modes yields the first-order correction as a function of $\{A,~ B, ~\gamma_2^{(0)}\}$:
\begin{eqnarray}
\gamma_4 \psi^{(1)} &=&\gamma_4 \sum\limits_{|j|+|k|=3} \psi_{jk}^{(1)} e_{jk}\nonumber\\
&=& -\frac{\gamma_4}{8 \pi^2 \gamma_2^{(0)}}( \frac{ A^3}{2  } e_{30}  +\frac{ B^3}{2 } e_{03} +{3 A^2 B} e_{21}  +{3 A  B^2 } e_{12} \nonumber\\
&&+{3  A^{*} B^2}  e_{-1,2} +{3 A^2 B^{*}} e_{2,-1})+ c.c..
\end{eqnarray}
The arbitrary $\psi_{\pm 1,0}^{(1)}$ and $\psi_{0,\pm1}^{(1)}$ are chosen to be zero. Again the sign of $\gamma_4$ determines whether correction sharpens or weakens cores. All the parameters $\{|A|,~|B|, ~\gamma_2^{(0)}, ~s, ~\gamma_4\}$ are fixed by the constraints. For the unidirectional flow in the $\bi{x}$-direction, $A = 0$ and the constraints yield
\begin{eqnarray}
|B|^2 &=& 4 \pi^2  E ,\\
\gamma_2^{(0)} &=&  \frac{2 \pi^2}{\Gamma_2^{f.g.} - 8 \pi^2 E},\\
s &=& \frac{3 \Gamma_2^{f.g.} (\Gamma_2^{f.g.} - 8 \pi^2 E)}{4 \pi^4},\\
\gamma_4 &=&   \frac{192 E^2 \pi^8 - 6 \pi^4 [\Gamma_2^{f.g.}]^2 + 8 \pi^6 \Gamma_4^{f.g.}}{
20736 E^4 \pi^8 - 7968 E^3 \pi^6 \Gamma_2^{f.g.} +  864 E^2 \pi^4 [\Gamma_2^{f.g.}]^2 - 3[ \Gamma_2^{f.g.}]^4}.
\label{gamma4Solution1}
\end{eqnarray}
 The unidirectional flow in the $\bi{y}$-direction has $B=0$ and $|A|^2 = 4 \pi^2  E $ and the same multipliers as flow in the $\bi{x}$-direction.
Symmetric dipole has $|A|^2  = |B|^2 = 2 \pi^2  E $, the same $\gamma_2^{(0)}$ and $s$ as the unidirectional flows, but different $\gamma_4$:
\begin{eqnarray}
\gamma_4  &=& \frac{96 E^2 \pi^8 - 6 \pi^4 [\Gamma_2^{f.g.}]^2 + 8 \pi^6 \Gamma_4^{f.g.}}{
2880 E^4 \pi^8 - 2280 E^3 \pi^6 \Gamma_2^{f.g.} +  432 E^2 \pi^4 [\Gamma_2^{f.g.}]^2 - 3[ \Gamma_2^{f.g.}]^4}.
\label{gamma4Solution3}
\end{eqnarray}
For fixed $\{ E, ~\Gamma_2^{f.g.}, ~\Gamma_4^{f.g.}\}$, three solutions up to any arbitrary translation are obtained. They are equally favored in the first-order perturbation theory, because the first-order correction to entropy vanishes for all three solutions like that on the sphere: 
\begin{eqnarray}
S[\rho] &=&  -2 \pi^2\ln\frac{2 \pi^2}{\Gamma_2^{f.g.} - 8 \pi^2 E}+2\pi^2( \ln\pi +1 )  +  {\cal O}(\gamma_4^2).
\end{eqnarray}

\section{The precession of angular momentum on a rotating sphere \label{app2}}

The total angular momentum $\bi{L}$ of a fluid as seen in the co-rotating reference frame of the sphere precesses with angular frequency $-\Omega$ about the $\bi{z}$-axis \cite{Herbert13}:
\begin{eqnarray}
\frac{\rmd L_x}{\rmd t} &=& \Omega L_y,\\
\frac{\rmd L_y}{\rmd t} &=& -\Omega L_x, \label{precessLy}\\
\frac{\rmd L_z}{\rmd t} &=& 0. \label{precessLz}
\end{eqnarray}
This can be seen by decomposing the EOM of vorticity field in terms of spherical harmonics and noting that the dynamics of $\ell =1$ modes are independent of $\ell >1$ modes \cite{Majda}. Since the hyperviscosity term does not act on $\ell = 1$ modes, it does not affect the evolution of angular momentum.  Alternatively the total angular momentum can be expressed in terms of the velocity field; the Navier-Stokes equation in a rotating frame determines the dynamics.  We follow the second route below. There is also a simpler way to derive the precession of angular momentum by realizing that the angular momentum is fixed as constant in the non-rotating reference frame \cite{Herbert13}.
 
The angular momentum vector on a unit sphere is  $\bi{L} = \int \rmd^2 \bi{r} (\bi{r} \times \bi{u})$ where velocity field $\bi{u} = u_{\theta} \hat{\theta} + u_{\phi} \hat{\phi}$. The components of $\bi{L}$ in the Cartesian basis are 
\begin{eqnarray}
L_x &=& \int \rmd^2 \bi{r} (- u_\theta \sin\phi - u_\phi \cos\theta \cos\phi), \label{L1}\\
L_y &=& \int \rmd^2 \bi{r} (u_\theta \cos\phi - u_\phi \cos\theta \sin\phi),\label{L2}\\
L_z &=& \int \rmd^2 \bi{r} (u_\phi \sin\theta). \label{L3}
\end{eqnarray}
Since $u_\theta = -(\partial \psi/\partial \phi)/\sin\theta$ and $u_\phi = \partial \psi/\partial \theta$, where $\psi(\theta, \phi)$ is the streamfunction, these components can be rewritten as
\begin{eqnarray}
L_x &=& -2 \int \rmd^2 \bi{r} \sin\theta \cos\phi ~\psi,\\
L_y &=& -2 \int \rmd^2 \bi{r} \sin\theta \sin\phi ~\psi \label{Ly},\\
L_z &=& -2 \int \rmd^2 \bi{r} \cos\theta ~\psi.
\end{eqnarray}
On the other hand, the 3D EOM for a barotropic incompressible inviscid flow in the rotating frame is 
\begin{eqnarray}
\frac{D \bi{u}_{3D}}{Dt} = -\frac{1}{\rho} \bi{\nabla} p + \Omega^2 \bi{R} -2 \bi{\Omega} \times \bi{u}_{3D},
\end{eqnarray}
where $\bi{R} \equiv \hat{r} r \sin\theta \sin\theta  + {\hat\theta} r \sin\theta \cos\theta$, $\rho$ is the density of the fluid, and $p$ is the pressure. By projection to the surface of the unit sphere using  $\bi{u}_{3D} = r \bi{u} = r (u_\theta \hat{\theta} + u_\phi \hat{\phi})$ and setting $r = 1$, the following equations of motion are obtained:
\begin{eqnarray}
\frac{D u_{\theta}}{Dt} &\equiv& \frac{\partial u_{\theta}}{\partial t} + [u_\theta \frac{\partial}{\partial \theta} + u_\phi \frac{1}{\sin\theta} \frac{\partial}{\partial \phi}]u_\theta 
\nonumber\\ 
&=& -\frac{1}{\rho} \frac{\partial p}{\partial \theta} + \Omega^2 \sin\theta \cos\theta + 2 \Omega \cos\theta u_\phi,\label{2dEOM1}\\
\frac{D u_{\phi}}{Dt} &\equiv& \frac{\partial u_{\phi}}{\partial t} + [u_\theta \frac{\partial}{\partial \theta} + u_\phi \frac{1}{\sin\theta} \frac{\partial}{\partial \phi}]u_\phi 
\nonumber \\
&=& -\frac{1}{\rho \sin\theta} \frac{\partial p}{\partial \phi} - 2 \Omega \cos\theta u_\theta,\label{2dEOM2}
\end{eqnarray}
with a third equation that determines the radial dependence of the pressure:
\begin{eqnarray}
0 &=& -\frac{1}{\rho} \frac{\partial p}{\partial r} + \Omega^2 \sin^2 \theta + 2 \Omega \sin\theta u_\phi.\label{2dEOM3}
\end{eqnarray}

The time evolution of the angular momentum may be found by combining equations \eref{L1} -- \eref{L3} with equations \eref{2dEOM1} and \eref{2dEOM2}.  Focusing first on the $\bi{x}$-component,
\begin{eqnarray}
\frac{\rmd L_x}{\rmd t} &=& \int \rmd^2 \bi{r} \left( - \frac{\partial u_\theta}{\partial t} \sin\phi -\frac{\partial  u_\phi}{\partial t} \cos\theta \cos\phi \right),
\end{eqnarray}
the partial derivatives $\partial u_\theta/\partial t$ and $\partial u_\phi/\partial t$ may be replaced by the material derivatives ${D u_\theta}/{D t}$ and ${D u_\phi}/{D t}$, because $J[\psi, \zeta]$ does not couple to the $\ell = 1$ modes \cite{Majda, Herbert13}.  We obtain:
\begin{eqnarray}
\frac{\rmd L_x}{\rmd t} &=&  \int \rmd^2 \bi{r} [( \frac{1}{\rho} \frac{\partial p}{\partial \theta} - \Omega^2 \sin\theta \cos\theta - 2 \Omega \cos\theta u_\phi)\sin\phi \nonumber\\
&&+ ( \frac{1}{\rho \sin\theta} \frac{\partial p}{\partial \phi} +2 \Omega \cos\theta u_\theta)\cos\theta \cos\phi].
\end{eqnarray}
The pressure terms contribute
\begin{eqnarray}
\int \rmd^2 \bi{r} \left[ \frac{1}{\rho} \left( \sin\phi \frac{\partial p}{\partial \theta} + \frac{\cos\theta \cos\phi}{\sin\theta}  \frac{\partial p}{\partial \phi} \right) \right]=0,
\end{eqnarray}
and the two terms cancel each other after integration by parts.  The term independent of velocity vanishes,
\begin{eqnarray}
\int \rmd^2 \bi{r} (-\Omega^2) \sin\theta \cos\theta \sin\phi = 0,
\end{eqnarray}
because $\int \rmd \phi \sin\phi =0$ and thus the time derivative of $L_x$ is given by
\begin{eqnarray}
\frac{\rmd L_x}{\rmd t} &=&  2 \Omega \int \rmd^2 \bi{r} [-u_{\phi} \cos\theta \sin\phi + u_\theta \cos^2 \theta \cos\phi]\nonumber\\
&=& -2\Omega  \int \rmd^2 \bi{r} \sin\theta \sin\phi ~\psi\nonumber\\
&=& \Omega L_y. 
\end{eqnarray} 
The equations of motion for $L_y$ and $L_z$, equations \eref{precessLy} and \eref{precessLz} are likewise obtained.


\end{document}